\documentclass[submission,copyright,creativecommons]{eptcs}

\usepackage{amsthm}

\usepackage{breakurl}             

\usepackage{graphicx}

\usepackage{url}

\usepackage[cmex10]{amsmath}
\usepackage[english]{babel}
\usepackage[latin1]{inputenc}
\usepackage{dsfont}
\usepackage{amssymb}
\usepackage{amsthm}
\usepackage{longtable}
\usepackage{color}
\usepackage{graphicx}
\usepackage{fontenc}
\usepackage{url}
\usepackage{psfrag}
\usepackage{paralist}
\usepackage{listings}

\usepackage{latexsym,epsfig}

\usepackage{times}
\usepackage[margin=20mm,top=20mm,bottom=30mm]{geometry} 

\newdimen\proofrulebreadth \proofrulebreadth=.05em
\newdimen\proofdotseparation \proofdotseparation=1.25ex
\newdimen\proofrulebaseline \proofrulebaseline=2ex
\newcount\proofdotnumber \proofdotnumber=3
\let\then\relax
\def\hfi{\hskip0pt plus.0001fil}
\mathchardef\squigto="3A3B
\newif\ifinsideprooftree\insideprooftreefalse
\newif\ifonleftofproofrule\onleftofproofrulefalse
\newif\ifproofdots\proofdotsfalse
\newif\ifdoubleproof\doubleprooffalse
\let\wereinproofbit\relax
\newdimen\shortenproofleft
\newdimen\shortenproofright
\newdimen\proofbelowshift
\newbox\proofabove
\newbox\proofbelow
\newbox\proofrulename
\def\shiftproofbelow{\let\next\relax\afterassignment\setshiftproofbelow\dimen0 }
\def\shiftproofbelowneg{\def\next{\multiply\dimen0 by-1 }%
\afterassignment\setshiftproofbelow\dimen0 }
\def\setshiftproofbelow{\next\proofbelowshift=\dimen0 }
\def\setproofrulebreadth{\proofrulebreadth}

\def\prooftree{
\ifnum  \lastpenalty=1
\then   \unpenalty
\else   \onleftofproofrulefalse
\fi
\ifonleftofproofrule
\else   \ifinsideprooftree
        \then   \hskip.5em plus1fil
        \fi
\fi
\bgroup
\setbox\proofbelow=\hbox{}\setbox\proofrulename=\hbox{}%
\let\justifies\proofover\let\leadsto\proofoverdots\let\Justifies\proofoverdbl
\let\using\proofusing\let\[\prooftree
\ifinsideprooftree\let\]\endprooftree\fi
\proofdotsfalse\doubleprooffalse
\let\thickness\setproofrulebreadth
\let\shiftright\shiftproofbelow \let\shift\shiftproofbelow
\let\shiftleft\shiftproofbelowneg
\let\ifwasinsideprooftree\ifinsideprooftree
\insideprooftreetrue
\setbox\proofabove=\hbox\bgroup$\displaystyle 
\let\wereinproofbit\prooftree
\shortenproofleft=0pt \shortenproofright=0pt \proofbelowshift=0pt
\onleftofproofruletrue\penalty1
}

\def\eproofbit{
\ifx    \wereinproofbit\prooftree
\then   \ifcase \lastpenalty
        \then   \shortenproofright=0pt  
        \or     \unpenalty\hfil         
        \or     \unpenalty\unskip       
        \else   \shortenproofright=0pt  
        \fi
\fi
\global\dimen0=\shortenproofleft
\global\dimen1=\shortenproofright
\global\dimen2=\proofrulebreadth
\global\dimen3=\proofbelowshift
\global\dimen4=\proofdotseparation
\global\count255=\proofdotnumber
$\egroup  
\shortenproofleft=\dimen0
\shortenproofright=\dimen1
\proofrulebreadth=\dimen2
\proofbelowshift=\dimen3
\proofdotseparation=\dimen4
\proofdotnumber=\count255
}

\def\proofover{
\eproofbit 
\setbox\proofbelow=\hbox\bgroup 
\let\wereinproofbit\proofover
$\displaystyle
}%
\def\proofoverdbl{
\eproofbit 
\doubleprooftrue
\setbox\proofbelow=\hbox\bgroup 
\let\wereinproofbit\proofoverdbl
$\displaystyle
}%
\def\proofoverdots{
\eproofbit 
\proofdotstrue
\setbox\proofbelow=\hbox\bgroup 
\let\wereinproofbit\proofoverdots
$\displaystyle
}%
\def\proofusing{
\eproofbit 
\setbox\proofrulename=\hbox\bgroup 
\let\wereinproofbit\proofusing
\kern0.3em$
}

\def\endprooftree{
\eproofbit 
  \dimen5 =0pt
\dimen0=\wd\proofabove \advance\dimen0-\shortenproofleft
\advance\dimen0-\shortenproofright
\dimen1=.5\dimen0 \advance\dimen1-.5\wd\proofbelow
\dimen4=\dimen1
\advance\dimen1\proofbelowshift \advance\dimen4-\proofbelowshift
\ifdim  \dimen1<0pt
\then   \advance\shortenproofleft\dimen1
        \advance\dimen0-\dimen1
        \dimen1=0pt
        \ifdim  \shortenproofleft<0pt
        \then   \setbox\proofabove=\hbox{%
                        \kern-\shortenproofleft\unhbox\proofabove}%
                \shortenproofleft=0pt
        \fi
\fi
\ifdim  \dimen4<0pt
\then   \advance\shortenproofright\dimen4
        \advance\dimen0-\dimen4
        \dimen4=0pt
\fi
\ifdim  \shortenproofright<\wd\proofrulename
\then   \shortenproofright=\wd\proofrulename
\fi
\dimen2=\shortenproofleft \advance\dimen2 by\dimen1
\dimen3=\shortenproofright\advance\dimen3 by\dimen4
\ifproofdots
\then
        \dimen6=\shortenproofleft \advance\dimen6 .5\dimen0
        \setbox1=\vbox to\proofdotseparation{\vss\hbox{$\cdot$}\vss}%
        \setbox0=\hbox{%
                \advance\dimen6-.5\wd1
                \kern\dimen6
                $\vcenter to\proofdotnumber\proofdotseparation
                        {\leaders\box1\vfill}$%
                \unhbox\proofrulename}%
\else   \dimen6=\fontdimen22\the\textfont2 
        \dimen7=\dimen6
        \advance\dimen6by.5\proofrulebreadth
        \advance\dimen7by-.5\proofrulebreadth
        \setbox0=\hbox{%
                \kern\shortenproofleft
                \ifdoubleproof
                \then   \hbox to\dimen0{%
                        $\mathsurround0pt\mathord=\mkern-6mu%
                        \cleaders\hbox{$\mkern-2mu=\mkern-2mu$}\hfill
                        \mkern-6mu\mathord=$}%
                \else   \vrule height\dimen6 depth-\dimen7 width\dimen0
                \fi
                \unhbox\proofrulename}%
        \ht0=\dimen6 \dp0=-\dimen7
\fi
\let\doll\relax
\ifwasinsideprooftree
\then   \let\VBOX\vbox
\else   \ifmmode\else$\let\doll=$\fi
        \let\VBOX\vcenter
\fi
\VBOX   {\baselineskip\proofrulebaseline \lineskip.2ex
        \expandafter\lineskiplimit\ifproofdots0ex\else-0.6ex\fi
        \hbox   spread\dimen5   {\hfi\unhbox\proofabove\hfi}%
        \hbox{\box0}%
        \hbox   {\kern\dimen2 \box\proofbelow}}\doll%
\global\dimen2=\dimen2
\global\dimen3=\dimen3
\egroup 
\ifonleftofproofrule
\then   \shortenproofleft=\dimen2
\fi
\shortenproofright=\dimen3
\onleftofproofrulefalse
\ifinsideprooftree
\then   \hskip.5em plus 1fil \penalty2
\fi
}

\newcommand{\leaf}[1]{{\begin{array}{c} #1 \end{array}}}

\newcommand{\namedruletree}[3]{{
     \prooftree 
       {\leaf{#2}}
     \justifies
       #3
     \using \rn{#1}
     \endprooftree
     }}

\newcommand{\andalso}{\quad\quad}

\newcommand{\infrule}[3]{\namedruletree{#1}{#2}{#3}\andalso}

\newcommand{\rn}[1]{\mbox{\textsc{#1}}}        

\newtheorem{theorem}{Theorem}[section]
\newtheorem{lemma}[theorem]{Lemma}
\newtheorem{proposition}[theorem]{Proposition}
\newtheorem{corollary}[theorem]{Corollary}

\newtheorem{definition}[theorem]{Definition}
\newtheorem{example}{Example}[section]
\newtheorem{exercise}{Exercise}[section]

\newcommand\Notation{\vspace{1ex}\noindent\textbf{\itshape Notation:}\hspace{1ex}}

\newcounter{case}

\newcommand\Base{\vspace{1ex}\textit{Base case}: }
\newcommand\Induction{\vspace{1ex}\textit{Inductive step}: }

\newcommand\valuation{\ensuremath{\mathcal{V}}}
\newcommand\labelc{\ensuremath{\lambda}}
\newcommand\labelPotential{\ensuremath{\tilde{\lambda}}}
\newcommand\extends{\ensuremath{\vartriangleright}}
\newcommand\labelH{\ensuremath{l}}
\newcommand\executing{\ensuremath{\frac{1}{2}}}
\newcommand\hystory[1]{\ensuremath{\downarrow\!\!#1}}
\newcommand\concdegup[1]{\ensuremath{|#1|_{uc}}}
\newcommand\concdegdown[1]{\ensuremath{|#1|_{dc}}}
\newcommand\closure[1]{\ensuremath{\mathcal{C}(#1)}}

\newcommand\constraints{\ensuremath{\mathbf{C}}}
\newcommand\startUniv[2]{\ensuremath{[\hspace{-2.5px}\{#1\}\hspace{-2.5px}]#2}}
\newcommand\terminateUniv[2]{\ensuremath{[ #1] #2}}
\newcommand\start[2]{\ensuremath{\{#1\}#2}}
\newcommand\terminate[2]{\ensuremath{\langle #1\rangle #2}}
\newcommand\terminatei[2]{\ensuremath{\langle \rangle^{#1} #2}}
\newcommand\starti[2]{\ensuremath{\{\}^{#1}#2}}
\newcommand\HDML{\ensuremath{\mathit{HDML}}}
\newcommand\HHDML{\ensuremath{\mathit{hHDML}}}
\newcommand\HDA{\ensuremath{\mathit{HDA}}}
\newcommand\HDAs{\ensuremath{\mathit{HDAs}}}
\newcommand\modelH{\ensuremath{\mathcal{H}}}
\newcommand\modelP[1]{\ensuremath{\modelH^{p}_{#1}}}

\newcommand\defequal{\ensuremath{\stackrel{\vartriangle}{=}}}
\newcommand\conflict{\ensuremath{\,\#\,}}
\newcommand\concurrel{\ensuremath{\,co\,}}

\newcommand\atomicformulas{\ensuremath{\Phi_{B}}}

\newcommand\constantprops{\ensuremath{\Phi_{B}}}
\newcommand\reachcell{\ensuremath{\rightarrow^{*}}}
\newcommand\splitpath{\ensuremath{\mathit{split}}}
\newcommand\splitequiv{\ensuremath{\approx_{s}}}
\newcommand\modalequiv{\ensuremath{\stackrel{\HDML}{\sim}}}

\newcommand\descentS{\ensuremath{\stackrel{s}{\rightarrow}}}
\newcommand\descentT{\ensuremath{\stackrel{t}{\leftarrow}}}
\newcommand\descentST{\ensuremath{\stackrel{st}{\longleftrightarrow}}}
\newcommand\descentSTstar{\ensuremath{\stackrel{st}{\longleftrightarrow^{*}}}}
\newcommand\descentSdash{\ensuremath{\stackrel{s}{\dashrightarrow}}}
\newcommand\descentTdash{\ensuremath{\stackrel{t}{\dashleftarrow}}}

\newcommand{\transition}[1]{\ensuremath{\stackrel{#1}{\longrightarrow}}}

\newcommand{\imply}{\ensuremath{\,\rightarrow\,}}
\newcommand{\equivalent}{\ensuremath{\,\leftrightarrow\,}}
\newcommand{\prove}{\ensuremath{\,\vdash}}
\newcommand{\bottom}{\perp}

\newcommand\until{\ensuremath{\, \mathcal{U} \,}}
\newcommand\untilC{\ensuremath{\, \mathcal{U}^{c} \,}}
\newcommand\untilL{\ensuremath{\, \mathcal{U}^{l} \,}}

\newcounter{feature}

\newcounter{axiom}
\renewcommand{\theaxiom}{(A\arabic{axiom})}
\newcommand{\axiom}{\refstepcounter{axiom}\theaxiom{}}
\newcounter{axiomprim}
\renewcommand{\theaxiomprim}{(A\arabic{axiom}')}
\newcommand{\axiomprim}{\refstepcounter{axiomprim}\theaxiomprim{}}
\newcounter{examp}

\newcommand{\refeq}[1]{(\ref{#1})}

\newcommand{\oomit}[1]{}

\newcommand{\cp}[1]{}
\newcommand{\private}[1]{}
\newcommand{\completeness}[1]{#1}

\newcounter{mytablecounter}
\renewcommand{\themytablecounter}{\arabic{mytablecounter}}
\newcommand{\mytableheader}{\par\nopagebreak\vspace{0.01in}\noindent\par\nopagebreak}

               {\mytableheader\noindent%
               \setlength{\parindent}{4em}%
               \setlength{\parskip}{6ex}%
               \refstepcounter{mytablecounter}%
               \textbf{Table \nopagebreak[2]\themytablecounter{}: }}{\par}

\begin{document}





\title{Higher Dimensional Modal Logic}

\author{Cristian Prisacariu
\institute{Dept. of Informatics, University of Oslo, \ -- \ P.O.\ Box 1080 Blindern, N-0316 Oslo, Norway.}
\email{cristi@ifi.uio.no}
}
\def\titlerunning{Higher Dimensional Modal Logic}
\def\authorrunning{C.~Prisacariu
}

%
%

\maketitle

\begin{abstract}

Higher dimensional automata (\HDA) are a model of concurrency that can express most of the traditional partial order models like Mazurkiewicz traces, pomsets, event structures, or Petri nets. Modal logics, interpreted over Kripke structures, are the logics for reasoning about sequential behavior and interleaved concurrency. Modal logic is a well behaved subset of first-order logic; many variants of modal logic are decidable. However, there are no modal-like logics for the more expressive \HDA\ models. In this paper we introduce and investigate a modal logic over \HDAs\ which incorporates two modalities for reasoning about ``during'' and ``after''. We prove that this general higher dimensional modal logic (\HDML) is decidable and we define 
an 
axiomatic system for it. We also show how, when the \HDA\ model is restricted to Kripke structures, a syntactic restriction of \HDML\ becomes the standard modal logic. Then we isolate the class of \HDAs\ that encode Mazurkiewicz traces and show how \HDML, with natural definitions of corresponding \textit{Until} operators, can be restricted to LTrL (the linear time temporal logic over Mazurkiewicz traces) or the branching time ISTL. We also study the expressiveness of the basic \HDML\ language wrt.\ bisimulations and conclude that \HDML\ captures the split-bisimulation.

\end{abstract}




\tableofcontents


\newpage
\section{Introduction}\label{sec:intro}
%



This paper extends \cite{P10concur} by adding all the proofs and some more explanations. Moreover, it corrects some essential errors that appeared in the \completeness{proofs of soundness and completeness of the }axiomatic system of \cite{P10concur}. 
The present paper also adds new results that steam from two comments that this work attracted. We discuss the expressive power of the basic logic wrt.\ bisimulations, concluding that it captures the split-bisimulation. We investigate more carefully the extension of the basic language with the \textit{Until} operator; we define precisely two kinds of \textit{Until}, and we use the LTL-like to encode the LTrL logic and the CTL-like to encode the ISTL logic. 

\textit{Higher dimensional automata} (\HDAs) are a general formalism for modeling concurrent systems \cite{pratt91hda,Glabbeek06HDA}. In this formalism concurrent systems can be modeled at different levels of abstraction, not only as all possible interleavings of their concurrent actions. \HDAs\ can model concurrent systems at any granularity level and make no assumptions about the durations of the actions, i.e., refinement of actions \cite{GlabbeekG01refinement} is well accommodated by \HDAs. Moreover, \HDAs\ are not constrained to only before-after modeling and expose explicitly the choices in the system. It is a known issue in concurrency models that the combination of causality, concurrency, and choice is difficult; in this respect, \HDAs\ and Chu spaces \cite{gupta94phd_chu} do a fairly good job \cite{Pratt03trans_cancel}.

Higher dimensional automata are more expressive than most of the models based on partial orders or on interleavings (e.g., Petri nets and the related Mazurkiewicz traces, or the more general partial order models like pomsets or event structures). Therefore, one only needs to find the right class of \HDAs\ in order to get the desired models of concurrency.

Work has been done on defining temporal logics over Mazurkiewicz traces \cite{MukundT96TrPTL} and strong results like decidability and expressive completeness are known \cite{DiekertG06,ThiagarajanW02LTrL}. For more general partial orders some temporal logics become undecidable \cite{AlurP99undecidable}. For the more expressive event structures there are fewer works; a modal logic is investigated in \cite{mukund90eventStructLogic}.

There is hardly any work on logics for higher dimensional automata \cite{Pratt03trans_cancel} and, as far as we know, there is no work on \emph{modal logics for \HDAs}. In practice, one is more comfortable with modal logics, like temporal logics or dynamic logics, because these are generally decidable (as opposed to full first-order logic, which is undecidable). 

That is why in this paper we introduce and develop a logic in the style of standard modal logic. This logic has \HDAs\ as models, hence, the name \textit{higher dimensional modal logic} (\HDML). 
This is our basic language to talk about general models of concurrent systems. For this basic logic we prove decidability using a form of filtration argument, and we show how compactness fails. Also, we provide an axiomatic system and prove it is sound \completeness{and complete } for the higher dimensional automata. 
\HDML\ in its basic variant is shown to become standard modal logic when the language and the higher dimensional models are restricted in a certain way.

\HDML\ contrasts with standard temporal/modal logics in the fact that \HDML\ can reason about \textit{what holds ``during'' some concurrent events are executing}. 
The close related logic for distributed transition systems of \cite{LodayaPRT95} is in the same style of reasoning only about what holds ``after'' some concurrent events have finished executing. As we show in the examples section, the ``after'' logics can be encoded in \HDML, hence also the logic of \cite{LodayaPRT95}.

\cp{Check the claim above again...}

The other purpose of this work is to provide a general framework for reasoning about concurrent systems at any level of abstraction and granularity, accounting also for choices and independence of actions. Thus, the purpose of the examples in Section~\ref{sec_examples} is to show that studying \HDML, and particular variants of it, is fruitful for analyzing concurrent systems and their logics. 
In this respect we study variants of higher dimensional modal logic inspired by temporal logic and dynamic logic. Already in Section~\ref{subsec_until} we add to the basic language two kinds of \textit{Until} operator, in the style of linear and branching time temporal logics. We show how this variant of \HDML, when interpreted over the class of \HDAs\ corresponding to Kripke structures, can be particularized just by syntactic restrictions to CTL \cite{ClarkeES83CTL}. A second variant, in Section~\ref{sec_partial_orders}, decorates the \HDML\ modalities with labels.
This multi-modal variant of \HDML\ together with the LTL-like \textit{Until} operator, when interpreted over the class of \HDAs\ that encodes Mazurkiewicz traces, becomes LTrL \cite{ThiagarajanW02LTrL} (the linear time temporal logic over Mazurkiewicz traces).



\section{Modal Logic over Higher Dimensional Automata}\label{sec_modal_logic}

In this section we define a higher dimensional automaton (\HDA) following the definition and terminology of \cite{Glabbeek06HDA,Pratt03trans_cancel}. Afterwards we propose \textit{higher dimensional modal logic} (\HDML) for reasoning about concurrent systems modeled as \HDAs. The semantic interpretation of the language is defined in terms of \HDAs\ (i.e., the \HDAs, with a valuation function attached, are the models we propose for \HDML).

\begin{figure}[tp]
\psfrag{q11}{$q_{0}^{1}$}
\psfrag{q12}{$q_{0}^{4}$}
\psfrag{q13}{$q_{0}^{3}$}
\psfrag{q14}{$q_{0}^{2}$}
\psfrag{q21}{$a$}
\psfrag{q22}{$b$}
\psfrag{q23}{$a$}
\psfrag{q24}{$b$}
\psfrag{q3}{$q_{2}$}
  \begin{center}
    \includegraphics[height=3.2cm]{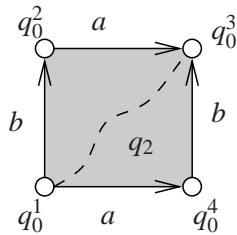}
  \end{center}
\vspace{-2ex}\caption{Example of a \HDA\ with two concurrent events labeled by $a$ and $b$.}
\label{fig_ex_hda}
\end{figure}

For an intuitive understanding of the \HDA\ model consider the standard example \cite{Pratt03trans_cancel,Glabbeek06HDA} pictured in Figure~\ref{fig_ex_hda}. It represents a \HDA\ that models two concurrent events which are labeled by $a$ and $b$ (one might have the same label $a$ for both events). The \HDA\ has four states, $q_{0}^{1}$ to $q_{0}^{4}$, and four transitions between them. This would be the standard picture for interleaving, but in the case of \HDA\ there is also a square $q_{2}$. Traversing through the interior of the square means that both events are executing. When traversing on the lower transition means that event one is executing but event two has not started yet, whereas, when traversing through the upper transition it means that event one is executing and event two has finished already. In the states there is no event executing, in particular, in state $q_{0}^{3}$ both events have finished, whereas in state $q_{0}^{1}$ no event has started yet.

In the same manner, \HDAs\ allow to represent three concurrent events through a cube, or more events through hypercubes. Causality of events is modeled by sticking such hypercubes one after the other. For our example, if we omit the interior of the square (i.e., the grey $q_{2}$ is removed) we are left with a description of a system where there is the choice between two sequences of two events, i.e., $a;b+b;a$.

\begin{definition}[higher dimensional automata]\label{def_hda}
A \emph{cubical set} $H=(Q,\overline{s},\overline{t})$ is formed of a family of sets $Q=\mathop{\bigcup}_{n=0}^{\infty}Q_{n}$ with all sets $Q_{n}$ disjoint, and for each $n$, a family of maps $s_{i}, t_{i}:Q_{n}\rightarrow Q_{n-1}$ with $1\leq i\leq n$ which respect the following \emph{cubical laws}:
\begin{equation}\label{eq_cubic_laws}
\hspace{-1ex}\alpha_{i}\circ\beta_{j}=\beta_{j-1}\circ\alpha_{i},\hspace{2ex}1\!\leq\!i\!<\!j\!\leq\!n\mbox{ and }\alpha,\beta\in\!\{s,t\}.
\end{equation}
In $H$, the $\overline{s}$ and $\overline{t}$ denote the collection of all the maps from all the families (i.e., for all $n$).
A \emph{higher dimensional structure} $(Q,\overline{s},\overline{t},\labelH)$ over an alphabet $\Sigma $ is a cubical set together with a \emph{labeling function} $\labelH:Q_{1}\rightarrow\Sigma $ which respects $\labelH(s_{i}(q))=\labelH(t_{i}(q))$ for all $q\in Q_{2}$ and $i\in\{1,2\}$.\footnote{Later, in Definition~\ref{def_general_labels_cells}, the labeling is extended naturally to all cells.} 
A \emph{higher dimensional automaton} $(Q,\overline{s},\overline{t},l,I,F)$ is a higher dimensional structure with two designated sets of \emph{initial} and \emph{final cells} $I\subseteq Q_{0}$ and $F\subseteq Q_{0}$.
\end{definition}

We call the elements of $Q_{0},Q_{1},Q_{2},Q_{3}$ respectively \textit{states}, \textit{transitions}, \textit{squares}, and \textit{cubes}, whereas the general elements of $Q_{n}$ are called n-dimensional cubes (or hypercubes). We call generically an element of $Q$ a \textit{cell} (also known as n-cell).
For a transition $q\in Q_{1}$ the $s_{1}(q)$ and $t_{1}(q)$ represent respectively its source and its target cells (which are \textit{states} from $Q_{0}$ in this case). Similarly for a general cell $q\in Q_{n}$ there are $n$ source cells and $n$ target cells all of dimension $n-1$. Intuitively, an n-dimensional cell $q$ represents a snapshot of a concurrent system in which $n$ events are performed at the same time, i.e., concurrently. A source cell $s_{i}(q)$ represents the snapshot of the system before the starting of the $i^{th}$ event, whereas the target cell $t_{i}(q)$ represents the snapshot of the system immediately after the termination of the $i^{th}$ event. A transition of $Q_{1}$ represents a snapshot of the system in which a single event is performed.

The cubical laws account for the geometry (concurrency) of the \HDAs; there are four kinds of cubical laws depending on the instantiation of $\alpha$ and $\beta$. For the example of Figure~\ref{fig_ex_hda} consider the cubical law where $\alpha$ is instantiated to $t$ and $\beta$ to $s$, and $i=1$ and $j=2$: $t_{1}(s_{2}(q_{2}))=s_{1}(t_{1}(q_{2}))$. In the left hand side, the second source cell of $q_{2}$  is, in this case, the transition $s_{2}(q_{2})=q_{1}^{1}=(q_{0}^{1},q_{0}^{2})$ and the first target cell of $q_{1}^{1}$ is $q_{0}^{2}$ (the only target cell because $s_{2}(q_{2})\in Q_{1}$); this must be the same cell when taking the right hand side of the cubical law, i.e., the first target cell is $t_{1}(q_{2})=q_{1}^{2}=(q_{0}^{2},q_{0}^{3})$ and the first source of $q_{1}^{2}$ is $q_{0}^{2}$.

We propose the language of \textit{higher dimensional modal logic} for talking about concurrent systems. \HDML\ follows the tradition and style of standard modal languages \cite{01modalLogicBook}.

\begin{definition}[higher dimensional modal logic]\label{def_hdml}
A formula $\varphi$ in \emph{higher dimensional modal logic} is constructed using the grammar below, from a set \atomicformulas\ of atomic propositions, with $\phi\in\atomicformulas$, which are combined using the Boolean symbols 
$\bottom$ and $\imply$ (from which all other standard propositional operations are generated),
 and using the modalities \start{}{}\ and \terminate{}{}.
$$
\begin{array}{rll}
\varphi &\ :=\ & \phi\mid \,\bottom\, \mid \varphi\imply\varphi\mid \start{}{\varphi} \mid \terminate{}{\varphi} 
\end{array}
$$
\end{definition}

We call \start{}{} the \emph{during modality} and \terminate{}{} the \emph{after modality}. The intuitive reading of $\start{}{\varphi}$ is: ``pick some event from the ones currently not running (must exist at least one not running) and start it; in the new configuration of the system (during which, one more event is concurrently executing) the formula $\varphi$ must hold''. The intuitive reading of $\terminate{}{\varphi}$ is: ``pick some event from the ones currently running concurrently (must exist one running) and terminate it; in the new configuration of the system the formula $\varphi$ must hold''. This intuition is formalized in the semantics of \HDML. 

The choice of our notation is biased by the intuitive usage of these modalities where the after modality talks about what happens after some event is terminated; in this respect being similar to the standard diamond modality of dynamic logic. Later, in Section~\ref{sec_partial_orders}, these modalities are decorated with labels. The during modality talks about what happens during the execution of some event and hence we adopt the notation of Pratt \cite{Pratt78exp-time.alg.pdl}.

The models of \HDML\ are higher dimensional structures together with a valuation function $\mathcal{V}:Q\rightarrow 2^{\atomicformulas}$ which associates a set of atomic propositions to each cell (of any dimension). This means that $\valuation$ assigns some propositions to each state of dimension 0, to each transition of dimension 1, to each square of dimension 2, to each cube of dimension 3, etc.
Denote a model of \HDML\ by $\modelH=(Q,\overline{s},\overline{t},l,\mathcal{V})$.
A \HDML\ formula is evaluated in a cell of such a model \modelH. 

One may see the \HDML\ models as divided into \textit{levels}, each level increasing the concurrency complexity of the system; i.e., level $Q_{n}$ increases the complexity compared to level $Q_{n-1}$ by adding one more event (to have $n$ events executing concurrently instead of $n-1$). One can see $Q_{0}$ as having concurrency complexity $0$ because there are no events executing there. The levels are linked together through the $s_{i}$ and $t_{i}$ maps. With this view in mind the during and after modalities should be understood as jumping from one level to the other; the $\start{}{}$ modality jumps one level up, whereas the $\terminate{}{}$ modality jumps one level down.

\begin{definition}[satisfiability]\label{def_satisfiability}
Table~\ref{table_HDMLsemantics} defines recursively the satisfaction relation $\models$ of a formula $\varphi$ wrt.\ a model \modelH\ in a particular n-cell $q$ (for some arbitrary $n$); denote this as $\modelH,q\models\varphi$. The notions of satisfiability and validity are defined as usual.

\begin{table}[t]
\begin{tabular}{@{\hspace{0ex}}r@{\hspace{0.5ex}}c@{\hspace{0.5ex}}l@{\hspace{1ex}}c@{\hspace{1ex}}l}
$\modelH,q$ & $\models$ & $\phi$ & iff & $\phi\in\mathcal{V}(q)$. \\
%
$\modelH,q$ & $\not\models$ & $\bottom$ &  & \\
%
$\modelH,q$ & $\models$ & $\varphi_{1}\imply\varphi_{2}$ & iff & when $\modelH,q\models\varphi_{1}$ then $\modelH,q\models\varphi_{2}$. \\
%
$\modelH,q$ & $\models$ & $\start{}{\varphi}$ & iff & assuming $q\in Q_{n}$ for some $n$,\\
& & &\multicolumn{2}{l}{$\exists q'\in Q_{n+1}\mbox{ s.t.\ } s_{i}(q')=q$ for some $1\leq i\leq n+1$, and $\modelH,q'\models\varphi$.}\\
%
$\modelH,q$ & $\models$ & $\terminate{}{\varphi}$ & iff & assuming $q\in Q_{n}$ for some $n$,\\
& & &\multicolumn{2}{l}{$\exists q'\in Q_{n-1}\mbox{ s.t.\ } t_{i}(q)=q'$ for some $1\leq i\leq n$, and $\modelH,q'\models\varphi$.}\\
\end{tabular}
  \caption{Semantics for \HDML.}
  \label{table_HDMLsemantics}
\end{table}
\end{definition}

Both modalities have an existential flavor.
In particular note that $\modelH,q_{0}\not\models\terminate{}{\varphi}$, for $q_{0}\in Q_{0}$ a state, because there is no event executing in a state, and thus no event can be terminated. Similarly, for the during modality, $\modelH,q_{n}\not\models\start{}{\varphi}$ for any n-cell $q_{n}\in Q_{n}$ when all sets $Q_{k}$, with $n<k$, are empty (i.e., the family of sets $Q$ is bounded by $n$). This says that there can be at most $n$ events running at the same time, and when reaching this limit one cannot start another event and therefore $\start{}{\varphi}$ cannot be satisfied.
 
The universal correspondents of \start{}{} and \terminate{}{} are defined in the usual style of modal logic. We denote these modalities by respectively $\startUniv{}{\varphi}$ and $\terminateUniv{\,}{\varphi}$; eg.\ $\startUniv{}{\varphi}\defequal\neg\start{}{\neg\varphi}$. 
The intuitive reading of $\terminateUniv{\,}{\varphi}$ is: ``pick any of the events currently running concurrently and after terminating it, $\varphi$ must hold in the new configuration of the system''. Note that this modality holds trivially for any state $q_{0}\in Q_{0}$, i.e., $\modelH,q_{0}\models\terminateUniv{\,}{\varphi}$.

In the rest of this section we prove that satisfiability for \HDML\ is decidable using a variation of the filtration technique \cite{01modalLogicBook}. 
Then we give an axiomatic system for \HDML\ and prove its soundness.


\subsection{Decidability of \HDML}

The filtration for the states is the same as in the standard modal logic, but for cells of dimension $1$ or higher we need to take care that the maps $t$ and $s$ in the filtration model remain maps and that they respect the cubical laws so that the filtration is still a \HDML\ model. This can be done, but the filtration model is bigger than what is obtained in the case of standard modal logic. On top, the proof of the small model property (Theorem~\ref{th_small_model}) is more involved due to the complexities of the definition of filtration given in Definition~\ref{def_filtration}.

\begin{definition}[subformula closure]\label{def_closure}
The \emph{subformula closure} of a formula $\varphi$ is the set of formulas $\closure{\varphi}$ defined recursively as:

\begin{tabular}{@{\hspace{0ex}}l@{\hspace{1ex}}c@{\hspace{1ex}}l@{\hspace{0ex}}}
$\closure{\phi}$ & \defequal & $\{\phi\}$, for $\phi\in\atomicformulas$\\
$\closure{\varphi_{1}\imply\varphi_{2}}$ & \defequal & $\{\varphi_{1}\imply\varphi_{2}\}\cup\closure{\varphi_{1}}\cup\closure{\varphi_{2}}$\\
$\closure{\start{}{\varphi}}$ & \defequal & $\{\start{}{\varphi}\}\cup\closure{\varphi}$\\
$\closure{\terminate{}{\varphi}}$ & \defequal & $\{\terminate{}{\varphi}\}\cup\closure{\varphi}$\\
\end{tabular}
\end{definition}

The \emph{size} of a formula (denoted $|\varphi|$) is calculated by summing the number of Boolean and modal symbols with the number of atomic propositions and $\bottom$ symbols that appear in the formula. (All instances of a symbol are counted.)

\begin{proposition}[size of the closure]\label{prop_size_of_closure}
The size of the subformula closure of a formula $\varphi$ is linear in the size of the formula; i.e., $|\closure{\varphi}|\leq |\varphi|$.
\end{proposition}

\begin{proof}
The proof is easy, using structural induction and observing that for the atomic formulas the size of the closure is exactly $1$, the size of the formula. For a compound formula like $\start{}{\varphi}$ the induction hypothesis says that $|\closure{\varphi}|\leq|\varphi|$ which means $1+|\closure{\varphi}|\leq 1+|\varphi|$.
\end{proof}

\begin{definition}[filtration]\label{def_filtration}
Given a formula $\varphi$, we define below a relation $\equiv$ (which is easily proven to be an equivalence relation) over the cells of a higher dimensional structure \modelH, where $q,q'\in Q_{i}$, for some $i\in\mathbb{N}$:
$$
q\equiv q'\mbox{ iff\ \ for any }\psi\in\closure{\varphi}\mbox{ then }(\modelH,q\models\psi\mbox{ iff }\modelH,q'\models\psi).
$$
A \emph{filtration model} $\modelH^{f}$ of some structure $\modelH$ through the closure set $\closure{\varphi}$ is the structure $(Q^{f},s^{f},t^{f},l^{f},\mathcal{V}^{f})$:

\begin{tabular}{@{\hspace{0ex}}l@{\hspace{1ex}}c@{\hspace{1ex}}l@{\hspace{0ex}}}
$Q_{n}^{f}$ & \defequal & $\{[q_{n}]\mid q_{n}\in Q_{n}\}$, where $[q_{n}]$ is\\
\hfill $[q_{0}]$ & \defequal & $\{q'\mid q_{0}\equiv q'\}$ when $q_{0}\in Q_{0}$, otherwise,\\
\hfill $[q_{n}]$ & \defequal & $\{q'\mid q_{n}\equiv q'\wedge t_{i}(q')\in[p_{i}]\wedge s_{i}(q')\in[p'_{i}]$ \\
& & \phantom{$\{q'\mid$} for all $1\leq i\leq n$ and for some fixed $[p_{i}],[p'_{i}]\in Q_{n-1}^{f}\}$.\\
$s_{i}^{f}([q_{n}])$ & \defequal & $[q_{n-1}]$\ \ iff\ \ for all $p\in[q_{n}]$, $s_{i}(p)\in[q_{n-1}]$.\\
$t_{i}^{f}([q_{n}])$ & \defequal & $[q_{n-1}]$\ \ iff\ \ for all $p\in[q_{n}]$, $t_{i}(p)\in[q_{n-1}]$.\\
$\mathcal{V}^{f}([q])$ & \defequal & $\mathcal{V}(q)$.\\
\end{tabular}
\end{definition}

\begin{lemma}\label{lemma_disjoint_equiv_classes}
 Any two sets $[p],[q]\in Q_{n}^{f}$, for some $n\in\mathbb{N}$, are disjoint.
\end{lemma}

\begin{proof}
By induction on $n$.

The base case for $n=0$ is easy as the definition of $Q_{0}^{f}$ results in the equivalence classes on $Q_{0}$ generated by the equivalence relation $\equiv$, which are disjoint.

\Induction{Consider $[p],[q]\in Q_{n}^{f}$, for which we assume that $\exists r\in Q_{n}$ with $r\in[p]$ and $r\in[q]$. From the definition we get (1) $q\equiv r\equiv p$ and, (2) for any $1\leq i\leq n$ and some fixed $[p'_{i}],[q'_{i}]\in Q_{n-1}^{f}$, $t_{i}(r)\in[p'_{i}]$ and $t_{i}(r)\in[q'_{i}]$. By the induction hypothesis we know that $[p'_{i}]$ and $[q'_{i}]$ are disjoint, which, together with (2) before, implies that $[p'_{i}]=[q'_{i}]$ for all $1\leq i\leq n$. Because of this and (1) it implies that $[q]=[p]$. Therefore we have proven that if two sets $[p],[q]\in Q_{n}^{f}$ have a cell in common then they must be the same. (Note that an analogous treatment of $s_{i}$ is needed.)
}
\end{proof}

\begin{lemma}\label{lemma_functions_start_term}~
\begin{enumerate}
 \item The definitions of $s_{i}^{f}$ and $t_{i}^{f}$ are that of \emph{maps} (as required in a higher dimensional structure).
\item The $s_{i}^{f}$ and $t_{i}^{f}$ respect the \emph{cubical laws} of a higher dimensional structure.
\end{enumerate}
\end{lemma}

\begin{proof}
For 1.\ we give the proof only for $t_{i}^{f}$, as the proof for $s_{i}^{f}$ is analogous. We use \textit{reductio ad absurdum} and assume, for some $[q]\in Q_{n}^{f}$, that $t_{i}^{f}([q])=[p]$ and $t_{i}^{f}([q])=[p']$ with $[p]\neq [p']$ and $[p],[p']\in Q_{n-1}^{f}$. From the definition we have that for all $q\in[q]$ both $t_{i}(q)\in[p]$ and $t_{i}(q)\in[p']$. From Lemma~\ref{lemma_disjoint_equiv_classes} we know that $[p]$ and $[p']$ are disjoint and we know that $t_{i}$ is a map (i.e., the outcome is unique), therefore we have the contradiction.

We have thus proven that for some input, $t_{i}^{f}$ returns a unique output. It now remains to show that $t_{i}^{f}$ is a total map; i.e., that for \textit{any} input $[q]\in Q_{n}^{f}$, with $n>0$, it returns some output $t_{i}^{f}([q])=[p]$. Since $[q]$ is not empty then it has at least one $q\in[q]$ and cf.~Definition~\ref{def_filtration}, $t_{i}(q)\in[q']$ for some fixed $[q']\in Q_{n-1}^{f}$. By Definition~\ref{def_filtration}, if there are other $q_{n}\in[q]$ then $t_{i}(q_{n})$ is also part of the fixed $[q']$. Thus, $\forall q_{n}\in[q]:t_{i}(q_{n})\in[q']$ meaning that $[q']$ is the outcome we are looking for $t_{i}^{f}([q])$. The same reasoning goes analogous for $s_{i}^{f}$.

For 2.\ we have to prove, for some arbitrary chosen $[q]\in Q_{n}^{f}$ and for any $1\leq i<j\leq n$ that 

\centerline{$t_{i}^{f}(t_{j}^{f}([q]))=t_{j-1}^{f}(t_{i}^{f}([q]))$.}

\noindent (Note that $t_{i}^{f}$ on the left side is different than the $t_{i}^{f}$ on the right side, as the left one is applied to elements of $Q_{n-1}^{f}$ whereas the right one is applied to elements of $Q_{n}^{f}$.) 
The other three kinds of cubical laws are treated analogous only that one needs to reason with the $s_{i}$ maps too.

Assume, wlog.\ because the opposite assumption would follow analogous reasoning, that $t_{i}^{f}(t_{j}^{f}([q]))=[q_{n-2}]$ with $[q_{n-2}]\in Q_{n-2}^{f}$. This leads to considering that $t_{j}^{f}([q])=[q_{n-1}]$ with $[q_{n-1}]\in Q_{n-1}^{f}$, and $t_{i}^{f}([q_{n-1}])=[q_{n-2}]$. From the definition we have both:\\
(1) $\forall q\in[q]:t_{j}(q)\in[q_{n-1}]$,\\
(2) $\forall q\in[q_{n-1}]:t_{i}(q)\in[q_{n-2}]$.\\
Therefore, from the two we have that \\
(3) $\forall q\in[q]:t_{i}(t_{j}(q))\in[q_{n-2}]$.

We want to prove that $[q_{n-2}]=t_{j-1}^{f}(t_{i}^{f}([q]))$, for which we can assume that $t_{i}^{f}([q])=[q'_{n-1}]$ for some $[q'_{n-1}]\in Q_{n-1}^{f}$. Therefore, it amounts to proving that $t_{j-1}^{f}([q'_{n-1}])=[q_{n-2}]$. For this it is enough to find some $p\in[q'_{n-1}]$ s.t.\ $t_{j-1}(p)\in[q_{n-2}]$, because by the Definition~\ref{def_filtration} (of the $t_{i}$ maps) it means that $\forall p\in[q'_{n-1}]$ it holds that $t_{j-1}(p)\in[q_{n-2}]$, i.e., our desired result.

From the assumption we have that $\forall q\in[q]:t_{i}(q)\in[q'_{n-1}]$. Pick one of these $t_{i}(q)$ and claim this to be the $p\in[q'_{n-1}]$ we are looking for. From the cubical laws for the initial \modelH\ model we know that for any $q\in[q]$, $t_{i}(t_{j}(q))=t_{j-1}(t_{i}(q))=t_{j-1}(p)$. Because of (3) we have that $t_{j-1}(p)\in[q_{n-2}]$, and thus our claim is proven; i.e, $t_{j-1}$ applied to the element $t_{i}(q)$ that we picked from $[q'_{n-1}]$, is in $[q_{n-2}]$.
\end{proof}

\begin{corollary}[filtration is a model]\label{cor_filtration_model}
 The filtration $\modelH^{f}$ of a model \modelH\ through a closure set $\mathcal{C}(\varphi)$ is a higher dimensional structure (i.e., is still a \HDML\ model).
\end{corollary}

\begin{proof}
Essentially, the proof amounts to showing that the definitions of $s_{i}^{f}$ and $t_{i}^{f}$ are that of \emph{maps} and that they respect the \emph{cubical laws} which were done in Lemma~\ref{lemma_functions_start_term}.
\end{proof}

\begin{lemma}[sizes of filtration sets]\label{lemma_size_of_filtration}
 Each set $Q_{n}^{f}$ of the filtration $\modelH^{f}$ obtained in Definition~\ref{def_filtration} has finite size which depends on the size of the formula $\varphi$ used in the filtration; more precisely each $Q_{n}^{f}$ is bounded from above by $2^{|\varphi|\cdot N}$ where $N=n!\cdot \sum_{k=0}^{n}\frac{2^{k}}{(n-k)!}$.
\end{lemma}

\begin{proof}
 The case for $0$ is simple as the number of equivalence classes of $Q_{0}$ can be maximum the number of subsets of the subformula closure $\mathcal{C}(\varphi)$ which is $2^{|\varphi|}$.

The case for $n=1$ is based on the size of $Q_{0}^{f}$. Each of the $2^{|\varphi|}$ equivalence classes in which $Q_{1}^{f}$ can be divided may have infinitely many cells. Any such equivalence class can still be broken into smaller subsets depending on the maps $t_{1}$ and $s_{1}$. Because $t_{1}$ can have outcome in any of the $[q_{0}]\in Q_{0}^{f}$, we get a first split into $2^{|\varphi|}$ subdivisions. For each of these we can still split it into $2^{|\varphi|}$ more subdivisions because of $s_{1}$. We thus get a maximum of $2^{|\varphi|}\cdot(2^{|\varphi|})^{2\cdot 1}$ for $Q_{1}^{f}$.
For the general case of $n$ we need to consider all maps $t_{i},s_{i}$, that means $2\cdot n$ maps. For each of these maps we split the $2^{|\varphi|}$ possible initial equivalence classes according to the size of $O_{n-1}^{f}$. Thus we get a maximum of $2^{|\varphi|}\cdot(|Q_{n-1}^{f}|)^{2\cdot n}$ subdivisions. Calculating this series gives the bound on the size of $Q_{n}^{f}$ as being $2^{|\varphi|\cdot N}$ where $N=n!\cdot \sum_{k=0}^{n}\frac{2^{k}}{(n-k)!}$.
\end{proof}

As a side remark, the size of $O_{n}^{f}$ is more than double exponential in the dimension $n$, but is less than triple exponential. More precisely, for $N$, the sum is bounded from above by $(n+1)\cdot 2^{n}$ which makes $N$ the order of $n!\cdot (n+1)\cdot 2^{n}$. We know that $n!$ grows faster than exponential, but not too fast; more precisely, using Stirling's approximation of $n!$ we have that $lg(n!)=\Theta(n\cdot lg(n))$ making $n!\cdot (n+1)\cdot 2^{n}=(n+1)\cdot 2^{n+lg(n!)}$ of order $(n+1)\cdot 2^{\Theta(n\cdot (lg(n)+1))}$. Therefore, $|O_{n}^{f}|$ is bounded by $2^{|\varphi|\cdot(n+1)\cdot 2^{\Theta(n\cdot (lg(n)+1))}}$ (where we consider $|\varphi|$ to be a constant, and hence, not contributing to the bound).\footnote{This discussion is for $n>0$ because $lg$ is undefined for $0$.}

\begin{lemma}[filtration lemma]\label{lemma_filtration}
 Let $\modelH^{f}$ be the filtration of $\modelH$ through the closure set $\mathcal{C}(\varphi)$, as in Definition~\ref{def_filtration}. For any formula $\psi\in\mathcal{C}(\varphi)$ and any cell $q\in\modelH$, we have $\modelH,q\models\psi$ iff $\modelH^{f},[q]\models\psi$.
\end{lemma}

\begin{proof}
 By induction on the structure of the formula $\psi$.

\Base{
For $\psi=\phi\in\atomicformulas$ is immediate from the definition of $\mathcal{V}^{f}$.
}

\Induction{
The case for \imply\ is straightforward making use of the induction hypothesis because the set $\mathcal{C}(\varphi)$ is closed under subformulas.

Take now $\psi=\terminate{}{\psi'}$ and we prove that $\modelH,q\models\terminate{}{\psi'}$ iff $\modelH^{f},[q]\models\terminate{}{\psi'}$. Considering the \textit{only if} implication we assume that (cf.\ definition of satisfiability from Table~\ref{table_HDMLsemantics}) $\exists q'\in Q_{n-1}:t_{i}(q)=q'\wedge q'\models\psi'$ for some $1\leq i\leq n$, and have to prove that $\exists [p]\in Q_{n-1}^{f}:t_{i}^{f}([q])=[p]\wedge [p]\models\psi'$.
Because $q\in[q]$ and $t_{i}(q)=q'$, using the definition of $[q]$ it implies that for all $q\in[q]$ is that $t_{i}(q)\in[q']$ which, by the definition of $t_{i}^{f}$, implies that $t_{i}^{f}([q])=[q']$. (Thus we have found the $[p]=[q']\in Q_{n-1}^{f}$.) From the induction hypothesis we have that $\modelH,q'\models\psi'$ implies that $\modelH^{f},[q']\models\psi'$. This ends the proof.

Consider now the \textit{if} implication and assume $\exists [p]\in Q_{n-1}^{f}:t_{i}^{f}([q])=[p]\wedge [p]\models\psi'$ for some $1\leq i\leq n$. From the definition of $t_{i}^{f}$ we have that $t_{i}(q)\in[p]$; which is the same as picking some $p'\in[p]$ with $t_{i}(q)=p'$. From the induction hypothesis we know that $\modelH^{f},[p]\models\psi'$ iff $\modelH,p\models\psi'$ for any $p\in[p]$ (in particular $\modelH,p'\models\psi'$). Thus $\exists p'\in Q_{n-1}:t_{i}(q)=p'\wedge \modelH,p'\models\psi'$ for some $1\leq i\leq n$, finishing the proof.

When we take $\psi=\start{}{\psi'}$ we use analogous arguments as in the proof of $\terminate{}{\psi'}$. In this case we work with the definition of $s_{i}^{f}$ and we look for cells of higher dimension (instead of lower dimension).
}
\end{proof}

We define two \emph{degrees of concurrency} of a formula $\varphi$: the \emph{upwards concurrency} (denoted $\concdegup{\varphi}$) and \emph{downwards concurrency} (denoted $\concdegdown{\varphi}$). The degree of upwards concurrency counts the maximum number of nestings of the during modality \start{}{} that are not compensated by a \terminate{}{} modality. (E.g., the formula $\start{}{\start{}{\phi}}\vee\start{}{\phi'}$ has the degree of upwards concurrency equal to $2$, the same as $\start{}{\terminate{}{\start{}{\start{}{\phi}}}}$.) The formal definition of \concdegup{\ } is:

\begin{tabular}{@{\hspace{0ex}}l@{\hspace{1ex}}c@{\hspace{1ex}}l@{\hspace{0ex}}}
$\concdegup{\!\bottom\!}$\hspace{1ex}\defequal\hspace{1ex}$\concdegup{\phi}$ & \defequal & $0$, for $\phi\in\atomicformulas$\\
$\concdegup{\varphi_{1}\imply\varphi_{2}}$ & \defequal & $max(\concdegup{\varphi_{1}},\concdegup{\varphi_{2}})$\\
$\concdegup{\start{}{\varphi}}$ & \defequal & $1+\concdegup{\varphi}$\\
$\concdegup{\terminate{}{\varphi}}$ & \defequal & $max(0,\concdegup{\varphi}-1)$\\
\end{tabular}\vspace{1ex}

The definition of the degree of downwards concurrency \concdegdown{\ } is symmetric to the one above in the two modalities; i.e., interchange the modalities in the last two lines.
Note that $\concdegup{\varphi}+\concdegdown{\varphi}\leq|\varphi|$. The next result offers a safe reduction of a model where we remove all cells which have dimension greater than some constant depending on the formula of interest.

\begin{lemma}[concurrency boundedness]\label{lemma_conc_bound}
 If a \HDML\ formula $\varphi$ is satisfiable, $\modelH,q\models\varphi$ with $q\in Q_{k}$, then it exists a model with all the sets $Q_{m}$, with $m>\concdegup{\varphi}+k$, empty, which satisfies the formula.
\end{lemma}

\begin{proof}
By induction on the structure of the formula $\varphi$.

\Base For $\phi\in\constantprops$ and $\bottom$ the evaluation is in the same cell $q$ and thus all the cells of dimension higher than $k$ are not important and can be empty.

\Induction For $\varphi_{1}\imply\varphi_{2}$ the semantics says that whenever $\modelH,q\models\varphi_{1}$ then $\modelH,q\models\varphi_{2}$. From the induction hypothesis we have that all cells of dimension greater than $k+\concdegup{\varphi_{1}}$ (respectively $k+\concdegup{\varphi_{2}}$) are not important for checking $\varphi_{1}$ (respectively $\varphi_{2}$). Thus it is a safe approximation to consider all the cells of at most dimension $max(k+\concdegup{\varphi_{1}},k+\concdegup{\varphi_{2}})=k+\concdegup{\varphi_{1}\imply\varphi_{2}}$ and all sets $Q_{m}$ of greater dimension can be empty.

For $\start{}{\varphi}$ the semantics says that we need to check the formula $\varphi$ in cells of dimension one greater, i.e., $q_{k+1}\models\varphi$. From the induction hypothesis we know that for checking $q_{k+1}\models\varphi$ it is enough to have only cells of most dimension $k+1+\concdegup{\varphi}=k+\concdegup{\start{}{\varphi}}$ (where all other cells can be removed).

For $\terminate{}{\varphi}$ the semantics says that we need to check $q_{k-1}\models\varphi$, that is, in cells of immediately lower dimension. For this, the induction hypothesis says that we need to consider cells of dimension at most $k-1+\concdegup{\varphi}$ which is the same as $k+(\concdegup{\varphi}-1)$. When $\concdegup{\varphi}=0$ then $k$ is a safe approximation and from the definition of the $\concdegup{\,}$ it is the same as $k+\concdegup{\terminate{}{\varphi}}$. Otherwise, when $\concdegup{\varphi}>0$, the definition of $\concdegup{\,}$ tells us that $k+(\concdegup{\varphi}-1)$ is exactly $k+\concdegup{\terminate{}{\varphi}}$.
\end{proof}

\Notation{
The formula $\terminate{}{\phi}\wedge\terminate{}{\neg\phi}$ expresses that there can be terminated at least two different events (in other words, the cell in which the formula is evaluated to true has dimension at least two). Similarly the formula $\terminate{}{(\phi\wedge\neg\phi')}\wedge\terminate{}{(\neg\phi\wedge\neg\phi')}\wedge\terminate{}{(\neg\phi\wedge\phi')}$ says that there are at least three events that can be terminated. For each $i\in\mathbb{N^{*}}$ one can write such a formula to say that there are at least $i$ events that can be terminated. Denote such a formula by $\terminate{}{i}$. Also define $\terminatei{i}{\varphi}$ as $i$ applications of the \terminate{}{} modality to $\varphi$ (i.e., $\terminate{}{\dots\terminate{}{\varphi}}$ where $\terminate{}{}$ appears $i$ times). Similar, for the during modality denote $\start{}{i}$ the formula that can start $i$ different events, and by $\starti{i}{\varphi}$ the $i$ applications of $\start{}{}$ to $\varphi$.
}

\begin{theorem}[small model property]\label{th_small_model}
 If a \HDML\ formula $\varphi$ is satisfiable then it is satisfiable on a finite model with no more than $\mathop{\sum_{n=0}^{|\varphi|}}2^{|\varphi|\cdot N}$ cells where $N=n!\cdot \sum_{k=0}^{n}\frac{2^{k}}{(n-k)!}$ .
\end{theorem}

\begin{proof}
%
%
Assume that there exists a model \modelH\ and a cell $q_{l}\in Q_{l}$ in this model for which $\modelH,q_{l}\models\varphi$. We can prove that there exists a (maybe different) model $\modelH'$ and a cell $q'_{l}$ that satisfy $\varphi$ but which $l < |\varphi|-\concdegup{\varphi}$. We do this by induction on the structure of $\varphi$.

\Base when $\varphi=\phi\in\atomicformulas$. The semantics needs to look only at the valuations, and by the assumption, the valuation of $q_{l}$ in \modelH\ satisfies $\varphi$. Hence we can just use one cell model where we attach this satisfying valuation to it. Therefore level $Q_{0}$ is enough; hence $l=0 < |\phi| - \concdegup{\phi}=1-0$.

\Induction when $\varphi=\varphi_{1}\imply\varphi_{2}$. By the semantics it means that whenever $\varphi_{1}$ is satisfied in $q_{l}$ also $\varphi_{2}$ is. But by the induction hypothesis it means that $l < |\varphi_{1}| - \concdegup{\varphi_{1}}$ and also $l < |\varphi_{2}| - \concdegup{\varphi_{2}}$. Therefore it is a safe approximation to take $l$ to be the maximum of the two: $l < max(|\varphi_{1}| - \concdegup{\varphi_{1}},|\varphi_{2}| - \concdegup{\varphi_{2}})$. We have to show that $l < |\varphi| - \concdegup{\varphi}$ and we do this by showing that $max(|\varphi_{1}| - \concdegup{\varphi_{1}},|\varphi_{2}| - \concdegup{\varphi_{2}}) < |\varphi| - \concdegup{\varphi}$. By expanding the definition on the right we get the inequality $max(|\varphi_{1}| - \concdegup{\varphi_{1}},|\varphi_{2}| - \concdegup{\varphi_{2}}) < |\varphi_{1}| + |\varphi_{2}| + 1 - max(\concdegup{\varphi_{1}},\concdegup{\varphi_{2}})$. This amounts to showing that $max(|\varphi_{1}| - \concdegup{\varphi_{1}},|\varphi_{2}| - \concdegup{\varphi_{2}}) + max(\concdegup{\varphi_{1}},\concdegup{\varphi_{2}}) < |\varphi_{1}| + |\varphi_{2}| + 1$. Denote the quantity $|\varphi_{1}| - \concdegup{\varphi_{1}} = A$ and $|\varphi_{2}| - \concdegup{\varphi_{2}} = B$ and hence have $|\varphi_{1}| = A + \concdegup{\varphi_{1}}$ and $|\varphi_{2}| = B + \concdegup{\varphi_{2}}$. Thus the inequality translates to $max(A,B) + max(\concdegup{\varphi_{1}},\concdegup{\varphi_{2}}) < A + \concdegup{\varphi_{1}} + B + \concdegup{\varphi_{2}} + 1$. Since both $A$ and $B$ (also the other quantities in the inequality) are positive the result is obvious as $max(A,B) < A + B$ (as being one of the summands) and $max(\concdegup{\varphi_{1}},\concdegup{\varphi_{2}}) < \concdegup{\varphi_{1}} + \concdegup{\varphi_{2}}$.

When $\varphi=\start{}{\varphi_{1}}$ the semantics says that exists $q_{l+1}\in Q_{l+1}$ where $\varphi_{1}$ holds. The inductive hypothesis says that $l+1 < |\varphi_{1}|-\concdegup{\varphi_{1}}$. This means that $l < |\varphi_{1}|-\concdegup{\varphi_{1}} - 1 = |\varphi_{1}|-\concdegup{\start{}{\varphi_{1}}} < |\varphi_{1}| + 1 -\concdegup{\start{}{\varphi_{1}}} = |\start{}{\varphi_{1}}|-\concdegup{\start{}{\varphi_{1}}}$.

When $\varphi=\terminate{}{\varphi_{1}}$ the semantics says that exists $q_{n-1}\in Q_{l-1}$ where $\varphi_{1}$ holds. From the inductive hypothesis we have $l-1 < |\varphi_{1}|-\concdegup{\varphi_{1}}$. This means that $l < |\varphi_{1}| + 1 -\concdegup{\varphi_{1}} = |\terminate{}{\varphi_{1}}|-\concdegup{\varphi_{1}}$. Because $max(0,\concdegup{\varphi_{1}}-1) < \concdegup{\varphi_{1}}$ it means that $|\terminate{}{\varphi_{1}}|-\concdegup{\varphi_{1}} < |\terminate{}{\varphi_{1}}| - max(0,\concdegup{\varphi_{1}}-1)$ hence $l < |\terminate{}{\varphi_{1}}| - \concdegup{\terminate{}{\varphi_{1}}}$.

From the above we can safely assume $l = |\varphi|-\concdegup{\varphi}$. 

From Lemma~\ref{lemma_conc_bound} we know that we need to consider only the sets $Q_{n}$ for $n\leq l+\concdegup{\varphi}= |\varphi|$, and all other sets of $Q$ are empty. From Lemma~\ref{lemma_filtration} we know that we can build a filtration model $\modelH^{f}$ s.t.\ the formula $\varphi$ is still satisfiable and, by Lemma~\ref{lemma_size_of_filtration}, we know that all the sets $Q_{n}^{f}$ have a finite number of cells. Thus we are safe if we sum up all the cells in all the $Q_{n}^{f}$, with $n\leq |\varphi|$.
\end{proof}

\begin{corollary}[decidability]\label{corr_decidability}
 Deciding the satisfiability of a \HDML\ formula $\varphi$ is done in space at most $\mathop{\sum_{n=0}^{|\varphi|}}2^{|\varphi|\cdot N}$ where $N$ is defined in Theorem~\ref{th_small_model}.
\end{corollary}


\subsection{Axiomatic system for \HDML}

\begin{table}[t]
\begin{tabular}{@{\hspace{0ex}}l@{\hspace{1ex}}l@{\hspace{2ex}}l@{\hspace{1ex}}l@{\hspace{0ex}}}
\multicolumn{4}{l}{\textbf{Axiom schemes:}}\\
\multicolumn{4}{l}{\axiom\label{ax_proptaut} \hspace{1ex} All instances of propositional tautologies.}  \\
\axiom\label{ax_modal1} & $\start{}{\bot}\equivalent\bottom$ &
\axiomprim\label{ax_modal11} & $\terminate{}{\bot}\equivalent\bottom$  \\
\axiom\label{ax_modal2} & $\start{}{(\varphi\vee\varphi')}\equivalent\start{}{\varphi}\vee\start{}{\varphi'}$  &
\axiomprim\label{ax_modal21} & $\terminate{}{(\varphi\vee\varphi')}\equivalent\terminate{}{\varphi}\vee\terminate{}{\varphi'}$  \\
\vspace{1ex}\axiom\label{ax_modal3} & $\startUniv{}{\varphi}\equivalent\neg\start{}{\neg\varphi}$  &
\axiomprim\label{ax_modal31} & $\terminateUniv{\,}{\varphi}\equivalent\neg\terminate{}{\neg\varphi}$ \\
\axiom\label{ax_HDML1} & $\terminate{}{i}\imply \terminatei{i}{\top}$\hspace{1ex} $\forall i\in\mathbb{N^{*}}$ &
\\
\axiom\label{ax_HDML2} & $\terminatei{2}{\top}\imply(\terminate{}{\terminateUniv{\,}{\varphi}}\imply\terminateUniv{\,}{\terminate{}{\varphi}})$ &
\\
\axiom\label{ax_HDML3} & $\start{}{\terminateUniv{\,}{\varphi}}\imply\terminateUniv{\,}{\start{}{\varphi}}$ &
\axiomprim\label{ax_HDML31} & $\terminate{}{\startUniv{}{\varphi}}\imply\startUniv{}{\terminate{}{\varphi}}$ \\
\axiom\label{ax_HDML4} & $\start{}{\terminatei{i}{\top}}\!\!\imply\!\startUniv{}{\terminatei{i}{\top}}$\hspace{1ex} $\forall i\in\mathbb{N}$ &
\axiomprim\label{ax_HDML41} & $\terminate{}{\terminatei{i}{\top}}\!\!\imply\!\terminateUniv{\,}{\terminatei{i}{\top}}$\hspace{1ex} $\forall i\in\mathbb{N}$ \\
\axiom\label{ax_HDML5} & $\terminatei{i}{\top}\!\!\imply\!\startUniv{}{\terminate{}{\terminatei{i}{\top}}}$\hspace{1ex} $\forall i\in\mathbb{N}$ &
\axiomprim\label{ax_HDML51} & $\start{}{\terminate{}{\terminatei{i}{\top}}}\!\!\imply\!\terminatei{i}{\top}$\hspace{1ex} $\forall i\in\mathbb{N}$ \vspace{1ex}\\
\axiom\label{ax_HDML6} & $\start{}{\start{}{\terminate{}{\varphi}}}\imply\start{}{\terminate{}{\start{}{\varphi}}}$\hspace{1.6ex} &
\axiomprim\label{ax_HDML61} & $\start{}{\terminate{}{\terminate{}{\varphi}}}\imply\terminate{}{\start{}{\terminate{}{\varphi}}}$ \vspace{1ex}\\
\multicolumn{4}{l}{\textbf{Inference rules:}}\\
(R1) & 
\infrule{\hspace{1ex}(MP)}{
    \varphi
    \andalso
    \varphi\imply\varphi'
    }{
      \varphi'
    }
\\
(R2) &
\infrule{\hspace{1ex}(D)}{
    \varphi\imply\varphi'
    }{
      \start{}{\varphi}\imply\start{}{\varphi'}
    }
 &
(R2') &
\infrule{\hspace{1ex}(D')}{
    \varphi\imply\varphi'
    }{
      \terminate{}{\varphi}\imply\terminate{}{\varphi'}
    }\vspace{2ex}
 \\
(R3) & Uniform variable substitution.\\
\end{tabular}
  \caption{Axiomatic system for \HDML.}
  \label{table_HDMLaxioms}
\end{table}

In the following we give an axiomatic system for \HDML\ and prove it to be sound. This system corrects the one in \cite{P10concur}. 
In Table~\ref{table_HDMLaxioms} we give a set of axioms and rules of inference for \HDML. If a formula is \textit{derivable} in this axiomatic system we write $\prove\varphi$. We say that a formula $\varphi$ is derivable from a set of formulas $S$ iff $\prove\psi_{1}\wedge\dots\wedge\psi_{n}\imply\varphi$ for some $\psi_{1},\dots,\psi_{n}\in S$ (we write equivalently $S\prove\varphi$). A set of formulas $S$ is said to be \textit{consistent} if $S\not\prove\bottom$, otherwise it is said to be \textit{inconsistent}. A consistent set $S$ is called \textit{maximal} iff all sets $S'$, with $S\subset S'$, are inconsistent.

\begin{proposition}[theorems]\label{prop_theorems}
The following are derivable in the axiomatic system of Table~\ref{table_HDMLaxioms}:

\begin{align}
\setcounter{equation}{0}
&\prove\start{}{(\varphi\imply\varphi')}\imply(\start{}{\varphi}\imply\start{}{\varphi'})\label{eq_th1}\\
&\prove\terminate{}{(\varphi\imply\varphi')}\imply(\terminate{}{\varphi}\imply\terminate{}{\varphi'})\label{eq_th2}\\
&\prove\terminatei{2}{\top}\imply(\terminate{}{\terminateUniv{\,}{\varphi}}\wedge\terminate{}{\terminateUniv{\,}{\neg\varphi}})\imply\bottom\label{eq_th3}\\
&\prove(\terminate{}{\terminate{}{\varphi}} \wedge\terminate{}{\terminateUniv{\,}{\neg\varphi}})\imply \terminatei{3}{\top}\label{eq_th18}\\
&\prove\terminateUniv{\,}{\terminateUniv{\,}{\!\bottom}}\imply(\terminate{}{\varphi}\imply\terminateUniv{\,}{\varphi})\label{eq_th17}\\
&\prove\terminate{}{\top}\imply(\start{}{\terminateUniv{\,}{\varphi}}\imply\terminate{}{\start{}{\varphi}})\label{eq_th6}\\
&\prove\start{}{\top}\imply(\terminate{}{\startUniv{}{\varphi}}\imply\start{}{\terminate{}{\varphi}})\label{eq_th7}\\
&\prove\startUniv{}{\terminate{}{\top}}\label{eq_th8}\\
&\prove\terminate{}{\startUniv{}{\bot}}\imply\startUniv{}{\bot}\label{eq_th4}\\
&\prove\start{}{\top}\imply\terminateUniv{\,}{\start{}{\top}}\label{eq_th5}\\
&\prove\start{}{\top}\wedge\terminate{}{\top}\imply\terminate{}{\start{}{\top}}\label{eq_th9}\\
&\prove\terminate{}{\top}\imply(\start{}{\terminate{}{\top}}\imply\terminate{}{\start{}{\top}})\label{eq_th10}\\
&\prove\start{}{(\terminate{}{\phi}\wedge\terminate{}{\neg\phi})}\imply(\terminate{}{\start{}{\phi}}\vee\terminate{}{\start{}{\neg\phi}})\label{eq_th11}\\
&\prove\start{}{\start{}{\terminate{}{\terminate{}{\phi}}}}\imply\start{}{\terminate{}{\start{}{\terminate{}{\phi}}}}\label{eq_th12}\\
&\prove\start{}{\start{}{\start{}{\terminate{}{\phi}}}}\imply\start{}{\terminate{}{\start{}{\start{}{\phi}}}}\label{eq_th13}\\
&\prove\start{}{\start{}{\terminate{}{\start{}{\phi}}}}\imply\start{}{\terminate{}{\start{}{\start{}{\phi}}}}\label{eq_th14}\\
&\prove\startUniv{}{\terminateUniv{\,}{\startUniv{}{\phi}}}\imply\startUniv{}{\startUniv{}{\terminateUniv{\,}{\phi}}}\label{eq_th15}\\
&\prove\terminateUniv{\,}{\startUniv{}{\terminateUniv{\,}{\phi}}}\imply\startUniv{}{\terminateUniv{\,}{\terminateUniv{\,}{\phi}}}\label{eq_th16}
\end{align}

Moreover, one can use the following derived rules:

\begin{tabular}{@{\hspace{0ex}}l@{\hspace{2ex}}l@{\hspace{0ex}}}
$\infrule{,}{
    \varphi
    }{
      \startUniv{}{\varphi}
    }$
 &
$\infrule{,}{
    \varphi
    }{
      \terminateUniv{\,}{\varphi}
    }$
\vspace{1ex}\\
$\infrule{,}{
    \varphi\imply\varphi'
    }{
      \startUniv{}{\varphi}\imply\startUniv{}{\varphi'}
    }$
 & 
$\infrule{.}{
    \varphi\imply\varphi'
    }{
      \terminateUniv{\,}{\varphi}\imply\terminateUniv{\,}{\varphi'}
    }$
\vspace{1ex}\\
\end{tabular}
\end{proposition}

\begin{proof}
The first two theorems are derivable as in standard modal logic only using the standard axioms \ref{ax_modal1}-\ref{ax_modal21}. The derived rules are also as in standard modal logic. The theorem \refeq{eq_th3} is a consequence of \ref{ax_HDML2}: $\terminatei{2}{\top}\imply\terminate{}{\terminateUniv{\,}{\varphi}}\wedge\terminate{}{\terminateUniv{\,}{\neg\varphi}}\stackrel{\ref{ax_HDML2}}{\imply}\terminatei{2}{\top}\imply\terminateUniv{\,}{\terminate{}{\varphi}}\wedge\terminate{}{\terminateUniv{\,}{\neg\varphi}}\stackrel{SML}{\imply}\terminate{}{(\terminate{}{\varphi}\wedge\terminateUniv{\,}{\neg\varphi})}\stackrel{SML}{\imply}\terminate{}{\terminate{}{(\varphi\wedge\neg\varphi)}}\stackrel{\ref{ax_modal11}}{\imply}\bot$\,.
The theorem \refeq{eq_th17} uses the contrapositive of axiom~\ref{ax_HDML1}: $\terminateUniv{\,}{\terminateUniv{\,}{\!\bottom}}\equivalent\neg\terminate{}{\terminate{}{\top}}\imply\neg(\terminatei{}{2})\equivalent\neg(\terminate{}{\varphi}\wedge\terminate{}{\neg\varphi})\equivalent (\terminate{}{\varphi}\imply\terminateUniv{\,}{\varphi})$.
The theorem \refeq{eq_th18} uses axiom~\ref{ax_HDML2}.
The theorem \refeq{eq_th6} is a consequence of \ref{ax_HDML3}: from propositional reasoning we have $\terminate{}{\top}\imply(\start{}{\terminateUniv{\,}{\varphi}}\imply\terminate{}{\start{}{\varphi}})\equivalent(\start{}{\terminateUniv{\,}{\varphi}}\wedge\terminate{}{\top}\imply\terminate{}{\start{}{\varphi}})$, and using \ref{ax_HDML3} we have  $\start{}{\terminateUniv{\,}{\varphi}}\wedge\terminate{}{\top}\stackrel{\ref{ax_HDML3}}{\imply}\terminateUniv{\,}{\start{}{\varphi}}\wedge\terminate{}{\top}\imply\terminate{}{\start{}{\varphi}}$. The theorem \refeq{eq_th7} is derivable in an analogous way as the one above only that we use axiom \ref{ax_HDML31}. The theorem \refeq{eq_th8} is just the instantiation of axiom~\ref{ax_HDML5} when $i=0$ (i.e., $\terminatei{0}{\top}\defequal\top$).
The theorem \refeq{eq_th4} is a consequence of \ref{ax_HDML31}: $\terminate{}{\startUniv{}{\bot}}\stackrel{\ref{ax_HDML31}}{\imply}\startUniv{}{\terminate{}{\bot}}\stackrel{\ref{ax_modal11}}{\imply}\startUniv{}{\bot}$. The theorem \refeq{eq_th5} is a consequence of the theorem \refeq{eq_th4}  by contraposition. The theorem \refeq{eq_th9} is derivable from theorem \refeq{eq_th8}. The theorem \refeq{eq_th10} is derivable from theorem \refeq{eq_th9}. The theorem \refeq{eq_th11} is derivable from theorem \refeq{eq_th9} after using axiom \ref{ax_HDML1} and axiom \ref{ax_HDML51} instantiate to $i=1$: $\start{}{(\terminate{}{\phi}\wedge\terminate{}{\neg\phi})}\imply\start{}{\terminate{}{2}}\stackrel{\ref{ax_HDML1}}{\imply}\start{}{\terminatei{2}{\top}}\equiv\start{}{\top}\wedge\start{}{\terminatei{2}{\top}}\stackrel{\ref{ax_HDML51}}{\imply}\start{}{\top}\wedge\terminate{}{\top}\stackrel{\refeq{eq_th9}}{\imply}\terminate{}{\start{}{\top}}\stackrel{prop}{\imply}\terminate{}{\start{}{(\phi\vee\neg\phi)}}\stackrel{SML}{\imply}\terminate{}{(\start{}{(\phi)}\vee\start{}{(\neg\phi)})}\stackrel{SML}{\imply}\terminate{}{\start{}{(\phi)}}\vee\terminate{}{\start{}{(\neg\phi)}}$. Theorem \refeq{eq_th12} follows either from axiom~\ref{ax_HDML6} by the D' rule or from axiom~\ref{ax_HDML61} by the D rule.  Theorem \refeq{eq_th14} is an instantiation of axiom~\ref{ax_HDML6}. Theorem \refeq{eq_th13} needs twice the application of axiom~\ref{ax_HDML6} and the D rule. We need here the application of the axiom two times because we move the $\terminate{}{}$ modality two times over $\start{}{}$, whereas for the other theorems we move the modality only once. The theorems \refeq{eq_th15} and \refeq{eq_th16} are just the contrapositives of axioms \ref{ax_HDML6} respectively \ref{ax_HDML61}.
\end{proof}

\begin{exercise}
A challenge is to prove the validity of:
\[
\terminate{}{(p \wedge \terminateUniv{\,}{\neg p})}
\wedge
\terminate{}{(\neg p \wedge \terminateUniv{\,}{\neg p})}
\wedge
\terminate{}{\terminate{}{p}}
\imply
\terminatei{4}{\top}
\]
This challenge is related to theorem \ref{prop_theorems}.\refeq{eq_th18}. A general version of this challenge should be possible, where one can deduce $\terminatei{i}{\top}$ from $\terminate{}{\terminate{}{p}}$ and $i-1$ distinct formulas $\terminate{}{(\phi_{i} \wedge \terminateUniv{\,}{\neg p})}$ which contradict on the $\phi_{i}$ components.
\end{exercise}

Before proving soundness we should have some intuition about the non-standard axioms \ref{ax_HDML1} to \ref{ax_HDML61}. First consider the axioms \ref{ax_HDML2} to \ref{ax_HDML31} which relate to the cubical laws. 
\begin{itemize}
\item Axiom~\ref{ax_HDML2} embodies the cubical law $t_{i}(t_{j}(q))=t_{j-1}(t_{i}(q))$ (i.e., the cubical law where $\alpha$ is instantiated to $t$ and $\beta$ to $t$). This axiom is to be checked only for cell of dimension 2 or higher (i.e., $\terminatei{2}{\top}$ holds).

\item The two axioms \ref{ax_HDML3} and \ref{ax_HDML31} relate to the cubical laws where $\alpha$ and $\beta$ are instantiated differently, one to $s$ and the other to $t$; e.g., $s_{i}(t_{j}(q))=t_{j-1}(s_{i}(q))$. We included both axioms \ref{ax_HDML3} and \ref{ax_HDML31} for symmetry reasons, but it is clear that one can be obtained from the other by contraposition.
\end{itemize}

The other axioms talk about the dimensions of the cells and about the division of the cells into layers $Q_{n}$.
\begin{itemize}
\item Axiom~\ref{ax_HDML1} $\terminate{}{i}\imply\terminatei{i}{\top}$ says that if in a cell there can be terminated at least $i$ different events then this means that this cell has dimension at least $i$ (i.e., one can go $i$ levels down by $\terminatei{i}{\top}$). This is natural because the dimension of a cell is given by the number of events that are currently executing concurrently.
\item Axiom~\ref{ax_HDML5} $\terminatei{i}{\top}\imply\startUniv{}{\terminate{}{\terminatei{i}{\top}}}$ has two purposes. In the basic variant (for $i=0$ it becomes $\startUniv{}{\terminate{}{\top}}$) it says that in any cell, however one starts an event then one can also terminate an event. In the general form the axiom says that from some level $i$ when going one level up (by starting an event) and then one level down (by terminating an event) we always end up on the same level $i$; i.e., we end in a cell of the same dimension like the cell that it started in. Axiom~\ref{ax_HDML51} intuitively finds out the level of the current cell. If one can start and then can terminate an event in a cell of at least dimension $i$ then the current cell also has dimension at least $i$.
\item Axiom~\ref{ax_HDML4} intuitively says that if from a cell we can start an event and reach a cell of some concurrency complexity (given by the $\terminatei{i}{\top}$) then any way of starting an event from this cell ends up in cells of the same complexity. Though similar in nature, axiom~\ref{ax_HDML41} can be seen intuitively as saying that if one $t$ map of the current cell ends up in a cell of dimension at least $i$ then all the $t$ maps end up in the same dimension. These two axioms relate with the part of the definition of the \HDA\ where all the $s_{i}$ and $t_{i}$ maps for some $n$ are defined on the same domain and codomain.
\item Axioms \ref{ax_HDML6} and \ref{ax_HDML61} are somehow related to the notion of homotopy (see eg.~\cite[ch.7.4]{Glabbeek06HDA}) or to the ways one can walk (i.e., the \textit{paths} on a \HDA, to be defined later) on the \HDAs\ using the \HDML\ modalities (or in other terms, these axioms are related to the histories of an event). One may reach a cell from another cell in a \HDA\ in different ways and the notion of homotopy says that all these ways are considered equivalent. Take the example of the square (cell of dimension 2) from Figure~\ref{fig_ex_hda} where the state in the upper-right corner can be reached from the cell in the lower-left corner in more than one way.

In this setting axioms \ref{ax_HDML6} and \ref{ax_HDML61} basically say that instead of going through the inside of a square one can go on one of it sides. In other words, instead of going through a cell of higher dimension one can go only through cells of lowed dimensions. Particular to our example from Figure~\ref{fig_ex_hda} the axiom \ref{ax_HDML6} says that when going from the lower-left corner through the inside of the square one can instead go through one of the lower or left sides and reach the same place. The other axiom \ref{ax_HDML61} says that for reaching the upper-right corner, instead of going through its inside one can just take one of its upper or right sides.

Note also the theorems \refeq{eq_th12}-\refeq{eq_th14} which involve four \HDML\ modalities stacked one on top of the other. These are theorems of the two axioms \ref{ax_HDML6} and \ref{ax_HDML61} which involve only three modalities. In particular note the converse implication of \refeq{eq_th12} which is not a theorem. This says intuitively that one cannot infer from just being able to walk on the edges of a square that the square is filled in, i.e., that true concurrency is present. This makes \HDML\ powerful enough for the distinction between true concurrency and interleaving.
\end{itemize}

Remark that a natural counterpart (using the $\start{}{}$ modality in place of $\terminate{}{}$) of the axiom~\ref{ax_HDML2} is $\start{}{\startUniv{}{\varphi}}\imply\startUniv{}{\start{}{\varphi}}$ (which appeared in the short paper version \cite{P10concur}). But this ``axiom'' is broken by the fact that \HDAs\ allow choices. This formula would be valid only when working inside a single full cube (i.e., no choices, just concurrency), as would be the case when representing Mazurkiewicz traces as \HDAs.

\begin{theorem}[soundness]\label{th_soundness}
 The axiomatic system of Table~\ref{table_HDMLaxioms} is sound; i.e., $\forall\varphi:\ \prove\varphi\ \Rightarrow\ \ \models\varphi$. 
\end{theorem}

\begin{proof}
For soundness of the axiomatic system it is enough to prove that the axioms \ref{ax_HDML1} to \ref{ax_HDML61} are valid. 

We start with axiom \ref{ax_HDML2} and assume $\modelH,q_{n}\models\terminate{}{\terminateUniv{\,}{\varphi}}$ for some $q_{n}\in Q_{n}$ and  $n\geq 2$ because of the assumption $\terminatei{2}{\top}$. This means that exists some $q_{n-1}\in Q_{n-1}$ s.t.\ $t_{k}(q_{n})=q_{n-1}$ for some $1\leq k\leq n$ with $\modelH,q_{n-1}\models\terminateUniv{\,}{\varphi}$, and from this it means that for any $1\leq l\leq n-1$, $\modelH,t_{l}(q_{n-1})\models\varphi$. We need to show that $\modelH,q_{n}\models\terminateUniv{\,}{\terminate{}{\varphi}}$. This means that for any $m\neq k$ we have to find a $1\leq m'\leq n-1$ s.t.\ $\modelH,t_{m'}(t_{m}(q_{n}))\models\varphi$.\footnote{We do not consider the $k$ because the case for $m=k$ is trivial from the assumption above, where we know that for $t_{k}$ and any $t_{l}$ it is the case that $\modelH,t_{l}(t_{k}(q_{n}))\models\varphi$; and because we are at least on the layer 2 it means that there exists at least one $t_{l}$.} This is easy by applying the cubical law, considering wlog.\ $m<k$, $t_{m}(t_{k}(q_{n}))=t_{k-1}(t_{m}(q_{n}))$.\footnote{We can apply the cubical laws because we are working with cells of dimension at least 2.\\For the other case of $m>k$ we get $m'=k$ by using a corresponding cubical law.} Thus, the $m'=k-1$ for which trivially $1\leq k-1\leq n-1$. From the assumption we showed that we have $\modelH,t_{m}(t_{k}(q_{n}))\models\varphi$ and hence $\modelH,t_{k-1}(t_{m}(q_{n}))\models\varphi$.

%

For axiom \ref{ax_HDML3} assume $\modelH,q_{n}\models\start{}{\terminateUniv{\,}{\varphi}}$ with $q_{n}\in Q_{n}$. This means that exists $q_{n+1}\in Q_{n+1}$ and $1\leq k\leq n+1$ s.t.\ $s_{k}(q_{n+1})=q_{n}$ and $\modelH,q_{n+1}\models\terminateUniv{\,}{\varphi}$. Further, this implies that for any $1\leq i\leq n+1$,  $\modelH,t_{i}(q_{n+1})\models\varphi$. We want to prove that $\modelH,q_{n}\models\terminateUniv{\,}{\start{}{\varphi}}$, which amounts to showing that for some arbitrary $1\leq m\leq n$ with $t_{m}(q_{n})=q_{n-1}$ we can find an $1\leq l\leq n$ and $q'_{n}\in Q_{n}$ s.t.\ $s_{l}(q'_{n})=q_{n-1}$ and $\modelH,q'_{n}\models\varphi$. We assume that it exists at least one $t_{m}$ to work with, for otherwise the formula $\terminateUniv{\,}{\start{}{\varphi}}$ holds trivially. We achieve the goal using the cubical laws: if $m<k$ then consider the cubical law $t_{m}(s_{k}(q_{n+1}))=s_{k-1}(t_{m}(q_{n+1}))$ and set $l=k-1$ and $q'_{n}=t_{m}(q_{n+1})$ for which we know from above that $\modelH,t_{m}(q_{n+1})\models\varphi$; otherwise if $k\leq m$ (which also means that $k\leq n$) then consider the cubical law $s_{k}(t_{m+1}(q_{n+1}))=t_{m}(s_{k}(q_{n+1}))$ and set $l=k$ and $q'_{n}=t_{m+1}(q_{n+1})$ (where $m+1\leq n+1$) for which we know that $\modelH,t_{m+1}(q_{n+1})\models\varphi$.

For \ref{ax_HDML31} we can just use propositional reasoning and argue its validity by contraposition with axiom \ref{ax_HDML3} above. Nevertheless, we want to also give here a model theoretic argument similar to the above. Thus, assume $\modelH,q_{n}\models\terminate{}{\startUniv{}{\varphi}}$ with $q_{n}\in Q_{n}$. This means that exists $1\leq k\leq n$ and $q_{n-1}$ s.t.\ $t_{k}(q_{n})=q_{n-1}$ and $\modelH,q_{n-1}\models\startUniv{}{\varphi}$, which means that for any $q'_{n}$ with $s_{i}(q'_{n})=q_{n-1}$ for some $1\leq i\leq n$ we have $\modelH,q'_{n}\models\varphi$. We want to prove that $\modelH,q_{n}\models\startUniv{}{\terminate{}{\varphi}}$ which amounts to showing that for some arbitrary $q_{n+1}$, with $s_{m}(q_{n+1})=q_{n}$ for some $1\leq m\leq n+1$, we can find an $1\leq l\leq n+1$ and a $q''_{n}$ s.t.\ $t_{l}(q_{n+1})=q''_{n}$ and $\modelH,q''_{n}\models\varphi$. We use the cubical laws: if $k<m$ then consider the cubical law $t_{k}(s_{m}(q_{n+1}))=s_{m-1}(t_{k}(q_{n+1}))$ and set $l=k$ and $q''_{n}=t_{k}(q_{n+1})$ for which we have said before that $\modelH,t_{l}(q_{n+1})\models\varphi$ because there is the $s_{m-1}$ that reaches a cell which satisfies $\startUniv{}{\varphi}$; otherwise if $m\leq k$ then consider the cubical law $s_{m}(t_{k+1}(q_{n+1}))=t_{k}(s_{m}(q_{n+1}))$ and set $l=k+1$ and $q''_{n}=t_{k+1}(q_{n+1})$ for which it holds that $\modelH,t_{l}(q_{n+1})\models\varphi$ because $\modelH,s_{m}(t_{l}(q_{n+1}))\models\startUniv{}{\varphi}$.

For axiom~\ref{ax_HDML5} assume $\modelH,q_{n}\models\terminatei{i}{\top}$ which means that $n\geq i$. Even more, $\terminatei{i}{\top}$ holds in any cell $q_{n}\in Q_{n}$ of dimension $n$. We need to prove that $\modelH,q_{n}\models\startUniv{}{\terminate{}{\terminatei{i}{\top}}}$. The proof is trivial when there is no $q_{n+1}$ with $s_{j}(q_{n+1})=q_{n}$. Therefore, we need to prove that for any $q_{n+1}$ with $s_{j}(q_{n+1})=q_{n}$, for some $1\leq j\leq n+1$, $\modelH,q_{n+1}\models\terminate{}{\terminatei{i}{\top}}$. Because $q_{n+1}\in Q_{n+1}$ then it must have at least one $t$ map that links it with some cell $q'_{n}\in Q_{n}$ on the lower level. In $q'_{n}$ the formula $\terminatei{i}{\top}$ holds and thus we finished the proof.

For axiom~\ref{ax_HDML51} assume $\modelH,q_{n}\models\start{}{\terminate{}{\terminatei{i}{\top}}}$ which means that exists $q_{n+1}\in Q_{n+1}$ with $s_{j}(q_{n+1})=q_{n}$ for some $1\leq j\leq n+1$ s.t.\ $\modelH,q_{n+1}\models\terminate{}{\terminatei{i}{\top}}$. This means that $n+1\geq i+1$ and thus $n\geq i$. Therefore, for any $q'_{n}\in Q_{n}$ the formula $\terminatei{i}{\top}$ holds because we can go at least $i$ levels down and find any cell satisfying $\top$, hence $\terminatei{i}{\top}$ holds also in $q_{n}\in Q_{n}$.

Axiom~\ref{ax_HDML4} can actually be derived from axioms~\ref{ax_HDML5} and \ref{ax_HDML51} as follows: for $i>1$ then $\start{}{\terminatei{i}{\top}}\stackrel{\ref{ax_HDML51}}{\imply}\terminatei{i-1}{\top}\stackrel{\ref{ax_HDML5}}{\imply}\startUniv{}{\terminatei{i}{\top}}$; whereas for $i=1$ it is just an instantiation of axiom~\ref{ax_HDML5} for $i=0$.
As we did for axiom~\ref{ax_HDML31} we leave these so that the reader has a more intuitive understanding of the apparent symmetries of these formulas.

Nevertheless, we give also a model-theoretic argument, hence assume $\modelH,q_{n}\models\start{}{\terminatei{i}{\top}}$. This means that exists $q_{n+1}$ and $1\leq j\leq n+1$ s.t.\ $s_{j}(q_{n+1})=q_{n}$ and $\modelH,q_{n+1}\models\terminatei{i}{\top}$. This means that the dimension of $q_{n+1}$ is greater than $i$, i.e.,  $n+1\geq i$. We want to prove that $\modelH,q_{n}\models\startUniv{}{\terminatei{i}{\top}}$ which amounts to showing that for any $q'_{n+1}\in Q_{n+1}$ with $s_{j}(q'_{n+1})=q_{n}$ for some $1\leq j\leq n+1$ we have $\modelH,q'_{n+1}\models\terminatei{i}{\top}$. But we know from before that the dimension of $q'_{n+1}$ is at least $i$; this means that we can go down at least $i$ levels and on the lowest level any cell models $\top$. Hence we have $\modelH,q'_{n+1}\models\terminatei{i}{\top}$.

For axiom~\ref{ax_HDML41} we use a similar argument as in the proof based on the semantics of $\terminate{}{}$ and $\terminateUniv{\,}{}$ this time.

For \ref{ax_HDML1} consider that $\modelH,q\models\terminate{}{i}$ which means that there exist $i$ different cells $q^{j}$ with $1\leq j\leq i$ which are the result of the application of a $t$ map to $q$. Because $t$ is a map it means that there exist at least $i$ different maps $t_{j}$ with $1\leq j\leq i$ that are applied to $q$. Therefore, $q$ is of dimension at least $i$ which means that we can go $i$ levels down (by using an inductive argument).  This makes the formula $\terminatei{i}{\top}$ true at $q$.

For \ref{ax_HDML6} assume $\modelH,q_{n}\models\start{}{\start{}{\terminate{}{\varphi}}}$ which by the definition of the semantics it means that $\exists q_{n+1}\in Q_{n+1},k\leq n+1:s_{k}(q_{n+1})=q_{n}$ and $\exists q_{n+2}\in Q_{n+2},i\leq n+2:s_{i}(q_{n+2})=q_{n+1}$ and $\exists q_{n+1}'\in Q_{n+1},j\leq n+1:t_{j}(q_{n+2})=q_{n+1}'$ and $\modelH,t_{j}(q_{n+2})\models\varphi$. We want to prove that $\modelH,q_{n}\models\start{}{\terminate{}{\start{}{\varphi}}}$. This amounts to finding three cells $q_{n+1}^{a}\in Q_{n+1}$, $q_{n}^{b}\in Q_{n}$, and $q_{n+1}^{c}\in Q_{n+1}$ s.t.\ $s_{l}(q_{n+1}^{a})=q_{n}$, $t_{m}(q_{n+1}^{a})=q_{n}^{b}$, and $s_{n}(q_{n+1}^{c})=q_{n}^{b}$ and $\modelH,q_{n+1}^{c}\models\varphi$. We treat three cases depending on $i$ and $j$.

Case when $j<i$ then choose $m=j$, $n=i-1$, $k=l$, and $q_{n+1}^{a}=q_{n+1}$ hence finding the cubical law $t_{m}(s_{i}(q_{n+2}))=s_{n}(t_{j}(q_{n+2}))$ which makes $t_{j}(q_{n+2})=q_{n+1}^{c}$ and hence, the desired $\modelH,q_{n+1}^{c}\models\varphi$ follows from the initial $\modelH,t_{j}(q_{n+2})\models\varphi$.

Case when $j>i$ then choose $m=j-1$, $n=i$, $k=l$, and $q_{n+1}^{a}=q_{n+1}$ hence finding the cubical law $s_{n}(t_{j}(q_{n+2}))=t_{m}(s_{i}(q_{n+2}))$ which makes $t_{j}(q_{n+2})=q_{n+1}^{c}$ and hence, the desired $\modelH,q_{n+1}^{c}\models\varphi$ follows as before.

Case when $i=j$ then it is not enough to work only with the $i$ and $j$ as the cubical laws do not apply any more. But there are ways depending on $k$. We need two cases. When $k<j$ consider $l=j-1$, $m=j-1$, $n=k$, and $q_{n+1}^{a}=s_{k}(q_{n+2})$ as coming from the cubical law $s_{k}(s_{j}(q_{n+2}))=s_{l}(s_{k}(q_{n+2}))$. Using a second cubical law $s_{k}(t_{j}(q_{n+2}))=t_{m}(s_{k}(q_{n+2}))=q_{n}^{b}$ we obtain $q_{n+1}^{c}=q_{n+1}'$ and hence the desired $\modelH,q_{n+1}^{c}\models\varphi$. Otherwise, when $k\geq j$ then choose $l=j$, $m=j$, $n=k$ and $q_{n+1}^{a}=s_{k+1}(q_{n+2})$ as coming from the cubical law $s_{l}(s_{k+1}(q_{n+2}))=s_{k}(s_{j}(q_{n+2}))$. Using as second cubical law $t_{m}(s_{k+1}(q_{n+2}))=s_{k}(t_{j}(q_{n+2}))=q_{n}^{b}$ we obtain $q_{n+1}^{c}=q_{n+1}'$ and hence the desired result as before.

For \ref{ax_HDML61} assume $\modelH,q_{n}\models\start{}{\terminate{}{\terminate{}{\varphi}}}$ which by the definition of the semantics it means that $\exists q_{n+1}\in Q_{n+1},i\leq n+1:s_{i}(q_{n+1})=q_{n}$ and $\exists q_{n}'\in Q_{n},j\leq n:t_{j}(q_{n+1})=q_{n}'$ and $\exists q_{n-1}\in Q_{n-1},k\leq n-1:t_{k}(q_{n}')=q_{n-1}$ and $\modelH,q_{n-1}\models\varphi$. We want to prove that $\modelH,q_{n}\models\terminate{}{\start{}{\terminate{}{\varphi}}}$. This amounts to finding three cells $q_{n-1}^{a}\in Q_{n-1}$, $q_{n}^{b}\in Q_{n}$, and $q_{n-1}^{c}\in Q_{n-1}$ s.t.\ $t_{m}(q_{n})=q_{n-1}^{a}$, $s_{n}(q_{n}^{b})=q_{n-1}^{a}$, and $t_{l}(q_{n}^{b})=q_{n-1}^{c}$, and $\modelH,q_{n-1}^{c}\models\varphi$. We again treat three cases depending on $i$ and $j$.

Case when $i<j$ then choose $m=j-1$, $n=i$, $l=k$, and $q_{n}^{b}=q_{n}'$ and get $q_{n-1}^{a}$ from the cubical law $s_{n}(t_{j}(q_{n+1}))=t_{m}(s_{i}(q_{n+1}))=q_{n-1}^{a}$. Since $q_{n-1}^{c}=t_{l}(q_{n}^{b})=t_{k}(q_{n}')=q_{n-1}$ we get our desired result $\modelH,q_{n-1}^{c}\models\varphi$.

Case when $i>j$ then choose $m=j$, $n=i-1$, $l=k$, and $q_{n}^{b}=q_{n}'$ and get $q_{n-1}^{a}$ from the cubical law $t_{m}(s_{i}(q_{n+1}))=s_{n}(t_{j}(q_{n+1}))=q_{n-1}^{a}$. We get our desired result $\modelH,q_{n-1}^{c}\models\varphi$ as before.

Case when $i=j$ requires two subcases after $k$ as the cubical laws are not applicable to $i$ and $j$ anymore. We follow a similar reasoning as we did for \ref{ax_HDML6}. When $k<j$ then choose $l=j-1$ and have $q_{n}^{b}=t_{k}(q_{n+1})$ and $q_{n-1}^{c}=q_{n-1}$ from the cubical law $q_{n-1}=t_{k}(t_{j}(q_{n+1}))=t_{l}(t_{k}(q_{n+1}))$. To connect everything consider the cubical law $t_{m}(s_{i}(q_{n+1}))=s_{n}(t_{k}(q_{n+1}))$ giving $m=k$ and $n=j-1$. When $k\geq j$ then choose $l=j$ and have $q_{n}^{b}=t_{k+1}(q_{n+1})$ and $q_{n-1}^{c}=q_{n-1}$ from the cubical law $t_{l}(t_{k+1}(q_{n+1}))=t_{k}(t_{j}(q_{n+1}))=q_{n-1}$. And all is connected right through the cubical law $s_{n}(t_{k+1}(q_{n+1}))=t_{m}(s_{j}(q_{n+1}))$ giving $m=k$ and $n=j$.
\end{proof}

\begin{theorem}[compactness failure]\label{th_compactness}
The \HDML\ with the semantics of Table~\ref{table_HDMLsemantics} does not have the compactness property. 
\end{theorem}

\begin{proof}
Compactness says that for any infinite set of formulas $\Gamma$ if all the finite subsets $S\subset\Gamma$ are satisfiable than the original $\Gamma$ is satisfiable.

The compactness failure for \HDML\ is witnessed by the following infinite set of formulas:
\[
\Gamma=\{\terminatei{i}{\top}\mid i\in\omega\}.
\]

Any finite subset $S=\{\terminatei{i}{\top}\mid i\leq n\}$ of $\Gamma$ is satisfiable on a model $\modelH_{n}$ which has $Q_{n}\neq\emptyset$ in any cell $q_{n}\in Q_{n}$ of dimension $n$; i.e., $\modelH_{n},q_{n}\models\terminatei{i}{\top}$ for all $\terminatei{i}{\top}\in S$.

On the other hand the infinite $\Gamma$ is not satisfiable on any pointed model, i.e., at a single point. For assume there exists a model $\modelH$ and some cell $q\in Q_{m}$ for some level $m$ where all formulas $\phi\in\Gamma$ are satisfiable $\modelH,q\models\phi$. But this is not possible as the formula $\terminatei{m+1}{\top}$ does not hold on any cell from level $Q_{m}$ or any level below. This is because when stripping off one $\terminate{}{}$ we go one level down cf.\ the semantics; and we cannot go down more than $m$ levels, cf.\ $q\in Q_{m}$ but we need to strip $m+1$ times the after operator $\terminate{}{}$.
No matter on which level we choose the point cell $q$ in a model there will always be a formula in $\Gamma$ that will not hold, because of the infiniteness of $\Gamma$ (also regardless of the infiniteness of the model that we choose). 

Intuitively, the compactness failure is due to the fact that the models of \HDML\ are bounded below in their levels and \HDML\ has a modality that goes down the levels (i.e., the after modality $\terminate{}{}$).
\end{proof}


\section{Examples of Encodings into Higher Dimensional Modal Logic}\label{sec_examples}

This section serves to exemplify ways of using \HDML. One may encode other logics for different concurrency models as restrictions of \HDML; in this respect we study the relation of \HDML\ with standard modal logic, with CTL, ISTL (a branching time temporal logic over configuration structures), and with linear time temporal logic over Mazurkiewicz traces LTrL. Another way of using \HDML\ is as a general logical framework for studying properties of concurrency models and their interrelation. This is done by finding the appropriate restrictions of \HDA\ and \HDML\ and investigating their relations and axiomatic presentations.


\subsection{Encoding standard modal logic into HDML}\label{sec_encoding_modal}

\begin{lemma}[Kripke structures]\label{lemma_Kripke}
The class of Kripke structures is captured by the class of higher dimensional structures where all sets $Q_{n}$, for $n>1$, are empty.
\end{lemma}

\begin{proof}
Essentially this result is found in \cite{Glabbeek06HDA}.
A \HDA\ $K=(Q_{0},Q_{1},s_{1},t_{1},\labelH)$ is a special case of \HDAs\ where all $Q_{n}=\emptyset$ for $n>1$. This is the class of \HDAs\ that encode Kripke frames. Because $Q_{2}$ (and all other cells of higher dimension) is empty there are no cubical laws applicable. Therefore, there is no geometric structure on $K$. Moreover, the restriction on the labeling function $l$ is not applicable (as $Q_{2}$ is empty). Add to such a \HDA\ a valuation function $\mathcal{V}$ to obtain a Kripke model $(Q_{0},Q_{1},s_{1},t_{1},\labelH,\mathcal{V})$. 
\end{proof}

\begin{proposition}[axiomatization of Kripke \HDAs]\label{prop_axiom_Kripke}
The class of higher dimensional structures corresponding to Kripke structures (from Lemma~\ref{lemma_Kripke}) is axiomatized by:
\vspace{-1ex}\begin{equation}
 \models \startUniv{}{\startUniv{}{\bot}}\label{ax_kripke}
\end{equation}
\end{proposition}

\begin{proof}
For any \HDA\ \modelH\ and any $q\in Q$ a cell of any dimension, we prove the double implication: 
$\modelH\models\startUniv{}{\startUniv{}{\bot}}$\hspace{1ex} iff\hspace{1ex} $\modelH$ is as in Lemma~\ref{lemma_Kripke}.

For the \textit{if} direction if $q\in Q_{1}$ then the axiom holds trivially because there are no cells on $Q_{2}$, hence $\modelH,q\models\startUniv{}{\startUniv{}{\bot}}$ holds and also $\startUniv{}{\bot}$. When $q\in Q_{0}$ the axiom holds because for any $q'\in Q_{1}$ with $s_{1}(q')=q$ it is the case that $\modelH,q'\models\startUniv{}{\bot}$ because there are no $q''\in Q_{2}$ cf. Lemma~\ref{lemma_Kripke}.

For the \textit{only if} direction consider a \modelH\ for which the axiom holds (i.e., for any cell $q\in Q$ then $\modelH,q\models\startUniv{}{\startUniv{}{\bot}}$); we need to show that any $Q_{n}$ with $n>1$ is empty. Assume the opposite, that there exists $q_{n}\in Q_{n}$  with $n>1$. This means that there is a sequence $s_{1}(\dots s_{i}(q_{n}))=q_{0}$ of source maps that ends in a cell $q_{0}\in Q_{0}$ of dimension $0$. But $\modelH,q_{0}\models\startUniv{}{\startUniv{}{\bot}}$, which means that there cannot be this sequence of source maps unless $q_{n}$ is of dimension at most $1$. This is a contradiction and hence the proof is finished.
%
\end{proof}

\begin{theorem}[standard modal logic]\label{th_modal_logic}
Consider the syntactic definition 
\[
\Diamond\varphi\defequal\start{}{\terminate{}{\varphi}}.
\]
The language of \emph{standard modal logic} uses only $\Diamond$ and is interpreted only over higher dimensional structures as defined in Lemma~\ref{lemma_Kripke} and only in cells of $Q_{0}$.
\end{theorem}

\begin{proof}
First we check that we capture exactly the semantics of standard modal logic; $\modelH,q_{0}\models\Diamond\varphi$ iff $\modelH,q_{0}\models\start{}{\terminate{}{\varphi}}$ iff $\exists q'\in Q_{1}$ s.t. $s_{1}(q')=q_{0}$ and $\modelH,q'\models\terminate{}{\varphi}$ iff $\exists q'_{0}\in Q_{0}$ s.t. $t_{1}(q')=q'_{0}$ and $\modelH,q'_{0}\models\varphi$. This is the same as $\exists q'_{0}\in Q_{0}$ reached in ``one transition'' from $q_{0}$ and $\modelH,q'_{0}\models\varphi$. (We go only through one transition cell $q'\in Q_{1}$.)

Clearly, with the axiom of Proposition~\ref{prop_axiom_Kripke}, $\modelH,q_{n}\not\models\Diamond\varphi$ for any $q_{n}\in Q_{n}$ for any $n\geq 1$. Therefore, $\Diamond\varphi$ makes sense only interpreted in states from $Q_{0}$.

Second we check that the axioms of standard modal logic for $\Diamond$ hold in our axiomatic system. Clearly $\Diamond\bot\equivalent\bot$; just apply \ref{ax_modal11} and then \ref{ax_modal1} to $\start{}{\terminate{}{\bot}}$. It is easy to see that $\Box\varphi\equivalent\neg\Diamond\neg\varphi$ as $\neg\start{}{\terminate{}{\neg\varphi}}\stackrel{\ref{ax_modal3}}{\equivalent}\startUniv{}{\neg\terminate{}{\neg\varphi}}\stackrel{\ref{ax_modal31}}{\equivalent}\startUniv{}{\terminateUniv{\,}{\varphi}}$ and the semantic of $\Box\varphi$ is the right one, i.e., for any $q'_{0}\in Q_{0}$, reached through some transition $q'\in Q_{1}$, is the case that $\modelH,q'_{0}\models\varphi$. We prove now that $\Diamond(\varphi\vee\varphi')\equivalent\Diamond\varphi\vee\Diamond\varphi'$. This is because $\start{}{\terminate{}{(\varphi\vee\varphi')}}\stackrel{\ref{ax_modal21}}{\equivalent}\start{}{(\terminate{}{\varphi}\vee\terminate{}{\varphi'})}\stackrel{\ref{ax_modal2}}{\equivalent}\start{}{\terminate{}{\varphi}}\vee\start{}{\terminate{}{\varphi'}}\stackrel{def}{\equivalent}\Diamond\varphi\vee\Diamond\varphi'$.

It is easy to see how we recover the corresponding inference rule for $\Diamond$. We thus have all the axiomatic system of standard modal logic and the proof is finished.
\end{proof}

Remark that the axioms \ref{ax_HDML1}-\ref{ax_HDML61} particular to \HDML\ are trivially satisfied for all states or transitions (i.e., cells of dimension 0 or 1). This means that for these cells these axioms do not impose any constraints. One can easily check that for each of the axioms \ref{ax_HDML1}-\ref{ax_HDML61}, which are implications, either the first formula does not hold or the second formula holds trivially. In fact, in the axiomatic system of Table~\ref{table_HDMLaxioms} with the new axiom \refeq{ax_kripke} added, one cannot prove formulas where the same existential modality is stacked twice or more (like $\start{}{\start{}{\dots}}$ or $\terminate{}{\terminate{}{\dots}}$). In fact, any such formula is provable unsatisfiable. This is also a reason for using the syntactic definition for the diamond from Theorem~\ref{th_modal_logic}.

\subsection{Adding an Until operator and encoding standard temporal logic}\label{subsec_until}

The \textit{basic} temporal logic is the logic with only the \textit{eventually} operator (and the dual \textit{always}). This language is expressible in the standard modal logic \cite{01modalLogicBook}. It is known that the \textit{Until} operator adds expressiveness (\textit{eventually} and \textit{always} operators can be encoded with \textit{Until} but not the other way around).

The \textit{Until} operator cannot be encoded in \HDML\ because of the local behavior of the during and after modalities; similar arguments as in modal logic about expressing \textit{Until} apply to \HDML\ too. The \textit{Until}  modality talks about the whole model (about all the configurations of the system) in an existential manner. More precisely, the \textit{Until} says that there must exist some configuration in the model, reachable from the configuration where \textit{Until} is evaluated, satisfying some property $\varphi$, and in all the configurations on all/some of the paths reaching the $\varphi$ configuration some other property $\psi$ must hold. Hence we need a notion of \textit{path} in a \HDA.

\begin{definition}[paths in \HDAs]\label{def_paths_HDA}
A \emph{simple step} in a \HDA\ is either $q_{n-1}\transition{s_{i}}q_{n}$ with $s_{i}(q_{n})=q_{n-1}$ or $q_{n}\transition{t_{i}}q_{n-1}$ with $t_{i}(q_{n})=q_{n-1}$, where $q_{n}\in Q_{n}$ and $q_{n-1}\in Q_{n-1}$ and $1\leq i\leq n$. A \emph{path} $\pi\defequal q^{0}\transition{\alpha^{0}}q^{1}\transition{\alpha^{1}}q^{2}\transition{\alpha^{2}}\dots$ is a sequence of single steps $q^{j}\transition{\alpha^{j}}q^{j+1}$, with $\alpha^{j}\in\{s_{i},t_{i}\}$. We say that $q\in\pi$ iff $q=q^{j}$ appears in one of the steps in $\pi$. The first cell in a path is denoted $st(\pi)$ and the ending cell in a finite path is $en(\pi)$. We call a cell $q'$ \emph{reachable} from some other cell $q$, and denote by $q\reachcell q'$, iff $\exists\pi:st(\pi)=q\wedge en(\pi)=q'$. Overload the notation $\pi\reachcell\pi'$ to mean that the path $\pi'$ extends $\pi$, with the usual definition.
\end{definition}

There are two main kinds of \textit{Until} operator that can be defined on a branching structure like \HDA: one is in the style of linear time temporal logic \cite{Pnueli77temporal_logic}; and the other in the style of computation tree logic (CTL). These two kinds are found defined also over Mazurkiewicz traces or configuration structures. There are proofs that the CTL style of defining the \textit{Until} yields undecidability both on traces \cite{Penczek92undecid} and on configuration structures \cite{AlurMP05decid_partial,AlurP99undecidable} and all these three proofs use different techniques, i.e., encoding a different undecidable problem. On the other hand the LTL style of definition of \textit{Until} over traces is decidable as part of LTrL \cite{ThiagarajanW02LTrL}; see also the related decidable definition part of the TrPTL logic \cite{MukundT96TrPTL}.

In the same spirit as done for temporal logic we boost the expressiveness of \HDML\ by defining an \textit{Until} operator over higher dimensional structures. We define both styles of \textit{Until} operators. We then show how the standard LTL logic (with its until operator interpreted over Kripke structures) is encoded into the \HDML\ framework. For the CTL-like definition we discuss if and how the details of the undecidability proofs over Mazurkiewicz traces can be done in the setting of \HDML. Note that the proofs in \cite{Penczek92undecid,AlurP99undecidable} lack many of the details. We concentrate on the proof using the Post correspondence problem from \cite{AlurP99undecidable}.

\begin{definition}[CTL-like Until operator]\label{def_untilCTL}
Define an \textit{Until} operator $\varphi\untilC\varphi'$, in the style of CTL, which is interpreted over a \HDA\ in a cell as below:
 
\vspace{0.5ex}\begin{tabular}{@{\hspace{0ex}}r@{\hspace{0.5ex}}c@{\hspace{0.5ex}}l@{\hspace{1ex}}c@{\hspace{1ex}}l}
$\modelH,q$ & $\models$ & $\varphi\untilC\varphi'$ & iff & $\exists\pi\in\modelH$ s.t.\ $st(\pi)=q \wedge en(\pi)=q'$,\\
& & &\multicolumn{2}{l}{$\modelH,q'\models\varphi'$, and \ $\forall q''\in\pi,q''\neq q'$ then $\modelH,q''\models\varphi$.}\\
\end{tabular}
\end{definition}

\begin{definition}[LTL-like Until operator]\label{def_untilLTL}
Define an \textit{Until} operator $\varphi\untilL\varphi'$, in the style of LTL, which is interpreted over a \HDA\ in a cell as below:
 
\vspace{0.5ex}\begin{tabular}{@{\hspace{0ex}}r@{\hspace{0.5ex}}c@{\hspace{0.5ex}}l@{\hspace{1ex}}c@{\hspace{1ex}}l}
$\modelH,q$ & $\models$ & $\varphi\untilL\varphi'$ & iff & $\exists q'\in\modelH$ s.t.\ $q\reachcell q' \wedge \modelH,q'\models\varphi'$ ,\\
& & &\multicolumn{2}{l}{and $\forall\pi\in\modelH,\forall q''\in\pi:st(\pi)=q \wedge en(\pi)=q'\wedge q''\neq q'$}\\
& & &\multicolumn{2}{l}{\phantom{and }then $\modelH,q''\models\varphi$.}\\
\end{tabular}
\end{definition}

The Definition~\ref{def_untilLTL} of \untilL\ is in the style of LTL in the sense that it looks only at one (concurrent) execution of the system ignoring choices (in the sense of \HDA). The Definition~\ref{def_untilCTL} of \untilC\ is more refined because it looks at a single linearization of a concurrent execution; and it is branching in the sense that it is not confined to one single concurrent execution, but the linearization may cross boundaries of concurrent runs, i.e., taking choices.

\begin{proposition}[modeling CTL Until]\label{prop_modeling_CTL}
The CTL \textit{Until} modality is encoded syntactically by $\varphi\,\exists\!\until\varphi'\defequal(\varphi\vee\terminate{}{\top})\untilC(\varphi'\wedge\neg\terminate{}{\top})$ when $\,\exists\!\until$ is interpreted only in states of Kripke \HDAs\ as in Lemma~\ref{lemma_Kripke}.
\end{proposition}

\begin{proof}
Essential for the proof is the fact that $\,\exists\!\until$ is interpreted over restricted \HDAs\ which model Kripke structures. Precisely, they have only cells of dimension $0$ (the states) and $1$ (the transitions), and moreover, we know which are states because the formula $\neg\terminate{}{\top}$ holds in all and only the cells of dimension $0$. Therefore, the right formula of the $\,\exists\!\until$ is evaluated only in states because $(\varphi'\wedge\neg\terminate{}{\top})$ can never hold in a cell of dimension greater than $0$.
Moreover, the transitions are not important for valuating the $\varphi$ because the formula $\terminate{}{\top}$ is always true in a transition (because any transition has a target state). On the other hand the formula $\terminate{}{\top}$ is never true in a state and hence the $\varphi$ has to be true so that the whole left part of the until to hold.

For this proof we only concentrate on showing that the semantics of the $\,\exists\!\until$ corresponds to the well known CTL semantics. Thus, we want to show that $\modelH,q_{0}\models\varphi\,\exists\!\until\varphi'$ is the same as saying that exists a finite sequence of states $q_{0}^{1},\dots,q_{0}^{k}\in Q_{0}$ with $q_{0}^{1}=q_{0}$, $\modelH,q_{0}^{k}\models\varphi'$, $\modelH,q_{0}^{i}\models\varphi$ for all $1\leq i<k$, and for any $1<i\leq k$ $q_{0}^{i}$ is reachable through a single transition from $q_{0}^{i-1}$. By th definition in the statement, $\modelH,q_{0}\models\varphi\,\exists\!\until\varphi'$ is the same as $\modelH,q_{0}\models(\varphi\vee\terminate{}{\top})\untilC(\varphi'\wedge\neg\terminate{}{\top})$. By the semantics of \untilC\ from Definition~\ref{def_untilCTL} we know that $\exists \pi$ a path in \modelH, which goes only through cells of dimension $0$ or $1$ because \modelH\ models a Kripke structure cf.\ Lemma~\ref{lemma_Kripke}, hence $\pi$ is of the form $q_{0},q_{1},q'_{0},\dots$; and moreover, we also have that $st(\pi)=q_{0} \wedge en(\pi)=q'$,\ $\modelH,q'\models(\varphi'\wedge\neg\terminate{}{\top})$, and $\forall q''\in\pi,q''\neq q'$ then $\modelH,q''\models(\varphi\vee\terminate{}{\top})$. Clearly $q'\in Q_{0}$ because $\neg\terminate{}{\top}$ must hold in $q'$ and hence $\varphi'$ holds in a state, i.e., $\modelH,q'\models\varphi'$. It remains to show that in all $q''$ which are states (i.e., those $q_{0}^{k}\in Q_{0}$) we have that $\modelH,q''\models\varphi$. But we know that $\modelH,q''\not\models\terminate{}{\top}$ because $q''$, being a cell of dimension $0$, has no $t$ map. Therefore, using $\modelH,q''\models(\varphi\vee\terminate{}{\top})$ from before, we have that $\modelH,q''\models\varphi$. 

Note that for the full CTL a universal correspondent of \untilC\ must be defined over \HDAs, but we do not go into these details here.
\end{proof}


\subsection{Partial order models and their logics in HDML}\label{sec_partial_orders}

This section is mainly concerned with Mazurkiewicz traces \cite{Mazurkiewicz88tracesTutorial} as a model of concurrency based on partial orders, because of the wealth of logics that have been developed for it \cite{MukundT96TrPTL,ThiagarajanW02LTrL}. Higher dimensional automata are more expressive than most of the partial orders models (like Mazurkiewicz traces, pomsets \cite{Pratt86pomsets},  or event structures \cite{NielsenPW79eventstructures}) as studied in \cite{Pratt00HDArev,Glabbeek06HDA}. 
In particular, an extensive part of \cite{Glabbeek06HDA} is devoted to showing how Petri nets are representable as some class of higher dimensional automata. 
The works of \cite{Pratt00HDArev,Pratt03trans_cancel,Glabbeek06HDA} show (similar in nature) how event structures can be encoded in higher dimensional automata. Mazurkiewicz traces are a particular class of event structures, precisely defined in \cite{RozoyT91MazAsEventStrucs}. We use this presentation, as a restricted partial order, of Mazurkiewicz traces. 

In the following we give definitions and standard results on partial orders, event structures, and Mazurkiewicz traces which are needed for the development of the higher dimensional modal logic for these models, in particular for Mazurkiewicz traces. In few words, we isolate the class of higher dimensional automata corresponding to Mazurkiewicz traces (and to partial orders or event structures in general) as the models of the \HDML. Then we restrict \HDML\ to get exactly the logics over Mazurkiewicz traces (we focus on the logics presented in \cite{ThiagarajanW02LTrL,DiekertG00LTL_traces}) and over the more general partial orders called communicating sequential agents in \cite{LodayaRT92csa} (like ISTL of \cite{AlurMP05decid_partial,AlurP99undecidable}).

\begin{definition}[partial orders]\label{def_partial_orders}
A partially ordered set (or \emph{poset}) is a set $E$ equipped with a partial order $\leq$, $(E,\leq)$. The \emph{history} of an element $e\in E$ (denoted $\hystory{e}$) is $\hystory{e}=\{e'\mid e'\leq e\}$. The notion of history is extended naturally to a set of elements $C\subseteq E$ (denoted $\hystory{C}$). A \emph{configuration} is a finite and history closed set of elements (i.e., $C=\hystory{C}$). Denote by $\mathcal{C}$ the set of all configurations.
(Obviously, $\emptyset$, and $\hystory{e}$, for any $e\in E$, are configurations.)  
The \emph{immediate successor} relation $\lessdot\subseteq E\times E$ is defined as $e\lessdot e'$ iff $e\neq e'$ and $e\leq e'$ and $\forall e''\in E$, $e\leq e''\leq e'$ implies $e=e''$ or $e'=e''$. A \emph{$\Sigma$-labeled poset} $(E,\leq,\labelc)$ is a poset with a labeling function $\labelc:E\rightarrow\Sigma$ which maps each element to a label from $\Sigma$. Define a \emph{transition relation} on the configurations of a labeled poset as $\transition{}\subseteq\mathcal{C}\times\Sigma\times\mathcal{C}$ given by $C\transition{a}C'$ iff $\exists e\in E$ s.t.\ $\labelc(e)=a$ and $e\not\in C$ and $C'=C\cup\{e\}$.
\end{definition}

When one sees the elements of $E$ as the \textit{events} of a system, the labels can be seen as the names of the actions that the events are instances of.

\begin{definition}[Mazurkiewicz traces]\label{def_maz_trace}
Consider a symmetric and irreflexive \emph{independence relation} $I\subseteq\Sigma\times\Sigma$ and its complement $D=\Sigma\times\Sigma\setminus I$, called the \emph{dependence relation}. Mazurkiewicz traces are labeled posets restricted by the independence relation as follows:

\begin{tabular}{@{\hspace{0ex}}l@{\hspace{1ex}}l@{\hspace{1ex}}}
$\forall e\in E$, & $\hystory{e}\mbox{ is finite}$,
 \\
$\forall e,e'\in E$, & $e\lessdot e'\Rightarrow(\labelc(e),\labelc(e'))\in D$, \\
$\forall e,e'\in E$, & $(\labelc(e),\labelc(e'))\in D\Rightarrow e\leq e'\mbox{ or }e'\leq e$. \\
\end{tabular}
\end{definition}

\begin{definition}[event structures]\label{def_event_struc}
Consider a symmetric and irreflexive relation $\conflict\subseteq E\times E$. This \emph{conflict relation} is added to a poset to form an \emph{event structure} $(E,\leq,\conflict)$ where the following restrictions apply:

\begin{tabular}{@{\hspace{0ex}}l@{\hspace{1ex}}l@{\hspace{2ex}}r@{\hspace{0ex}}}
$\forall e,e',e''\in E$, & $e\conflict e'$ and $e'\leq e''$ implies $e\conflict e''$, & \\
$\forall e,e'\in E$, & $e\in C$ and $e\conflict e'$ implies $e'\not\in C$. & \\
\end{tabular}

\noindent An event structure is called \emph{finitary} iff \ $\forall e\in E,\,\hystory{e}$ is finite.
\end{definition}

The second constraint on event structures says that the configurations of an event structure are conflict-free. Define the relation of concurrency for an event structure to be: 

\centerline{$\concurrel=E\times E\setminus (\conflict\cup\leq\cup\leq^{-1})$.}

\begin{proposition}[families of configurations]\label{prop_fam_config}
 A finitary event structure $(E,\leq,\conflict)$ is uniquely determined by its family of configurations $\mathcal{C}_{E}$ (denoted $(E,\mathcal{C}_{E})$).
\end{proposition}

\begin{proof}
This result is found in \cite{Pratt03trans_cancel}. We summarize here the results leading to it.

The two relations $e\leq e'$ and $e\conflict e'$ are mutually exclusive, because, otherwise, the set $\hystory{e'}$ would not be a configuration (because of the second constraint of Definition \ref{def_event_struc}).

If two events $e,e'$ do not appear together in any configuration of $\mathcal{C}_{E}$ then $e\conflict e'$ ($e\conflict e'$ iff $\nexists C\in\mathcal{C}_{E}$ s.t.\ $e,e'\in C$).

If in any configuration where $e'$ exists, $e$ exists too then $e\leq e'$ ($e\leq e'$ iff $\forall C\in\mathcal{C}_{E},e'\in C\Rightarrow e\in C$).
\end{proof}

We usually use a labeled poset and work with labeled event structures $(E,\leq,\conflict,\labelc)$, or $(E,\mathcal{C}_{E},\labelc)$ when using their corresponding family of configurations.

\begin{proposition}[traces as event structures]\label{prop_traces_as_event_strucs}
Any Mazurkiewicz trace, as in Definition~\ref{def_maz_trace}, corresponds to a \emph{trace configuration structure}, which is a labeled event structure $(E,\mathcal{C}_{E},\labelc)$ with an empty conflict relation that respects the following restriction:\vspace{0.5ex}

\centerline{\vspace{0.5ex}$\labelc$ is \emph{a nice labeling} and \emph{context-independent},}

\noindent where nice labeling means

$\forall e,e'\in E,\ \labelc(e)=\labelc(e')\Rightarrow e\leq e'\mbox{ or }e'\leq e$

\noindent and context-independent means

$\forall a,b\in\Sigma,$ $(\labelc^{-1}(a)\times\labelc^{-1}(b))\cap\concurrel\neq\emptyset\hspace{1ex}\Rightarrow\hspace{1ex}(\labelc^{-1}(a)\times\labelc^{-1}(b))\cap\lessdot =\emptyset$ .

\end{proposition}

\begin{proof}
This result is essentially found in \cite{MukundT96TrPTL,RozoyT91MazAsEventStrucs}. We remind how one gets the independence relation of a Mazurkiewicz trace from a trace configuration structure:

\hfill $I=\{(a,b)\mid(\labelc^{-1}(a)\times\labelc^{-1}(b))\cap\concurrel\neq\emptyset \}$.
\end{proof}

One can view a configuration as a valuation of events $E\rightarrow\{0,1\}$, and thus we can view an event structure as a valuation $f_{E}:2^{E}\rightarrow\{0,1\}$, which selects only those configurations that make the event structure.

The terminology that we adopt now steams from the Chu spaces representation of \HDAs\ \cite{Pratt00HDArev,Pratt03trans_cancel}. We fix a set $E$, which for our purposes denotes events. Consider the class of \HDAs\ which have a single hypercube of dimension $|E|$, hence each event represents one dimension in the \HDA. This hypercube is denoted $3^{E}$, in relation to $2^{E}$, because in the \HDA\ case each event may be in three phases, \textit{not started}, \textit{executing}, and \textit{terminated} (as opposed to only terminated or not started). The valuation from before becomes now $E\rightarrow\{0,\executing,1\}$, where \executing\  means executing. The set of three values is linearly ordered $0<\executing<1$ to obtain an \emph{acyclic} \HDA\ \cite{Pratt03trans_cancel}, and all cells of $3^{E}$ are ordered by the natural lifting of this order pointwise. The dimension of a cell is equal to the number of $\executing$ in its corresponding valuation.

\Notation{
In the context of a single hypercube $3^{E}$ we denote the cells of the cube by lists of $|E|$ elements $e_{1}e_{2}\dots e_{|E|}$ where each $e_{i}$ takes values in $\{0,\executing,1\}$ and represents the status of the $i^{th}$ event of the \HDA.
}

With the above conventions, the cells of dimension $0$ (i.e., the states of the \HDA) are denoted by the corresponding valuation restricted to only the two values $\{0,1\}$; and correspond to the configurations of an event structure. The set of states of such a \HDA\ is partially ordered by the order $<$ we defined before. 
In this way, from the hypercube $3^{E}$ we can obtain any family of configurations $\mathcal{C}_{E}$ by removing all $0$-dimensional cells that represent a configuration $C\not\in\mathcal{C}_{E}$.\footnote{We remove also all those cells of higher dimension that are connected with the 0-dimensional cells that we have removed.} 
By Proposition~\ref{prop_fam_config} we can reconstruct the event structure.

In Definition~\ref{def_satisfiability} the interpretation of the during and after modalities of \HDML\ did not take into consideration the labeling of the \HDA. The labeling was used only for defining the geometry of concurrency of the \HDA. Now we make use of this labeling function in the semantics of the labeled modalities of Definition~\ref{def_labeled_modal}. But first we show how the labeling extends to cells of any dimension.

\begin{definition}[general labeling]\label{def_general_labels_cells}
Because of the condition $l(s_{i}(q))=l(t_{i}(q))$ for all $q\in Q_{2}$, all the edges $e_{1}\dots e_{i-1}\executing\,e_{i+1}\,\dots e_{|E|}$, with $e_{j}\in\{0,1\}$ for $j\neq i$, have the same label. Denote this as the label $l_{i}$. The label of a general cell $q\in Q_{n}$ is the multiset of $n$ labels $l_{j_{1}}\dots l_{j_{n}}$ where the $j$'s are exactly those indexes in the representation of $q$ for which $e_{j}$ has value \executing.
\end{definition}

As is the case with multi-modal logics or propositional dynamic logics \cite{harel00dynamicLogic}, we extend \HDML\ to have a multitude of modalities indexed by some alphabet $\Sigma$ (the alphabet of the \HDA\ in our case). This will be the same alphabet as that of the Mazurkiewicz trace represented by the \HDA. 
In propositional dynamic logic there is an infinite number of modalities because they are indexed by an alphabet consisting of the regular expressions; yet these can be expressed in terms of a finite number of basic modalities (indexed by only the basic expressions). In our case we consider only an unstructured alphabet $\Sigma$ which is considered finite.

\begin{definition}[labeled modalities]\label{def_labeled_modal}
Consider two labeled modalities \emph{during} $\start{a}{\varphi}$ and \emph{after} $\terminate{a}{\varphi}$ where $a\in\Sigma$ is a label from a fixed alphabet. The interpretation of the labeled modalities is given as:

\begin{tabular}{@{\hspace{0ex}}r@{\hspace{0.5ex}}c@{\hspace{0.5ex}}l@{\hspace{1ex}}c@{\hspace{1ex}}l}
$\modelH,q$ & $\models$ & $\start{a}{\varphi}$ & iff & assuming $q\in Q_{n}$ for some $n$, $\exists q'\in Q_{n+1}$ s.t.\ \\
& & &\multicolumn{2}{l}{\hspace{2ex}$s_{i}(q')=q$ for some $1\leq i\leq n$, $l(q')=l(q)a$\ \ and $\modelH,q'\models\varphi$.}\\
%
$\modelH,q$ & $\models$ & $\terminate{a}{\varphi}$ & iff & assuming $q\in Q_{n}$ for some $n$, $\exists q'\in Q_{n-1}$ s.t.\\
& & &\multicolumn{2}{l}{\hspace{2ex}$t_{i}(q)=q'$ for some $1\leq i\leq n$, $l(q)=l(q')a$\ \ and $\modelH,q'\models\varphi$.}\\
\end{tabular}
\end{definition}

Having the labeled modalities one can get the unlabeled variants as a disjunction over all labels 
\[
\start{}{\varphi}\defequal\bigvee_{a\in\Sigma}\start{a}{\varphi}\label{def_unLabeled}
\]

The same as in Proposition~\ref{prop_axiom_Kripke} we captured axiomatically in the basic \HDML\ language the Kripke models, the question now is whether we can capture in the basic \HDML\ language with labeled modalities the Mazurkiewicz traces.
The initial results in Lemma~\ref{lemma_trace_HDA} cast the restrictions on labeled event structures of Proposition~\ref{prop_traces_as_event_strucs} into the \HDA\ setting in the view discussed above. Nevertheless, the context-independence property of the labeling function $\lambda$ is special and we discuss it afterwards.

\begin{lemma}[trace restrictions in \HDA]\label{lemma_trace_HDA}
\ 

The notion of \emph{empty conflict relation} from Definition~\ref{def_event_struc} is captured in \HDML\ by the axiom:
\begin{equation}
a\neq b\models (\start{a}{\top}\wedge\start{b}{\top})\imply(\start{a}{\start{b}{\top}}\wedge \start{b}{\start{a}{\top}})\label{axiom_empty_conflict}
\end{equation} 

The notion of \emph{nice labeling} from Proposition~\ref{prop_traces_as_event_strucs} is captured in \HDML\ by the axiom:
\begin{equation}
\models \terminate{a}{\top}\imply\neg\start{a}{\top}\label{axiom_niceLabel}
\end{equation} 

The notion of \emph{dependent actions} $a$ and $b$ from Definition~\ref{def_maz_trace} is captured in \HDML\ by the axiom:
\begin{equation}
\models \terminate{a}{\top}\imply\neg\start{b}{\top}\label{axiom_nonConcurAct}
\end{equation} 
\end{lemma}

\begin{proof}
Mazurkiewicz traces do not employ the notion of conflict relation of the event structures. In other words, traces are encoded as event structures with an empty conflict relation. To such event structures the two restrictions of Definition~\ref{def_event_struc} do not apply, being vacuously satisfied. Therefore, the Mazurkiewicz traces become, in this view, just configuration structures with the labeling function restricted as in Proposition~\ref{prop_traces_as_event_strucs}. Because the conflict relation is what captures choices in event structures and in higher dimensional automata, the Mazurkiewicz traces are just linear models, unable to capture choices.

The axiom \refeq{axiom_empty_conflict} restricts \HDAs\ to not have choices. Essentially the axiom says that if in some cell one can start two different events (with different labels) then these two events are concurrent, i.e., the two during modalities can be stacked one on top of the other. Note that the axiom talks only about different labels. Choices between events with the same label are still allowed. To remove this form of nondeterminism we just need to add the modal axiom for determinism: $\models \start{a}{\varphi}\imply\startUniv{a}{\varphi}$.

Such restricted \HDAs\ still allow for \textit{autoconcurrency} which is not the case in Mazurkiewicz traces. The \textit{nice labeling} axiom \refeq{axiom_niceLabel} removes autoconcurrency. It basically says that two events with the same label cannot be concurrent; i.e., if an event labeled with $a$ has been started then no other event labeled with $a$ can start. Note that this axiom is meaningful on transitions and cells of higher dimension, but not in states; i.e., it is meaningful during the execution of the already started $a$-labeled events, not before starting them.

The last axiom \refeq{axiom_nonConcurAct} models the fact that two dependent actions are not concurrent, which is the last restriction in the Definition~\ref{def_maz_trace} of Mazurkiewicz traces. Note that the nice labeling restriction says that the dependence relation is reflexive, as required in Definition~\ref{def_maz_trace}.
\end{proof}

We could not capture the context-independent  restriction on the labeling because it does not have just a universal presentation, so that we can capture it with axioms. This restriction is existential in nature, looking through all the higher dimensional automaton for some particular events. In fact it has a mixture of existential and universal assertions. Precisely, a labeling being context-independent is as saying that: if there exists throughout the \HDA\ two events labeled with $a$ and $b$ which are concurrent, then all the pairs of events from the same \HDA\ that are labeled with $a$ and $b$ must be concurrent. Or we can characterize it otherwise with the notion of \textit{not-concurrent} as: if there exists throughout the \HDA\ two events labeled with $a$ and $b$ which are not concurrent, then all the pairs of events from the same \HDA\ that are labeled with $a$ and $b$ must not be concurrent. We can also have another view on this property, using two validities: either all the pairs of events labeled with $a$ and $b$ are not concurrent (i.e., axiom \refeq{axiom_nonConcurAct}) or all the pairs of events labeled with $a$ and $b$ are concurrent.

We conjecture that the context-independent restriction on the labeling function cannot be captured just in the basic \HDML\ language, but the more expressive temporal operators are needed, which can talk about the whole \HDA\ structure in an existential manner. Maybe just the \textit{eventually} temporal modality is enough, instead of the stronger \textit{Until} operator. Yet another question is whether just the LTL-like \textit{Until} operator from Definition~\ref{def_untilLTL} is enough.

In the remainder of this section we show how the LTrL logic of \cite{ThiagarajanW02LTrL} and the ISTL logic of \cite{AlurMP05decid_partial,AlurP99undecidable} is captured in the higher dimensional framework. These logics, as well as those presented in \cite{MukundT96TrPTL,DiekertG00LTL_traces}, are interpreted in some particular configuration of a Mazurkiewicz trace (or of a restricted partial order). We take the view of Mazurkiewicz traces as restricted labeled posets from Proposition~\ref{def_maz_trace} but we use their representation using their corresponding family of configurations as in Proposition~\ref{prop_traces_as_event_strucs}. Therefore, we now interpret \HDML\ over restricted \HDAs\ as we discussed above.

\begin{proposition}[encoding LTrL]\label{prop_encoding_LTrL}
The language of LTrL consists of the propositional part of \HDML\ together with the following two definitions: 
\begin{itemize}
\item of the \textit{Until} operator \hfil $\varphi\,\overline{\until}\varphi'\defequal(\varphi\vee\terminate{}{\top})\untilL(\varphi'\wedge\neg\terminate{}{\top})$;
\item and the next step operator, for $a\in\Sigma$, \hfil $\overline{\langle a\rangle}\varphi\defequal\start{a}{\terminate{a}{\varphi}}$.
\end{itemize} 
When interpreted only in the states of a \HDA\ representing a Mazurkiewicz trace this language has the same behavior as the one presented in \cite{ThiagarajanW02LTrL}
\end{proposition}

\begin{proof}
The states of the \HDA\ are the configurations of the Mazurkiewicz trace. Thus, our definition of the LTrL language is interpreted in one trace at one particular configuration; as is done in \cite{ThiagarajanW02LTrL}. The original semantics of LTrL uses transitions from one configuration to another labeled by an element from the alphabet $\Sigma$ of the trace. It is easy to see that our syntactic definition of $\overline{\langle a\rangle}\varphi$ has the same interpretation as the corresponding one in \cite{ThiagarajanW02LTrL}. The proof is similar to the proof of Theorem~\ref{th_modal_logic}. In particular, when $\overline{\langle a\rangle}\varphi$ is interpreted in some state of the \HDA, i.e., in a configuration of the trace, then the formula $\varphi$ must hold in the state reached by going through a transition labeled with $a$. This means that we just made a single step, cf.\ the definition of \cite{ThiagarajanW02LTrL}, from the initial configuration to a new one where one new event labeled by $a$ has been added.

The \textit{Until} operator of \cite{ThiagarajanW02LTrL} has the same definition as the one in standard LTL but adapted to the Mazurkiewicz traces setting; thus, in the syntactic definition of $\overline{\until}$ we use the LTL-like $\untilL$ from Definition~\ref{def_untilLTL}.
\end{proof}

The ISTL logic is interpreted over communicating sequential agents (CSA), which are a restricted form of partial orders that still allows choices (as opposed to Mazurkiewicz traces). ISTL interprets the CTL until operator in configurations of a CSA. Therefore, we first need to find the exact restriction of \HDAs\ modeling CSA and then just use the syntactic definition $\exists\!\until$ of Proposition~\ref{prop_modeling_CTL}. We do not go into details here but discuss the undecidability results for $\exists\!\until$.

In \cite{Penczek92undecid} the $\exists\!\until$ is interpreted only over Mazurkiewicz traces and an undecidability proof is given using a simple trace that looks like a grid, with only two labels that are independent. The proof of \cite{AlurP99undecidable} uses a simple CSA but which allows choices. Intuitively, \cite{AlurP99undecidable} builds infinitely many grids as in \cite{Penczek92undecid}. Both these proofs work with infinite partial orders (i.e., infinitely many events): \cite{Penczek92undecid} works on an infinite grid; whereas \cite{AlurP99undecidable} works with infinitely many finite grids. There are two stages in these algorithms: the first is to encode all and only these infinite structures with some formula (for which the \textit{Until} definitions are not even needed, but only their weaker forms like $\exists G$ are enough); the second stage is to encode the actual tests in the undecidability problem (the tiling problem in \cite{Penczek92undecid} and the Post correspondence problem in \cite{AlurP99undecidable}). The first stage can be seen as setting the board for the undecidable problem. 

We do not pursue further here investigation into the (un)decidability of \HDML\ with the \textit{Until} operator.

%
%
%

\section{Expressiveness in terms of bisimulations}\label{sec_express}

There are various ways of investigating the expressiveness of a logic. One way that we explored in the previous section is to see what other logics can be syntactically encoded into the studied logic and to isolate the exact restriction of the studied logic (and its models) that belongs to the encoded logic.

Another way of looking at the expressiveness of a modal logic is by investigating the kind of bisimulation that it captures. In this section we do this for \HDML, with the aim to get more insights into the distinguishing power of the basic language of \HDML. By distinguishing power we mean what kind of (two) models can be distinguished by a single \HDML\ formula and what models are indistinguishable. The notion of \textit{indistinguishable} is given through an appropriate bisimulation; i.e., if the two models are bisimilar (for some specific notion of bisimulation) then an observer cannot distinguish them. The observer, in our case, has only the power to test logical \HDML\ formulas on the two models.
Since we will  refer to works that consider labeled transition systems, we will use the labeled versions of the \HDML\ modalities as in Definition~\ref{def_labeled_modal}.

Other expressiveness results for modal (temporal) logics include investigations into what exact subset of first (or second) order logic they capture, as is done for linear time temporal logic \cite{kamp68phd} (see \cite{emerson90temporal.logic} for an overview) or for the LTrL \cite{ThiagarajanW02LTrL}. We do not pursue this line of research here.

\HDML\ captures precisely the split-bisimulation and is strictly coarser than ST-bisimulation or history preserving bisimulation. Therefore, we confine our presentation here to only split-bisimulation, and discuss shortly the reasons that make \HDML\ less expressive than the other bisimulations on \HDAs.

\begin{definition}[split-bisimulation]\label{def_splitbisim}
The \splitpath\ of a finite path $\pi$ in a \HDA\ is the sequence  $\splitpath(\pi)\defequal\sigma_{1}\dots\sigma_{n}$ where $\sigma_{i}=l_{i}(q^{i})^{+}$ if $\alpha^{i}=s$ and $\sigma_{i}=l_{i}(q^{i})^{-}$ if $\alpha^{i}=t$ for $1\leq i\leq n$. Two higher dimensional automata $(\modelH_{A},q_{A}^{0})$ and $(\modelH_{B},q_{B}^{0})$ (with $q_{A}^{0}$ and $q_{B}^{0}$ two initial cells) are \emph{split-bisimulation equivalent} if there exists a binary relation $R$ between their paths starting at $q_{A}^{0}$ respectively $q_{B}^{0}$ that respects the following:
\begin{enumerate}
\item if $\pi_{A}R\pi_{B}$ then $\splitpath(\pi_{A})=\splitpath(\pi_{B})$;
\item if $\pi_{A}R\pi_{B}$ and $\pi_{A}\reachcell\pi_{A}'$ then $\exists\pi_{B}'$ with $\pi_{B}\reachcell\pi_{B}'$ and $\pi_{A}'R\pi_{B}'$;
\item if $\pi_{A}R\pi_{B}$ and $\pi_{B}\reachcell\pi_{B}'$ then $\exists\pi_{A}'$ with $\pi_{A}\reachcell\pi_{A}'$ and $\pi_{A}'R\pi_{B}'$;
\end{enumerate}
Denote this as $(\modelH_{A},q_{A}^{0})\splitequiv (\modelH_{B},q_{B}^{0})$.
\end{definition}

The ST-bisimulation replaces the first requirement with equality between ST-traces of the two paths. Intuitively, the ST-trace of a path is like the split-trace only that the end labels $l_{i}(\cdot)^{-}$ are keeping count of which start label they match with; i.e., $l_{i}(\cdot)^{j}$ where at the $j^{th}$ point the corresponding event has been started. Therefore, ST-traces know exactly which event ends; whereas the split-traces may confuse this. History preserving bisimulation is defined using the notions of adjacency and homotopy for \HDA\ and intuitively, for some cell in the \HDA\ we have a grip on its history also. Thus, history preserving bisimulation has access to the whole partially ordered history of the current executing events, ST-bisimulation has access only to some point from the past (i.e., the origin of some event), whereas the split-bisimulation has only a notion of previous step on the path. We come back to these intuitions throughout this section.

A modal logic is said to capture some equivalence relation $\sim$ if for any two models $\modelH$ and $\modelH'$, they are equated by the $\sim$ relation iff they are \emph{modally equivalent}.

\begin{definition}[modal equivalence]\label{def_modal_equiv}
Define the \emph{\HDML\ modal equivalence} as the relation $\modalequiv$ s.t.:
\[
(\modelH,q)\modalequiv(\modelH',q') \mbox{ iff } \forall\varphi:\modelH,q\models\varphi\Leftrightarrow\modelH',q'\models\varphi.
\]
\end{definition}

To keep the presentation simple we will work with frames instead of models; i.e., with no propositional constants. Before presenting the formal result note that \HDML\ can distinguish branching points, as is the case with bisimulations opposed to trace equivalences; the standard example in process algebras ($a(b+c)$ vs.\ $ab+ac$) is distinguished by the \HDML\ formula $\startUniv{a}{\terminateUniv{a}{(\start{b}{\top}\wedge\start{c}{\top})}}$. \HDML\ also distinguishes between interleaving and split-2 concurrency, where the standard example of $a||b$ vs.\ $ab+ba$ is distinguished by the formula $\start{a}{\start{b}{\top}}$ which holds only for $a||b$.

\begin{proposition}[\HDML\ captures split-bisimulation]\label{prop_splitHDML}\ 

The relations \modalequiv\ and\ \ \splitequiv\ coincide.
\end{proposition}

\begin{proof}
Proving the inclusion $\splitequiv\,\subseteq\,\modalequiv$ is simple. Use induction on the structure of the formula and use the last two conditions for \splitequiv\  with a smallest extension of the paths, i.e., when only one simple step is added to the path. The split-traces give the label and the $s$ or $t$ needed (when working with $\start{}{}$ respectively $\terminate{}{}$).

Proving the other inclusion $\modalequiv\,\subseteq\,\splitequiv$ needs the standard assumptions of finite nondeterminism (or image-finite as it is also known) and finite concurrency. This proof uses \textit{reductio ad absurdum} to show that the relation \modalequiv\ is respecting the three conditions of Definition~\ref{def_splitbisim}. Showing these conditions for all the paths is inductive, starting with the empty path and making only simple steps of extending the paths in the conditions 2 and 3, because this is enough to get the general form of these conditions.

For the empty paths the condition 1 is trivially satisfied. We work here with simple steps that extend the path with $s$ maps labeled by some $a$; and the other map $t$ is treated analogous. Consider the initial cells $q_{A}\modalequiv q_{B}$, and that $s_{i}(q_{A}^{1})=q_{A}$ labeled by $a$ (i.e., we extend the empty split-trace with $a^{+}$). We will assume that there is no way of extending (with a single step) the empty path in $\modelH_{B}$ cf.\ condition 2 of Definition~\ref{def_splitbisim}: i.e., $\not\exists q_{B}^{1}$
s.t.\ $s_{i}(q_{B}^{1})=q_{B}$, for some $i$, and labeled with $a$, and modal equivalent $q_{B}^{1}\modalequiv q_{A}^{1}$. If the assumption holds because there is no way of starting an $a$-labeled event then the modal formula $\startUniv{a}{\bottom}$ holds in $q_{B}$. But because in $q_{A}$ holds $\start{a}{\top}$ and $q_{A}\modalequiv q_{B}$ then  we get a contradiction because $q_{B}\models\start{a}{\top}\wedge\startUniv{a}{\bottom}$. Because of the finite nondeterminism and finite concurrency, the set of cells $\{q_{B}^{1},\dots,q_{B}^{n}\}$ reachable by an $s$ map labeled by $a$ from $q_{B}$, is finite. It remains to check the modal equivalence of the new cells. Clearly the split-traces of the new paths are the same because we extend with the same $s$ map labeled with the same $a$. Assume that for each cell $q_{B}^{i}$ there exists some formula $\varphi^{i}$ that holds in $q_{A}^{1}$ but not in $q_{B}^{i}$. Hence, $q_{A}\models\start{a}{(\varphi^{1}\wedge\dots\wedge\varphi^{n})}$ but $q_{B}\not\models\start{a}{(\varphi^{1}\wedge\dots\wedge\varphi^{n})}$, which is a contradiction with the fact that $q_{A}$ and $q_{B}$ are modal equivalent (i.e., model the same formulas).
\end{proof}

Because split-bisimulation can distinguish choices, then \HDML\ can distinguish all the examples of \cite{GlabbeekV97splitting} that were meant there to distinguish between the many trace-based equivalences. In particular, \HDML\ distinguishes the $X_{n}^{\mathit{odd}}$ and $X_{n}^{\mathit{even}}$ pomset processes (in their \HDA\ representation) which are meant to distinguish the split-$n+1$ from the split-$n$ trace equivalences (e.g., the formula $\start{1}{(\start{2}{\top}\wedge \terminate{1}{(\start{0}{\terminate{0}{\start{2}{\terminate{2}{\start{2}{\top}}}}} \wedge \startUniv{2}{\terminateUniv{2}{\startUniv{0}{\terminateUniv{0}{\neg\start{1}{\top}}}}})})}$ distinguishes the two examples in \cite[Figure 2]{GlabbeekV97splitting} because it holds on $X_{2}^{\mathit{even}}$ but not on $X_{2}^{\mathit{odd}}$). Also, \HDML\ can distinguish the examples in \cite[Figure 3]{GlabbeekV97splitting} because the formula $\startUniv{a}{\startUniv{b}{\terminateUniv{b}{\terminateUniv{a}{\start{c}{\top}}}}}$ holds in the pomset process $Y$ but not in $X$ (in their \HDA\ presentation). This example is meant in \cite{GlabbeekV97splitting} to distinguish the ST-trace equivalence from all the split-$n$ trace equivalences because the two pomset processes are indistinguishable by any of the split-$n$ trace equivalences.

Nevertheless, when it comes to bisimulation equivalences \HDML\ captures only split-bisimulation. Intuitively, the examples above can be distinguished by \HDML\ because they have different branching points before the problematic autoconcurrency square. \HDML\ becomes stuck when it has to deal with autoconcurrency; i.e., when in a concurrency square with both sides labeled the same, \HDML\ cannot distinguish which of the two events it finishes. But ST-bisimulation and history preserving bisimulation can distinguish the two events by looking at the history. In particular, \HDML\ is unable to distinguish any of the ``owl'' examples of \cite{GlabbeekV97splitting} which are meant to separate the split-$n$-bisimulations. 

In conclusion, \HDML\ sits pretty low in the equivalences spectrum of van Glabbeek and Vaandrager \cite{GlabbeekV97splitting}, capturing only split-bisimulation. An interesting question for future work is what is a minimal extension to \HDML\ that captures ST-bisimulation, or history preserving bisimulation?

\private{
\section{History Higher Dimensional Modal Logic}

In this section we extend \HDML\ with two existential modalities that can be seen as the opposite of the during and after modalities of \HDML. We call this new logic \textit{History Higher Dimensional Modal Logic} (and denote \HHDML) because the two new modalities essentially look in the history to see where the current situation of the system is coming from.

}

\section{Conclusion}\label{sec_conclusion}

We have investigated a modal logic called \HDML\ which is interpreted over higher dimensional automata. The language of \HDML\ is simple, capturing both the notions of ``during'' and ``after''. The associated semantics is intuitive, accounting for the special geometry of the \HDAs. An adaptation of the filtration method was needed to prove decidability. We have associated to \HDML\ an axiomatic system which incorporates the standard modal axioms and has a few natural axioms extra, which are related to the cubical laws and to the dimensions of \HDAs.

We isolated axiomatically the class of \HDAs\ that encode Kripke structures and shown how standard modal logic is encoded into \HDML\ when interpreted only over these restricted \HDAs. We then showed how to extend the expressiveness of \HDML\ using the \textit{Until} operator by defining two kinds of \textit{Until} over \HDAs: one \untilL\ in the LTL style and one \untilC\ in the CTL style. Using the \untilC\ we showed how to encode syntactically the CTL $\exists\!\until$ into \HDML\ when interpreted over the Kripke \HDAs. We also showed how weaker concurrency models like Mazurkiewicz traces or (restrictions of) event structures can be encoded in \HDML\ and how some of their specific properties can be captured axiomatically only in the basic language of \HDML. We also looked at encoding specific logics for these restricted models (particularly the LTrL and ISTL) in the extensions of \HDML\ with the \textit{Until} operators.

In the last technical section we investigated the distinguishing power of \HDML\ and isolated the basic language of \HDML\ as capturing exactly the split-bisimulation. Nevertheless, the power to distinguish different branching points allowed \HDML\ to distinguish all the examples of \cite{GlabbeekV97splitting} that were meant there to separate the split-n-trace equivalences and the ST-trace equivalence. In this respect we gave some discussions trying to identify the weak points of \HDML\ compared to ST-bisimulation or history preserving bisimulation.

Interesting further work is to look more into the relation of \HDML\ (and its temporal extensions) with other logics for weaker models of concurrency like with the modal logic of \cite{mukund90eventStructLogic} for event structures or other logics for Mazurkiewicz traces. Particularly interesting is to give details of how or if the undecidability results of \cite{AlurP99undecidable,AlurMP05decid_partial} are applicable to our setting.

When investigating deeper the extensions of \HDML\ wrt.\ the captured bisimulations, the work of \cite{BaldanC10concur} is of particular relevance and comparisons with the logics presented there worth wild.


\vspace{3ex}
\noindent\textbf{Acknowledgements:}\hspace{2ex} I would like to thank Martin Steffen and Olaf Owe for their useful comments, as well as to the anonymous reviewers of previous drafts of this work.



%
%
%
%
%


\begin{thebibliography}{10}
\expandafter\ifx\csname url\endcsname\relax
  \def\url#1{\texttt{#1}}\fi
\expandafter\ifx\csname urlprefix\endcsname\relax\def\urlprefix{URL }\fi
\expandafter\ifx\csname href\endcsname\relax
  \def\href#1#2{#2} \def\path#1{#1}\fi

\bibitem{P10concur}
C.~Prisacariu, {Modal Logic over Higher Dimensional Automata}, in: P.~Gastin,
  F.~Laroussinie (Eds.), 21st International Conference on Concurrency Theory
  (CONCUR10), Vol. 6269 of {LNCS}, Springer, 2010, pp. 494--508.

\bibitem{pratt91hda}
V.~R. Pratt, Modeling concurrency with geometry, in: Principles of Programming
  Languages (POPL'91), 1991, pp. 311--322.

\bibitem{Glabbeek06HDA}
R.~J. van Glabbeek, {On the Expressiveness of Higher Dimensional Automata},
  Theoretical Computer Science 356~(3) (2006) 265--290.

\bibitem{GlabbeekG01refinement}
R.~J. van Glabbeek, U.~Goltz, Refinement of actions and equivalence notions for
  concurrent systems, Acta Informatica 37~(4/5) (2001) 229--327.

\bibitem{gupta94phd_chu}
V.~Gupta, {Chu Spaces: A Model of Concurrency}, Ph.D. thesis, Stanford
  University (1994).

\bibitem{Pratt03trans_cancel}
V.~R. Pratt, {Transition and Cancellation in Concurrency and Branching Time},
  Mathematical Structures in Computer Science 13~(4) (2003) 485--529.

\bibitem{MukundT96TrPTL}
M.~Mukund, P.~S. Thiagarajan, {Linear Time Temporal Logics over Mazurkiewicz
  Traces}, in: Mathematical Foundations of Computer Science (MFCS'96), Vol.
  1113 of {LNCS}, Springer, 1996, pp. 62--92.

\bibitem{DiekertG06}
V.~Diekert, P.~Gastin, {From local to global temporal logics over Mazurkiewicz
  traces}, Theoretical Computer Science 356~(1-2) (2006) 126--135.

\bibitem{ThiagarajanW02LTrL}
P.~S. Thiagarajan, I.~Walukiewicz, {An Expressively Complete Linear Time
  Temporal Logic for Mazurkiewicz Traces}, Information and Computation 179~(2)
  (2002) 230--249.

\bibitem{AlurP99undecidable}
R.~Alur, D.~Peled, Undecidability of partial order logics, Information
  Processing Letters 69~(3) (1999) 137--143.

\bibitem{mukund90eventStructLogic}
K.~Lodaya, M.~Mukund, R.~Ramanujam, P.~S. Thiagarajan, {Models and Logics for
  True Concurrency}, Tech. Rep. IMSc-90-12, Inst. Mathematical Science, Madras,
  India (1990).

\bibitem{LodayaPRT95}
K.~Lodaya, R.~Parikh, R.~Ramanujam, P.~S. Thiagarajan, {A Logical Study of
  Distributed Transition Systems}, Information and Computation 119~(1) (1995)
  91--118.

\bibitem{ClarkeES83CTL}
E.~M. Clarke, E.~A. Emerson, A.~P. Sistla, Automatic verification of finite
  state concurrent systems using temporal logic specifications, in: Principles
  of Programming Languages (POPL'83), 1983, pp. 117--126.

\bibitem{01modalLogicBook}
P.~Blackburn, M.~de~Rijke, Y.~Venema, Modal Logic, Vol.~53 of Cambridge Tracts
  in Theoretical Computer Science, Cambridge Univ. Press, 2001.

\bibitem{Pratt78exp-time.alg.pdl}
V.~R. Pratt, {A Practical Decision Method for Propositional Dynamic Logic:
  Preliminary Report}, in: Symposium on Theory of Computing (STOC'78), ACM
  Press, 1978, pp. 326--337.

\bibitem{Pnueli77temporal_logic}
A.~Pnueli, The temporal logic of programs, in: Symposium on Foundations of
  Computer Science (FOCS'77), IEEE Computer Society Press, 1977, pp. 46--57.

\bibitem{Penczek92undecid}
W.~Penczek, {On Undecidability of Propositional Temporal Logics on Trace
  Systems}, Information Processing Letters 43~(3) (1992) 147--153.

\bibitem{AlurMP05decid_partial}
R.~Alur, K.~L. McMillan, D.~Peled, {Deciding Global Partial-Order Properties},
  Formal Methods in System Design 26~(1) (2005) 7--25.

\bibitem{Mazurkiewicz88tracesTutorial}
A.~W. Mazurkiewicz, Basic notions of trace theory., in: REX Workshop, Vol. 354
  of {LNCS}, Springer, 1988, pp. 285--363.

\bibitem{Pratt86pomsets}
V.~R. Pratt, {Modeling Concurrency with Partial Orders}, Journal of Parallel
  Programming 15~(1) (1986) 33--71.

\bibitem{NielsenPW79eventstructures}
M.~Nielsen, G.~D. Plotkin, G.~Winskel, Petri nets, event structures and
  domains., in: Semantics of Concurrent Computation, Vol.~70 of {LNCS},
  Springer, 1979, pp. 266--284.

\bibitem{Pratt00HDArev}
V.~R. Pratt, Higher dimensional automata revisited, Mathematical Structures in
  Computer Science 10~(4) (2000) 525--548.

\bibitem{RozoyT91MazAsEventStrucs}
B.~Rozoy, P.~S. Thiagarajan, Event structures and trace monoids, Theoretical
  Computer Science 91~(2) (1991) 285--313.

\bibitem{DiekertG00LTL_traces}
V.~Diekert, P.~Gastin, {LTL Is Expressively Complete for Mazurkiewicz Traces},
  in: International Colloquium on Automata, Languages and Programming
  (ICALP'00), Vol. 1853 of {LNCS}, Springer, 2000, pp. 211--222.

\bibitem{LodayaRT92csa}
K.~Lodaya, R.~Ramanujam, P.~S. Thiagarajan, {Temporal Logics for Communicating
  Sequential Agents: I}, International Journal on Foundations of Computer
  Science 3~(2) (1992) 117--159.

\bibitem{harel00dynamicLogic}
D.~Harel, D.~Kozen, J.~Tiuryn, Dynamic Logic, MIT Press, 2000.

\bibitem{kamp68phd}
H.~Kamp, {Tense Logic and the Theory of Linear Orders}, Ph.D. thesis, UCLA
  (1968).

\bibitem{emerson90temporal.logic}
E.~A. Emerson, {Temporal and Modal Logic}, in: Handbook of Theoretical Computer
  Science, Volume B, 1990, pp. 995--1072.

\bibitem{GlabbeekV97splitting}
R.~J. van Glabbeek, F.~W. Vaandrager, {The Difference between Splitting in n
  and n+1}, Information and Computation 136~(2) (1997) 109--142.

\bibitem{BaldanC10concur}
P.~Baldan, S.~Crafa, A logic for true concurrency, in: P.~Gastin,
  F.~Laroussinie (Eds.), 21st International Conference on Concurrency Theory
  (CONCUR10), Vol. 6269 of {LNCS}, Springer, 2010, pp. 147--161.

\end{thebibliography}

\newpage

\appendix

\section{Completeness}

This section is not finished. It presents the main ideas and steps needed to prove the completeness of the axiomatic system for \HDML\ from Table~\ref{table_HDMLaxioms}; but still some details need to be fit into place. Any comments on this proof are welcome. The complications and details of this completeness proof are as such because of the intricate geometrical structure of the \HDA\ model that we work with. Some of the inductive reasoning that is needed does not follow standard patters, and makes arguments more involved.

\cp{For the editor: The paper includes a 10 pages appendix with the technical details of proof of completeness for the logic. The body of the paper includes a section where the method and the main steps that are used in the proof of completeness are described; the new concepts/techniques involved are given more intuition. I do not know if this proof should be included in the main body of the paper or not.}

We first fix some terminology and notation. Because of the finite model property for \HDML\ from Theorem~\ref{th_small_model} and because compactness fails cf.\ Theorem~\ref{th_compactness}, we are inclined to use atoms in the proof of completeness for \HDML\ and build finite canonical models (instead of using maximal consistent sets and standard canonical model). 

\begin{definition}[atoms]
Recall from Definition~\ref{def_closure} that\, $\closure{\varphi}$ is the subformula closure set of some given formula $\varphi$. 
Denote by $\neg\closure{\varphi}=\closure{\varphi}\cup\{\neg\varphi'\mid\varphi'\in\closure{\varphi}\}$ the set of subformulas together with their negated forms. 
A set of formulas $A$ is called an \emph{atom} for $\varphi$ if $A$ is a maximal consistent subset of $\neg\closure{\varphi}$. Denote $At(\varphi)$ the set of all atoms for $\varphi$.
For an atom $A\in At(\varphi)$ denote by $\hat{A}$ the formula obtained as $\phi_{1}\wedge\dots\wedge\phi_{n}$ with $\phi_{i}\in A$.
\end{definition}

Intuitively, atoms are sets of formulas which are free of immediate propositional inconsistencies (like $\phi\wedge\neg\phi$).

\begin{lemma}[properties on atoms]\label{lemma_atoms_prop}
Standard results for atoms tell us that for some formula $\varphi$ and any atom $A\in At(\varphi)$ is the case that:

\vspace{-1ex}\begin{enumerate}
\renewcommand{\theenumi}{(\roman{enumi})}
\item for all $\psi\in\neg\closure{\varphi}$ then only one of $\psi$ or $\neg\psi$ are in $A$;\label{atoms1}

%

\item for all $\psi\imply\psi'\in\neg\closure{\varphi}$ then  $\psi\imply\psi'\in A$ iff whenever $\psi\in A$ then $\psi'\in A$; \label{atoms6}

\item if $\psi\in\neg\closure{\varphi}$ and $\psi$ is consistent then there exists an $A\in At(\varphi)$ s.t.\ $\psi\in A$; (This is an analog of Lindenbaum's Lemma.)\label{atoms4}

\item any consistent set of formulas $S\subseteq\neg\closure{\varphi}$ can be grown to an atom $A_{S}\in At(\varphi)$.\label{atoms5}
\end{enumerate}
\end{lemma}

\begin{definition}[canonical saturated \HDA]\label{def_canonic_saturate}
A \HDA\ is called \emph{canonical for the formula $\varphi$} if a \emph{canonical labeling} $\labelc:Q\rightarrow At(\varphi)$ can be attached to the \HDA. A labeling function is \emph{canonical} if the following conditions hold:
\begin{enumerate}
%
%
\item for any $q_{n}\in Q_{n},q_{n-1}\in Q_{n-1}$, with $n>0$, and $\forall 0\leq i\leq n$, if $s_{i}(q_{n})=q_{n-1}$ then $\forall\psi\in\neg\closure{\varphi}$ if $\startUniv{}{\psi}\in\labelc(q_{n-1})$ then $\psi\in\labelc(q_{n})$,

\item for any $q_{n}\in Q_{n},q_{n-1}\in Q_{n-1}$, with $n>0$, and $\forall 0\leq i\leq n$, if $t_{i}(q_{n})=q_{n-1}$ then $\forall\psi\in\neg\closure{\varphi}$ if $\terminateUniv{\,}{\psi}\in\labelc(q_{n})$ then $\psi\in\labelc(q_{n-1})$.
\end{enumerate}

A canonical \HDA\ is called \emph{saturated} if:
\begin{enumerate}
\item whenever $\start{}{\psi}\in\labelc(q_{n-1})$ then $\exists q_{n}\in Q_{n}$ and $\exists 0\leq i\leq n$ s.t.\ $s_{i}(q_{n})=q_{n-1}$ and $\psi\in\labelc(q_{n})$,

\item whenever $\terminate{}{\psi}\in\labelc(q_{n})$ then $\exists q_{n-1}\in Q_{n-1}$ and $\exists 0\leq i\leq n$ s.t.\ $t_{i}(q_{n})=q_{n-1}$ and $\psi\in\labelc(q_{n-1})$.
\end{enumerate}

\end{definition}

\begin{lemma}[truth lemma]\label{lemma_truth}
In a canonical saturated \HDA\ \modelH\ for a formula $\varphi$, with the valuation defined as $\mathcal{V}(q_{n})=\{\phi\in\atomicformulas\mid\phi\in\labelc(q_{n})\}$, it holds that $\modelH,q_{n}\models\psi$ iff $\psi\in\labelc(q_{n})$, for any $\psi\in\neg\closure{\varphi}$.
\end{lemma}

\begin{proof}
By induction on the structure of $\psi$.

\Base 
$\psi=\phi\in\atomicformulas$. From the definition we have $\modelH,q_{n}\models\phi$ iff $\phi\in\mathcal{V}(q_{n})$ iff $\phi\in\labelc(q_{n})$.

\Induction 
The case for the Boolean connectives follows easily from the properties on atoms. Finally we treat cases for the two modalities. Consider the during modality. The left to right direction is based on the canonicity of \modelH.

We prove $\modelH,q_{n}\models\start{}{\varphi}\Rightarrow\start{}{\varphi}\in\labelc(q_{n})$. From the definition we know that $\exists q'\in Q_{n+1}$ and $\exists 0\leq i\leq n+1$ s.t.\ $s_{i}(q')=q_{n}$ and $\modelH,q'\models\varphi$. From the induction hypothesis we have that $\modelH,q'\models\varphi$ iff $\varphi\in\labelc(q')$. Together with the canonicity of \modelH\ we have that $\start{}{\varphi}\in\labelc(q_{n})$. Proof finished.

For the right to left direction we use the saturation of \modelH. 
We prove $\start{}{\varphi}\in\labelc(q_{n})\Rightarrow \modelH,q_{n}\models\start{}{\varphi}$. Using the saturation of \modelH\ we have that $\exists q_{n+1}\in Q_{n+1}$ and $\exists 0\leq i\leq n+1$ s.t.\ $s_{i}(q_{n+1})=q_{n}$ and $\varphi\in\labelc(q_{n+1})$. By the induction hypothesis it implies that $\modelH,q_{n+1}\models\varphi$. Thus, by the definition we have that $\modelH,q_{n}\models\start{}{\varphi}$.

The proof for the $\terminate{}{}$ modality is symmetric using the second conditions of canonicity and saturation of \modelH.
\end{proof}


For modal logics over complex structures like \HDAs\ the step-by-step method of proving completeness is a first candidate. But we cannot use it in the standard way with maximal consistent sets as the cells of the \HDA. Instead we will use atoms, i.e., finite maximal consistent sets. On the other hand, the standard way of using atoms in completeness proofs is to build a finite canonical model and show that it respects the required special structure. This is not easy in our case. In consequence we use a step-by-step method for building a finite model and label the cells with atoms. This model will have the right \HDA\ structure and will respect canonicity properties required by a truth lemma. In this sense we adapt and combine the two methods of step-by-step and atoms-based finite canonical models to \HDML. On the other hand the main difficulty of our proof is the construction method which is rather involved.
Note that we prove a weak completeness result, which is normal because a strong completeness is out of reach because of the compactness failure.

\cp{may remove depending on the proof which seams to make the whole model canonical...}

A first attempt to prove completeness is to try to build a canonical saturated model for any consistent formula. This fails, partly because \HDML\ is a forward looking logic but the special cubical geometry of the \HDAs\ require to construct the backwards part of the \HDA\ (that which is not reachable through the two modalities of \HDML). But it is not possible to guarantee the canonicity for this part. Nevertheless, the following notions and results tell us that we can ignore canonicity for this part of the model. Therefore when doing the actual step-by-step construction of the required \HDA\ for some arbitrary consistent formula we will concentrate on respecting canonicity only for the relevant (cf.\ the results below) part of the structure. 

\begin{definition}[pseudo \HDA]\label{def_pseudoModel}
For a \HDA\ $\modelH$ and a cell $q\in\modelH$ we call \emph{the forward generated pseudo \HDA\ for the cell $q$} (or \emph{pseudo \HDA} for short) the structure $\modelP{q}=(Q',\overline{s'},\overline{t'},l')$ obtained from \modelH\ by the following generative definition:
\begin{itemize}
\item $q\in Q'$;
\item if $q\in Q'$ then $\forall q_{s}\in Q$ if it exists $i$ s.t.\ $s_{i}(q_{s})=q$ then $q_{s}\in Q'$;
\item if $q\in Q'$ then $\forall q_{t}\in Q$ if it exists $i$ s.t.\ $t_{i}(q)=q_{t}$ then $q_{t}\in Q'$;
\item no other cell of $Q$ is in $Q'$;
\end{itemize}
and where $\overline{s'}\defequal \overline{s}|_{Q'}$, $\overline{t'}\defequal \overline{t}|_{Q'}$, and $l'\defequal l|_{Q'}$ are the respective restriction to this new set of cells $Q'$.
\end{definition}

Intuitively, the pseudo \HDAs\ are similar to the idea of a point-generated submodel in standard modal logic. The following lemma intuitively says that \HDML\ satisfaction is invariant under pseudo model construction.

\begin{lemma}[invariance under pseudo \HDAs]\label{lemma_pseudoInvariance}
For a \HDA\ \modelH\ and a cell $q\in\modelH$ for which $\modelP{q}$ denotes the pseudo \HDA\ for $q$, then for any \HDML\ formula $\varphi$ and any cell $q^{p}\in\modelP{q}$, we have:
\[
\modelH,q^{p}\models\varphi \mbox{\ \  iff\ \  }\modelP{q},q^{p}\models\varphi.
\]
\end{lemma}

\begin{proof}
The proof uses induction on the structure of the formula $\varphi$.
Since the pseudo \HDA\ does not change the valuation then the base case for propositional constants and the inductive cases for the Boolean operators are trivial, as we have to look at the same cell.

It remains to treat the inductive cases for the two \HDML\ modalities; we will treat only $\varphi=\start{}{\psi}$, and the other modality is treated analogous. Since the set of cells of the pseudo \HDA\ is just a subset of the original \modelH, i.e., $Q'\subseteq Q$, then the right-to-left implication is immediate (using induction on the subformula $\psi$). For the left-to-right implication consider that $\modelH,q^{p}\models\varphi$ and we show that $\modelP{q},q^{p}\models\varphi$. From the semantic definition we have that it exists $s_{i}(q_{n+1})=q^{p}$, for some $i$, with $\modelH,q_{n+1}\models\psi$. From the pseudo \HDA\ Definition~\ref{def_pseudoModel}, since $q^{p}\in\modelP{q}$ then also $q_{n+1}\in\modelP{q}$ and $s'_{i}(q_{n+1})=q^{p}$. From $\modelH,q_{n+1}\models\psi$, by the induction hypothesis we have that $\modelP{q},q_{n+1}\models\psi$. These imply the desired result $\modelP{q},q^{p}\models\start{}{\psi}$.
\end{proof}

\begin{definition}[pseudo canonicity]\label{def_pseudoCanonical}
We call a \HDA\ \emph{pseudo canonical for $q$} if its pseudo \HDA\ for $q$ is canonical (cf.\ Definition~\ref{def_canonic_saturate}). A pseudo canonical \HDA\ is called \emph{saturated} if its pseudo \HDA\ is saturated.
\end{definition}

\begin{lemma}[truth lemma for pseudo canonical \HDAs]\label{lemma_truth_Pseudo}
In a \HDA\ \modelH\ which is pseudo canonical for $q$ and saturated wrt.\ a formula $\varphi$, with the valuation defined as $\mathcal{V}(q_{n})=\{\phi\in\atomicformulas\mid\phi\in\labelc(q_{n})\}$, then for any $\psi\in\neg\closure{\varphi}$ and $q_{n}\in\modelP{q}$ it holds that
\[
\modelH,q_{n}\models\psi\mbox{\ iff\ }\psi\in\labelc(q_{n}).
\]
\end{lemma}

\begin{proof}
The proof follows from the Truth Lemma~\ref{lemma_truth}.
\end{proof}

To prove completeness of the axiomatic system all that remains is to show that for any consistent formula $\varphi$ we can build such a pseudo canonical saturated \HDA. During the step-by-step construction process we constantly struggle to saturate the \HDA\ (that we work with) while respecting the pseudo canonicity. Such not saturated \HDAs\ are called \emph{defective}, as they may have defects, which we formally define below. But important is that any of these defects can be repaired. This is what the repair lemma does, using the two \textit{enriching} and \textit{lifting} constructions. The completeness theorem then shows that while starting with a minimal pseudo canonical \HDA\ we can incrementally build a defect free pseudo canonical \HDA, i.e., a pseudo canonical and saturated \HDA.

\begin{definition}[defects]\label{def_defects}
There are two types of defects for \modelH\ (corresponding to a violation of a saturation condition): 
\begin{itemize}
\item a \emph{D1 defect} of \modelH\ is a cell $q_{n}\in Q_{n}$ with $\start{}{\psi}\in\labelc(q_{n})$ for which there is no $q_{n+1}\in Q_{n+1}$ and no $1\leq i\leq n+1$, with $s_{i}(q_{n+1})=q_{n}$ and $\psi\in\labelc(q_{n+1})$;

\item a \emph{D2 defect} of \modelH\ is a cell $q_{n}\in Q_{n}$ with $\terminate{}{\psi}\in\labelc(q_{n})$ for which there is no $q_{n-1}\in Q_{n-1}$ and no $1\leq i\leq n-1$, with $t_{i}(q_{n})=q_{n-1}$ and $\psi\in\labelc(q_{n-1})$.
\end{itemize}

\end{definition}

During the construction of the model we cannot work with atoms directly because we will revisit the same cell several times, each time needing to add more restrictions to its label. We are still working with atoms as labels, only that we consider all possible atoms that respect such properties (eg., all the atoms that could extend some consistent set of formulas). In the end of the construction we just pick one, to obtain the pseudo canonical saturated model we are seeking. The key result here is that all the constraints that are gathered during the construction should allow for the existence of at least one respecting atom. We use the following definitions.

\begin{definition}[potential labeling function]\label{def_potentialLabeling}
We define a \emph{potential labeling function} $\labelPotential:Q\rightarrow 2^{\constraints}$ which for any cell $q\in Q$ returns a set of constraints from $\constraints$. A constraint $c\in\constraints$ can be either a consistent set of formulas $S\in\closure{\varphi}$ (call this a \emph{set constraint}) or a formula $\start{}{\hat{A}}$ or $\terminate{}{\hat{A}}$, with $A\in At(\varphi)$ an atom, (call these \emph{existential constraints}). A potential labeling is called \emph{potential canonical} iff there exists some labeling function $\labelc$ s.t.\  for any cell $q\in Q$, $\labelc(q)$ is consistent with all the constraints $\labelPotential(q)$.
\end{definition}

\begin{lemma}\label{lemma_potentialNotCanon}
A potential labeling is \emph{not canonical} iff any of the following is the case:
\begin{itemize}
\item for some cell $q$ the union of all the set constraints in $\labelPotential(q)$ is inconsistent;
\item for some cell $q$ there exists a formula $\varphi\in A$ with $A$ appearing in one of the existential constraints as $\start{}{\hat{A}}\in\labelPotential(q)$ (or as $\terminate{}{\hat{A}}\in\labelPotential(q)$) for which there exists a corresponding formula $\startUniv{\,}{\neg\varphi}$ (respectively $\terminateUniv{\,}{\neg\varphi}$) in one of the set constraints of $\labelPotential(q)$.
\end{itemize}

\end{lemma}

\begin{proof}

\end{proof}

\begin{definition}
For two \HDAs, $\modelH_{1}$ and $\modelH_{2}$, with corresponding potential canonical labellings $\labelPotential^{1}$ respectively $\labelPotential^{2}$, we say that \emph{$\modelH_{2}$ extends $\modelH_{1}$} (written $\modelH_{2}\extends\modelH_{1}$) iff $\modelH_{2}$ has all the cells and maps of $\modelH_{1}$ and possibly some new cells and maps (i.e., some extra structure), and for all the old cells $q\in\modelH_{1}$ the constraints may only increase, i.e., $\labelPotential^{1}(q)\subseteq\labelPotential^{2}(q)$.
\end{definition}

Note that increasing the number of constraints means that there is less uncertainty about the ultimate atom that is going to label a cell.

The two constructions that we give below are working on pseudo canonical \HDAs, where the minimal such \HDA\ contains only one cell; this is the starting pseudo canonical \HDA\ in the completeness Theorem~\ref{th_completeness}.

For a D1 defect, i.e., a cell $q$ as in Definition~\ref{def_defects}, the \textit{enriching construction} adds one new cell that has $q$ as one of its sources and is labeled with an atom containing $\varphi$. Moreover, all the other maps of this new cell need to be added, together with all the necessary new cells, respecting the cubical laws. The new enriched \HDA\ will not have $q$ as a D1 defect any  more.

\begin{lemma}[enriching construction]\label{lemma_enriching}
For an \modelH\ with an associated potential canonical labeling \labelPotential, and for a defective cell $q$ (i.e., $\start{}{\varphi}\in\labelPotential(q)$) the following construction, which we call \emph{enriching of the \modelH\ wrt.\ $q$ and $\start{}{\varphi}$} builds an $\modelH'$ which extends \modelH\ (i.e., $\modelH'\extends\modelH$) and does not contain the defect of $q$ nor introduces new defects for $q$.
\end{lemma}
\vspace{-1ex}{\footnotesize
\begin{lstlisting}[numbers=left,numberstyle=\tiny,numbersep=5pt,mathescape=true,escapechar=\%,morekeywords={return,function,for,if,else}]
function enrich(n,q,$\varphi$){
 $Q_{n+1}$ := $Q_{n+1}\cup \{q_{n+1}\}$; //fresh cell
 update map $s_{n+1}$ %s.t.% $s_{n+1}(q_{n+1})=q$; 
 add constraints $\labelPotential(q_{n+1})=\labelPotential(q_{n+1})\cup\{\varphi\}\cup\{\psi\mid \startUniv{}{\psi}\in\labelPotential(q)\}$;
 add constraints $\labelPotential(q)=\labelPotential(q)\cup\start{}{\widehat{\labelc(q_{n+1})}}$;
 addSourceMaps(n+1,$q_{n+1}$,0,$\emptyset$);
 addTargetMaps(n+1,$q_{n+1}$,0,$\emptyset$);
}
function addSourceMaps($k$,$q$,$m$,$q'$){
 if(k>=1){
  $Q_{k-1}$ := $Q_{k-1}\cup \{q_{k-1}^{1},\dots, q_{k-1}^{k-1-m}\}$; //fresh cells
  for($l$=1 to m){ 
   update map $s_{k-l}$ %s.t.% $s_{k-l}(q)=s_{k-m}(s_{k-l+1}(q'))$; 
   add constraints $\labelPotential(q)=\labelPotential(q)\cup\{\psi\mid \startUniv{}{\psi}\in\labelPotential(s_{k-m}(s_{k-l+1}(q')))\}$;
   add constraints $\labelPotential(s_{k-l}(q))=\labelPotential(s_{k-l}(q))\cup\start{}{\widehat{\labelc(q)}}$;
  }
  for(i=k-1-m to 1){
   update map $s_{i}$ %s.t.% $s_{i}(q)=q_{k-1}^{i}$;
   update map $s_{k-1}$ %s.t.% $s_{k-1}(q_{k-1}^{i})=s_{i}(s_{k}(q))$;
   add constraints $\labelPotential(q_{k-1}^{i})=\labelPotential(q_{k-1}^{i})\cup\{\psi\mid \startUniv{}{\psi}\in\labelPotential(s_{i}(s_{k}(q)))\}\cup\start{}{\widehat{\labelc(q)}}$;
   add constraints $\labelPotential(s_{k-1}(q_{k-1}^{i}))=\labelPotential(s_{k-1}(q_{k-1}^{i}))\cup\start{}{\widehat{\labelc(q_{k-1}^{i})}}$;
   addSourceMaps(k-1,$q_{k-1}^{i}$,k-1-m-i,q);
   addTargetMaps(k-1,$q_{k-1}^{i}$,0,$\emptyset$);
   add constraints $\labelPotential(q)=\labelPotential(q)\cup\{\psi\mid \startUniv{}{\psi}\in\labelPotential(q_{k-1}^{i})\}$;
  }}}
function addTargetMaps($k$,$q$,$m$,$q'$){
 if(k >= 1){
  $Q_{k-1}$ := $Q_{k-1}\cup \{q_{k-1}^{1},\dots, q_{k-1}^{k-m}\}$; //fresh cells
  for($l$=0 to m-1){ 
   update map $t_{k-l}$ %s.t.% $t_{k-l}(q)=t_{k+1-m}(t_{k-l+1}(q'))$; 
   add constraints $\labelPotential(t_{k-l}(q))=\labelPotential(t_{k-l}(q))\cup\{\psi\mid \terminateUniv{\,}{\psi}\in\labelPotential(q)\}\label{addTM1loop}$;
   add constraints $\labelPotential(q)=\labelPotential(q)\cup\terminate{}{\widehat{\labelc(t_{k-l}(q))}}$;
  }
  for(i=k-m to 1){
   update map $t_{i}$ %s.t.% $t_{i}(q)=q_{k-1}^{i}$;
   add constraints $\labelPotential(q_{k-1}^{i})=\{\psi\mid \terminateUniv{\,}{\psi}\in\labelPotential(q)\}\label{addTM2loop}$;
   add constraints $\labelPotential(q)=\labelPotential(q)\cup\terminate{}{\widehat{\labelc(q_{k-1}^{i})}}$;
   if(k > 1){
    for(j=1 to k-1){ //add k-1 maps $s_{j}$ to $q_{k-1}^{i}$ cf. cubical laws
     if(j<i){ 
      update map $s_{j}$ %s.t.% $s_{j}(t_{i}(q))=t_{i-1}(s_{j}(q))$;
      add constraints $\labelPotential(q_{k-1}^{i})=\labelPotential(q_{k-1}^{i})\cup\{\psi\mid \startUniv{\,}{\psi}\in\labelPotential(t_{i-1}(s_{j}(q)))\}\label{addTM3loopIf}$;
      add constraints $\labelPotential(t_{i-1}(s_{j}(q)))=\labelPotential(t_{i-1}(s_{j}(q)))\cup\start{}{\widehat{\labelc(q_{k-1}^{i})}}$;
     }else {   
      update map $s_{j}$ %s.t.% $t_{i}(s_{j+1}(q))=s_{j}(t_{i}(q))$;
      add constraints $\labelPotential(q_{k-1}^{i})=\labelPotential(q_{k-1}^{i})\cup\{\psi\mid \startUniv{\,}{\psi}\in\labelPotential(t_{i}(s_{j+1}(q)))\}\label{addTM3loopElse}$;
      add constraints $\labelPotential(t_{i}(s_{j+1}(q)))=\labelPotential(t_{i}(s_{j+1}(q)))\cup\start{}{\widehat{\labelc(q_{k-1}^{i})}}$;
    }}
   addTargetMaps(k-1,$\,q_{k-1}^{i}$,k-m-i,q);
   }}}}
\end{lstlisting}
}

\begin{proof}
The proof has five stages. The first three are meant to show that the enriched model is an extension of the old model (i.e., $\modelH'\extends\modelH$): 1) we first show that the structure of the old \HDA\ is untouched, i.e., all old cells and maps are in place; 2) we then show that all set constraints are still consistent sets; 3) the third step shows that . Basically the steps two and three are corresponding to Lemma~\ref{lemma_potentialNotCanon} to show that the new potential labeling is still potential canonical, i.e., that there still exists a way of instantiating the constraints to atoms. The fourth stage shows that $\modelH'$ is a model indeed, i.e., that all the maps are in place and all necessary cubical laws are respected. The last stage shows that the enriched model does not have the old defect  and that no new defects are introduced in the potential labeling of the initial cell $q$.

First remark that we do not change the initial shape of the original \modelH; we only add fresh cells and fresh maps for these cells; we also add maps to old cells connected to new cells. This concludes the first stage in proving that $\modelH'\extends\modelH$. 
The second stage is proven as Lemma~\ref{lemma_enrich_s2}, whereas the third stage is proven as Lemma~\ref{lemma_enrich_s3}. Therefore, $\modelH'\extends\modelH$.

We show next that we indeed construct a higher dimensional structure. A careful reading of the enriching construction should answer this question in affirmative. We need to make sure that to each new cell we add all the $s$ and $t$ maps according to its dimension and that we link these maps correctly according to the cubical laws. 

The enriching construction proceeds as follows. It takes our initial cell $q$ and its dimension $n$ and the formula that gives the D1 defect. It adds a new cell $q_{n+1}$ of dimension one greater than $q$ and links this with $q$ through the $s_{n+1}$ map. It labels the new cell s.t.\ the defect of $q$ is repaired. The labeling is not important for our current argument but it is used in the argument for canonicity. 
To have the new cell $q_{n+1}$ correctly added we need to add $n$ more $s$ maps (i.e., the $s_{i}$ maps with $1\leq i\leq n$) and $n+1$ more $t$ maps to it. The $s$ maps are added by the \lstinline[mathescape=true,escapechar=\$]!addSourceMaps! and the $t$ maps are added by the \lstinline[mathescape=true,escapechar=\$]!addTargetMaps!. 

Consider now the \lstinline[mathescape=true,escapechar=\$]!addSourceMaps! function which takes as arguments the cell to which it must add the maps and the dimension of this cell, together with two other arguments used for bookkeeping of the cubical laws that need to be added for each cell. More precisely, the $m$ argument records how many cubical laws the $q$ cell enters into. Note that this function (the same as \lstinline[mathescape=true,escapechar=\$]!addTargetMaps!) adds maps only if the dimension of the cell is greater than $0$, because, by definition, states in a \HDA\ have no maps. \lstinline[mathescape=true,escapechar=\$]!addSourceMaps! adds only $k-1$ maps to its cell argument because one $s$ map has already been added before (e.g., for $q_{n+1}$ we have added the map $s_{n+1}$ and it remains to add the other maps from $s_{1}$ to $s_{n}$). All these maps link to new cells of dimension one lower (i.e., dimension $k-1$). Actually there are less new cells because some of the $s$ maps must link to already existing cell so to respect the cubical laws. The $m$ argument tells how many $s$ maps should come only from cubical laws and hence, we add only $k-1-m$ new cells. The next loop adds these maps respecting the cubical laws; e.g., for the cell $q_{n}^{n-1}=s_{n-1}(q)$ we add the map $s_{n-1}(q_{n}^{n-1})$ as the result of $s_{n-1}(s_{n}(q))$ (which are cells that have already been added) because of the cubical law $s_{n-1}(s_{n}(q))=s_{n-1}(s_{n-1}(q))$. In fact, for the cell $q_{n}^{1}$ each of its $s$ maps links to some existing cell, thus no new cells are added. 

Each of the $k-1-m$ new cells are linked with $q$ by the corresponding $s_{i}$ map. It is also added the $s_{k-1}$ map (i.e., the map with greatest index among the $k-1$ maps that the cell needs). This is done so to respect the cubical laws $s_{i}(s_{k}(q))=s_{k-1}(s_{i}(q))$. We now need to recursively add the required $s$ and $t$ maps for the new cell. We call the  \lstinline[mathescape=true,escapechar=\$]!addSourceMaps! for this cell $q_{k-1}^{i}$ of dimension $k-1$ and, depending on the index $i$ in the loop, we specify that $k-1-m-i$ maps should be added directly through the cubical laws and not by using new cells. We must also carry along the node $q$ to which the cubical laws link. We also add the $t$ maps for $q_{k-1}^{i}$ by calling the \lstinline[mathescape=true,escapechar=\$]!addTargetMaps! function.

The function \lstinline[mathescape=true,escapechar=\$]!addTargetMaps! adds all the $t$ maps of the cell (not one less as the \lstinline[mathescape=true,escapechar=\$]!addSourceMaps! is doing). \lstinline[mathescape=true,escapechar=\$]!addTargetMaps! also tries to respect the cubical laws first, and thus the $m$ argument tells which maps come only from a cubical law like $t_{i}(t_{j}(q))=t_{j-1}(t_{i}(q))$. For a cell $q$ of dimension $k$ \lstinline[mathescape=true,escapechar=\$]!addTargetMaps! adds $k-m$ new cells of dimension $k-1$ and links each of these cells through a corresponding $t_{i}$ map. For each new cell a recursive call to \lstinline[mathescape=true,escapechar=\$]!addTargetMaps! is needed to add all the necessary $t$ maps. The $s$ maps of the new cells are added in the end taking care that all the cubical laws of the form $s_{i}(t_{j}(q))=t_{j-1}(s_{i}(q))$ are respected. All these $s$ maps are linked to cells which come from $t$ maps that have been added by the \lstinline[mathescape=true,escapechar=\$]!addSourceMaps! function before.
\end{proof}

\begin{lemma}\label{lemma_enrich_s2}
The new sets of formulas that are added by the enrich algorithm of Lemma~\ref{lemma_enriching} (i.e., at lines 4, 14, 20, 24, 31, 36, 42, 46) are consistent sets.
\end{lemma}

\begin{proof}
This lemma is essentially the second stage in the proof of the correctness of the enrich construction from Lemma~\ref{lemma_enriching}.

Consider only the first set that we construct at line 4. The proof for all the other sets is analogous and simpler. Assume that this set is not consistent, which means two cases: 1) $\psi_{1}\wedge\dots\wedge\psi_{k}\imply\bot$, for $\psi_{i}\in\labelPotential(q_{n+1})$ and $\startUniv{}{\psi_{i}}\in\labelPotential(q)$ with $1\leq i\leq k$, and 2) $\psi_{1}\wedge\dots\wedge\psi_{k}\imply\neg\varphi$, for $\psi_{i}\in\labelPotential(q_{n+1})$, $\startUniv{}{\psi_{i}}\in\labelPotential(q)$, and $\start{}{\varphi}\in\labelPotential(q)$. For case 1) we know from modal logic that $\startUniv{}{\psi_{1}}\wedge\dots\wedge\startUniv{}{\psi_{k}}\imply\startUniv{}{(\psi_{1}\wedge\dots\wedge\psi_{k})}$ which, together with the assumption, it means that $\startUniv{}{\psi_{1}}\wedge\dots\wedge\startUniv{}{\psi_{k}}\imply\startUniv{}{\bot}$. This means that $\startUniv{}{\bot}\in\labelPotential(q)$ which is a contradiction with the fact that $\labelPotential(q)$ contains an existential modality, namely, $\start{}{\varphi}$.\footnote{For the same argument we could have used the existential constraint that is imposed on $q$ at line 5, which implies that the set constraint (i.e., any atom for $q$ containing the universal modalities $\startUniv{}{\psi_{i}}$) must be consistent with $\start{}{\top}$; which results in a contradiction with the deduced $\startUniv{}{\bottom}$.} For case 2) we follow a similar argument to obtain $\startUniv{}{\psi_{1}}\wedge\dots\wedge\startUniv{}{\psi_{k}}\imply\startUniv{}{(\psi_{1}\wedge\dots\wedge\psi_{k})}\imply\startUniv{}{\neg\varphi}\imply\neg\start{}{\varphi}$. But this is a contradiction because $\labelPotential(q)$ already contains $\start{}{\varphi}$ and hence would make $\labelPotential(q)$ inconsistent, contradicting the assumption of potentially canonical labeling of the old \HDA.

Note that throughout the rest of the paper when we write $\varphi\in\labelPotential(q)$ we mean that the formula $\varphi$ is part of one of the set constraints in $\labelPotential(q)$; we use the same notation for the fact that the formula is part of a single constraint when this is clear from the context, as is the case in the paragraph above where we consider only the set constraint build at line 4.

Therefore, we do not need to wary about inconsistencies coming from inside one of the new set constraints that the algorithm builds. It remains to see if any of the formulas in the new set constraints is inconsistent with some formula already existing in the potential label of the cell where the new set constraint is added. This cannot happen at line 4 because the cell $q_{n+1}$ is fresh and has at this point no label attached. Inconsistencies may come from the \lstinline[mathescape=true,escapechar=\$]!addSourceMaps! function 
that is called recursively in a depth-first manner.

We explain now how this function works and how the source maps are added by the enrich function.

Starting with the defective cell of dimension $n$ the enrich function adds a new cell $q_{n+1}$ of dimension $n+1$ and adds its highest $s$ map, i.e., $s_{n+1}$. Then it calls the function \lstinline[mathescape=true,escapechar=\$]!addSourceMaps! to add the rest $n$ source maps.
This one works in a depth-first manner and adds source maps starting with the highest one. The first call, at line 6 in the body of the enrich itself, adds one less $s$ map, but the other recursive calls add all the maps. Because of the cubical laws, some the the $s$ maps will reach cells that already exist. This is the reason for going in a decreasing order adding first the highest $s$ map, and adding $s_{1}$ last. In fact $s_{1}$ will have each of its $s$ maps connected to some existing cell. For all the fresh cells that are added, the highest $s$ map will connect to a cell from the old \modelH. The rest of the cells connect to other fresh cells. This is part of the reason for which \lstinline[mathescape=true,escapechar=\$]!addSourceMaps! works in a depth-first manner when adding the labeling constraints.
At the deepest level $Q_{1}$ the new cell will connect its only $s_{1}$ map to an old cell from $\modelH$ and its new set constraint will be build from the potential canonical old label. When closing the recursions and going up the levels, the function builds the new set constraints from all these lower cells that are connected through the $s$ maps (one of them is from the old \modelH, remember). Therefore, the set constraints of all the fresh cells are eventually build only from the labels of old cells.

Consider any of the fresh cells added by the \lstinline[mathescape=true,escapechar=\$]!addSourceMaps! function,
i.e., except the $q_{n+1}$ cell added in the body of the enrich. Assume that two formulas $\varphi$ and $\neg\varphi$ come from two different sources containing each a universal formula $\startUniv{}{\varphi}$ respectively $\startUniv{}{\neg\varphi}$ (as these cannot come from the same source). Because we build these set constraints only from other set constraints from lower level cells reached through $s$ maps it is clear that eventually we reach one of the old cells (from the old \modelH) which contains both $\startUniv{}{^{k}\varphi}$ and $\startUniv{}{^{k}\neg\varphi}$ (we denoted by $\startUniv{}{^{k}}$ the application of $n$ times of the $\startUniv{}{}$ modality) with $k\leq n$. This results in $\startUniv{}{^{k}\bottom}$. But because the original $q$ contains $\start{}{\varphi}$ and each of the old cells reached through an $s$ map has an existential constraint it implies that any of these old cells is consistent with $\start{}{\top}$, and also the problematic one that by assumption above would have the formula $\startUniv{}{\bottom}$. Thus we get inconsistency in the old \modelH, and hence a contradiction.
For the first fresh cell $q_{n+1}$ the same argument holds only that we need to treat the case when we actually reach the initial defective formula $\start{}{\varphi}$. This immediately exhibits the inconsistency with the formula $\startUniv{}{\neg\varphi}$, hence the contradiction with the fact that the set constraints of the old \modelH\ are consistent.

There is no other way of inconsistencies to creep in the new set constraints for the fresh cells added by the \lstinline[mathescape=true,escapechar=\$]!addSourceMaps! function.
We continue the argument for the \lstinline[mathescape=true,escapechar=\$]!addTargetMaps! function.

If in \lstinline[mathescape=true,escapechar=\$]!addSourceMaps! 
function the accumulation of the set constraints was done in a bottom up fashion after settling the lower cells, i.e., at line 24; now in \lstinline[mathescape=true,escapechar=\$]!addTargetMaps! 
function the collection is done in a top down fashion, for $t$ maps collecting from all reachable cells that were previously labeled, i.e., at line 31. 
The \lstinline[mathescape=true,escapechar=\$]!addTargetMaps! function
works on adding the new $t$ maps also starting with the highest one and always finishing with $t_{1}$. But many of the maps reach already existing cells: the first \lstinline[mathescape=true,escapechar=\$,morekeywords={for}]!for! loop
takes care of such $t$ maps, whereas in the second loop all the $s$ maps reach cells that have been added in the \lstinline[mathescape=true,escapechar=\$]!addSourceMaps! 
function. For the old cells in the first loop, the function updates the already existing potential labeling by adding new set constraints. For the fresh cells the second loop, at line 36, adds a completely new potential label with one set constraint, and in the next line adds also an existential constraint. All these fresh cells reached through the $t$ maps have their labels updated when the $s$ maps are added. These connect to already existing cells, from where all the boxes have to be accumulated in the label of the new cell, i.e., these are the contents at lines~\ref{addTM3loopIf} and \ref{addTM3loopElse}. Note that for some fresh cell the algorithm adds one set constraint coming from each of its source maps.

After this intuitive presentation it is easy to identify where the inconsistencies in the set constraints can come from: 
\begin{enumerate}
\item\label{inconsistency1} either in the first loop at line~\ref{addTM1loop} when collecting a new box constraint having a box formula $\terminateUniv{\,}{\varphi}$ where the formula $\neg\varphi$ may already be in the potential label $\labelPotential(t_{k-l}(q))$ as coming from before from some formula $\terminateUniv{\,}{\neg\varphi}$ in a box constraint of some cell connected to $t_{k-l}(q)$ through a $t$ map;

\item\label{inconsistency2} in the second loop when $\varphi$ comes in the label $\labelPotential(q_{k-1}^{i})$ from the box formula $\terminateUniv{\,}{\varphi}$ of a cell linked to $q_{k-1}^{i}$ through a $t$ map and the formula $\neg\varphi$ comes from a box formula $\startUniv{}{\neg\varphi}$ at lines~\ref{addTM3loopIf} or \ref{addTM3loopElse} coming from the box constraints of some cell that is connected to $q_{k-1}^{i}$ through a $s$ map;

\item\label{inconsistency3} or when two box formulas $\startUniv{}{\varphi}$ and $\startUniv{}{\neg\varphi}$ are in the labels of two cells connected to $q_{k-1}^{i}$ through $s$ maps; i.e., in the third loop at lines ~\ref{addTM3loopIf} or \ref{addTM3loopElse}.
\end{enumerate}

For \ref{inconsistency1} it means we are in the first loop, at line 31, in the setting of the cubical law $t_{k+1-m}(t_{k-l+1}(q'))=t_{k-l}(t_{k+1-m}(q'))$, where $q=t_{k+1-m}(q')$ and $q'$ was introduced before by either a previous application of \lstinline[mathescape=true,escapechar=\$]!addTargetMaps! 
or is one of the cells added by the \lstinline[mathescape=true,escapechar=\$]!addSourceMaps!.
Because the two formulas $\terminateUniv{\,}{\psi}$ and $\terminateUniv{\,}{\neg\psi}$ may come only from set constraints then the potential label of $q'$ contains both $\terminateUniv{\,}{\terminateUniv{\,}{\psi}}$ and $\terminateUniv{\,}{\terminateUniv{\,}{\neg\psi}}$. We may assume that $q'$ is not added  by \lstinline[mathescape=true,escapechar=\$]!addTargetMaps!,
but comes from the other two functions; otherwise, we just need to stack several times the $\terminateUniv{\,}{}$ modality until we reach such a cell, and the reasoning would carry over verbatim. As we argued before, there exists a cell $q''$ in the old \modelH\ which contains $\startUniv{}{^{k}\terminateUniv{\,}{\terminateUniv{\,}{\psi}}}$ and $\startUniv{}{^{k}\terminateUniv{\,}{\terminateUniv{\,}{\neg\psi}}}$, or in the case when we work with the initial defective formula then $\labelPotential(q'')$ contains $\startUniv{}{^{k}\terminateUniv{\,}{\terminateUniv{\,}{\psi}}}$ and $\startUniv{}{^{k-1}\start{}{\terminateUniv{\,}{\terminateUniv{\,}{\neg\psi}}}}$, where $k\geq 2$. Because $q''$ is from the old \modelH\ it means that its labeling is potential canonical and hence has no inconsistencies (i.e., there exist atoms to respect its constraints). But $\startUniv{}{^{k}\terminateUniv{\,}{\terminateUniv{\,}{\psi}}}\wedge\startUniv{}{^{k}\terminateUniv{\,}{\terminateUniv{\,}{\neg\psi}}}\imply\startUniv{}{^{k}\terminateUniv{\,}{\terminateUniv{\,}{\bottom}}}$ which contradicts (hence the inconsistency) with the fact that any atom is consistent with $\startUniv{}{^{k}\terminate{\,}{\terminate{\,}{\top}}}$. This is because of axiom \ref{ax_HDML5} applied $k$ times to get $\startUniv{}{^{k}\terminate{\,}{^{k}\top}}$ which implies $\startUniv{}{^{k}\terminate{\,}{\terminate{\,}{\top}}}$. For the other formulas $\labelPotential(q'')$ contains $\startUniv{}{^{k}\terminateUniv{\,}{\terminateUniv{\,}{\psi}}}$ and $\startUniv{}{^{k-1}\start{}{\terminateUniv{\,}{\terminateUniv{\,}{\neg\psi}}}}$ use the same axiom \ref{ax_HDML5} and infer $\startUniv{}{^{k-1}\start{}{\terminate{\,}{\terminate{\,}{\bottom}}}}$ which is inconsistent with the existential constraint in $\labelPotential(q'')$ that essentially says that the atom should be consistent also with $\start{}{\start{}{\varphi}}$.

For 2) we are in the second loop and the formula $\psi$ was added at line 36 as coming from $\terminateUniv{\,}{\psi}\in\labelPotential(q)$ and the other formulas is added at line 42 (or at line 46 for the same argument) as coming from $\startUniv{}{\neg\psi}\in\labelPotential(t_{i-1}(s_{j}(q)))$; i.e., we are in the setting of a cubical law $s_{j}(t_{i}(q))=t_{i-1}(s_{j}(q))$. Assume that $q$ is the defective cell, for otherwise we have less cases to wary about as $q$ would be one of the cells added by \lstinline[mathescape=true,escapechar=\$]!addSourceMaps!
and we would have several $\terminateUniv{\,}{}$ stacked on top of the formulas and the argument would be analog to the one we give below. The fact that $\startUniv{}{\neg\psi}\in\labelPotential(t_{i-1}(s_{j}(q)))$ means that it comes from a set constraint of $s_{j}(q)$, i.e., $\terminateUniv{\,}{\startUniv{}{\neg\psi}\in\labelPotential(s_{j}(q))}$. If $s_{j}(q))$ is not part of the original \modelH\ then there is a cell in \modelH\ which would have the formula $\startUniv{}{^{k}\terminateUniv{\,}{\startUniv{}{\neg\psi}}}$, for some $k\geq 1$. We again use the fact that the old \modelH\ has a potential canonical labeling and hence is consistent, using the existential constraints, with the formula $\start{}{^{k}\start{}{\terminateUniv{\,}{\psi}}}$. These two last formulas are inconsistent. 
Putting them together we obtain $\start{}{^{k}(\terminateUniv{\,}{\startUniv{}{\neg\psi}}}\wedge\start{}{\terminateUniv{\,}{\psi}})$ which by axiom \ref{ax_HDML3} we get $\start{}{^{k}(\terminateUniv{\,}{\startUniv{}{\neg\psi}}}\wedge\terminateUniv{\,}{\start{}{\psi}})\imply\start{}{^{k}(\terminateUniv{\,}{\bottom}})$. But this contradicts with the $\startUniv{}{^{k}\terminate{}{^{k}\top}}$ coming from several applications of axiom \ref{ax_HDML5}.

For 3) the two formulas $\startUniv{}{\psi}$ and $\startUniv{}{\neg\psi}$ come from two cells introduced by 
\lstinline[mathescape=true,escapechar=\$]!addSourceMaps!
which contain $\terminateUniv{\,}{\startUniv{}{\psi}}$ respectively $\terminateUniv{\,}{\startUniv{}{\neg\psi}}$; they cannot come from the old \modelH.
But both these cells reach some cell in the old \modelH\ that will contain $\startUniv{}{^{k}\terminateUniv{\,}{\startUniv{}{\psi}}}$ and $\startUniv{}{^{k}\terminateUniv{\,}{\startUniv{}{\neg\psi}}}$ and moreover this is consistent with the formula $\start{}{^{k+1}\top}$, cf.\ the existential constraints. The two formulas together imply $\startUniv{}{^{k}\terminateUniv{\,}{\startUniv{}{\bottom}}}$ which together with $\startUniv{}{^{k}\terminate{}{^{k}\top}}$ implies $\startUniv{}{^{k}\terminate{\,}{\startUniv{}{\bottom}}}$, which by using axiom \ref{ax_HDML31} implies $\startUniv{}{^{k}\startUniv{}{\bottom}}$ leading to an inconsistency with $\start{}{^{k+1}\top}$.

Both functions always take care to add the existential constraints for any map that is added.
\end{proof}

\begin{definition}[descents]\label{def_descents}
Define the relation $\descentS\subseteq Q\times Q$ as $q\descentS q'$ iff $\exists s_{i}:s_{i}(q)=q'$. Define $\descentT\subseteq Q\times Q$ as $q\descentT q'$ iff $\exists t_{i}:t_{i}(q')=q$. Define $\descentST=\descentS\cup\descentT$ (and call its elements \emph{descent steps}), and $\descentSTstar$ as their reflexive transitive closure. We call a sequence (i.e., composition of relations) from $\descentSTstar$ a \emph{descent chain}. A descent chain is \emph{maximal} if no more descent steps can be added.
\end{definition}

Descent chains are somehow the opposite of \textit{paths} in \HDAs, cf.\ Definition~\ref{def_paths_HDA}.

\begin{lemma}\label{lemma_descents_1}
For the enrich algorithm for any of the new cells that are added, for any of its immediate starting descends it will eventually end up descending in one of the old cells of \modelH. Formally: $\forall q'\in\modelH'\setminus\modelH,\forall q'\descentST,\exists q\in\modelH:q'\descentST\circ\descentSTstar q$.
\end{lemma}

\begin{proof}
The first fresh cell is added at line 2 and is directly linked through $s_{n+1}$, at line 3, to the original defective cell from the old \modelH.

It remains to check that all other $s$ and $t$ maps of this cell are eventually reaching the old model \modelH. We do this inductively by going down the recursion calls until we find the minimal single descent steps. In particular we have to check only the $s$ maps because from this initial fresh cell $q_{n+1}$ only \descentS\ steps are possible. 

If $s_{n+1}(q_{n+1})$ is linked to the cell $q$ in the old \modelH, then for all the other $s_{i}(q_{n+1})$ their $s_{n}$ map is linked to the $s_{i}$ map of $q$, hence reaching in one \descentS\ step \modelH. We used the cubical law $s_{n}(s_{n+1}(q_{n+1})=s_{n}(s_{n}(q_{n+1}))$.

We use the similar law $s_{n-1}(s_{n}(s_{n}(q_{n+1})))=s_{n-1}(s_{n-1}(s_{n}(q_{n+1})))$ to argue that taking the descent step using the source $s_{n-1}$ reaches the old \modelH\ in two steps. 
The same reasoning is carried over inductively until the last recursion call.

In conclusion, all fresh cells added by the 
\lstinline[mathescape=true,escapechar=\$]!addSourceMaps!
reach the old \modelH\ through any of their sources by following a descent chain formed only of descent steps from \descentS, which, depending on the index of the source, take longer or shorter to reach \modelH.

The other fresh cells are added in the 
\lstinline[mathescape=true,escapechar=\$]!addTargetMaps!,
at line 28.
It is easy to see that all their immediate \descentT\ possible steps can eventually reach \modelH. An easy inductive reasoning suffices for this argument. Start with the cells added in the 
\lstinline[mathescape=true,escapechar=\$]!addSourceMaps! function or at line 7 in the body of the enrich function itself.
All these reach in one \descentT\ step a cell that we argued before that it can reach \modelH\ through a chain of only $\descentS$.
For the other cells added at deeper recursion calls inside 
\lstinline[mathescape=true,escapechar=\$]!addTargetMaps!
we can reach the cells from before, which have a descent chain to \modelH.

It remains to show that for all the fresh cells added in the 
\lstinline[mathescape=true,escapechar=\$]!addTargetMaps! 
their $s$ maps also lead to \modelH, i.e., that starting also with a \descentS\ step also leads eventually to a descent chain to \modelH. This is done also inductively starting with the cells that are added in the first call to 
\lstinline[mathescape=true,escapechar=\$]!addTargetMaps! function, 
and not inside its body (i.e., this step also considers the first calls inside the 
\lstinline[mathescape=true,escapechar=\$]!addSourceMaps! function).
Therefore, we consider some cell $q'$ which we proved that it eventually reaches \modelH; this cell has a target, say $t_{k}$, to the fresh cell $q$ that we are concerned with and itself has a source to some other cell $s_{j}(q)=q''$. Depending on $k,j$ we use the following cubical laws: if $j<k$ then $s_{j}(t_{k}(q'))=t_{k-1}(s_{j}(q'))$; if $k\leq j$ then $t_{k}(s_{j+1}(q'))=s_{j}(t_{k}(q'))$. Thus, in any case we can have a \descentT\ step from $q''$ to $s_{j}$ (or $s_{j+1}$ depending on the case), but these cells can reach \modelH, because they are reached from the initial $q'$ through a \descentS\ step. This base case is finished.

For cells added at deeper recursion calls, inside 
\lstinline[mathescape=true,escapechar=\$]!addTargetMaps!,
we use the same cubical laws and reach cells that we just proven in the step before that can reach \modelH. Depending on the indexes of the $s$ maps, the descending chains are longer or shorter.
\end{proof}

Note that in the proof of Lemma~\ref{lemma_enrich_s2} we made heavy use of the fact that we could go down a descent chain that was made of only \descentS\ steps. Because of this we were stacking up $\startUniv{}{}$ modalities. We will shortly make precise this method of stacking modalities depending on the descent chain and we will see more use of it and in more varied settings.

\begin{corollary}\label{cor_descents}\ 

\begin{enumerate}
\item For any fresh cell added by the enrich algorithm there exists a maximal descending chain and this one reaches a cell in the old \modelH\ that can make no \descentS\ and no \descentT\ steps. 
\item For any fresh cell, any descent chain that reaches \modelH\ can be completed to the maximal descent chain. 
\item For any fresh cell that has one descent chain starting with \descentT\ and one starting with \descentS, both these descent chains eventually reach the same cell in the old \modelH\ and use the same number of descent steps and the same number of \descentS\ steps (hence the same number of \descentT\ steps also).
\item For two cells connected as $s_{i}(q)=q'$ then the maximal descending chain of $q$ is one greater then the maximal descending chain of $q'$.
\end{enumerate}
\end{corollary}

\begin{proof}

\end{proof}

\begin{lemma}\label{lemma_enrich_s3}
For the enrich algorithm of Lemma~\ref{lemma_enriching} all the new existential constraints that are added to the fresh cells or to cells from the old \HDA\ are consistent with the set constraints of that cell.
\end{lemma}

\begin{proof}
This lemma essentially makes the third stage of the proof that the \HDA\ built by the enrich construction extends the old \HDA. This proof is based on the fact that the set constraints of each cell are consistent, cf.\ Lemma~\ref{lemma_enrich_s2}. We will make use of the fact that we enrich an old \modelH\ which has a potential canonical labeling, as we did in the proof of Lemma~\ref{lemma_enrich_s2}. The proof is by \textit{reductio ad absurdum} and assumes for an existential constraint $\start{}{\labelc(q)}$ in the $\labelPotential(q')$ there exists a formula $\startUniv{}{\neg\psi}$ in the set constraints of $\labelPotential(q')$ for which the formula $\psi$ is in some set constraint of $\labelPotential(q)$. 

These two formulas come from other set constraints being under some $\startUniv{}{}$ or $\terminateUniv{\,}{}$ box modality; and these bigger formulas in turn come from other set constraints added by the algorithm. And so on until we reach the old \modelH. From here we only know that the labeling in potential canonical; and we will use this in the proof.
Therefore there are many ways that the assumed formulas may have come from, and we need to find a way to treat all these different ways.

First we checked all the cases by hand for the particular application of enrich when the old cell is of dimension 2 and hence the new cell that needs to be added (with all its maps) is of dimension 3. Finding a clear pattern in these cases with varied length of the constraint propagation was tedious.\footnote{Even if for the cells or maps added/reached at the most inner recursion depth in the enrich algorithm the cases were easy to check or trivial, it is not possible to use an inductive reasoning in this way because at outer recursion levels the mesh of maps that connect to some particular cell becomes too complex.} The definitions and results above about descent chains are the basis of the general proof pattern that we will develop now further. These chains relate to the histories of a cell and to the notions of adjacency and homotopy.

We take any $s$ map introduced by the algorithm, i.e., $s_{i}(q)=q'$, and make the assumption from above. We will arrive at an inconsistency in the old \modelH, hence the contradiction. 
(The same proof method works for the $t$ maps and an analog assumption as above only that we use $\terminateUniv{\,}{}$ instead of $\startUniv{}{}$.)
Note that from cell $q$ we can make a \descentS\ descent step to reach $q'$.
From Corollary~\ref{cor_descents} we know that from the cell $q'$, hence also from $q$, there exists some descent chain reaching \modelH. If the descent chain is empty, i.e., $q'\in\modelH$, and if all the descent chains of $q$ consist in the single descent step to $q'$ then the result is trivial. This case corresponds to when the algorithm is at the most deep recursion call.
If there are other descent steps starting from $q$ then we are in a nontrivial case.

From Corollary~\ref{cor_descents} we know that any two different descent chains starting from $q$ will eventually end up in the same cell. Moreover, we know that any such descent chain eventually reaches the \modelH\ or the newly added cell $q_{n+1}$ at line 2 in the algorithm. We can argue that is enough to consider reaching this cell instead of reaching the original \modelH.

The proof method takes two such descent chains starting from $q$: one going first through $q'$ and the other going through some other different cell. For each of these chains we stop at the first cell from \modelH\ or $q_{n+1}$. The idea is that until there we are walking through the fresh cells added by the algorithm, and hence we collect boxes on the way.

\begin{example}
Take the example of $q$ which has one descending chain $q\descentT q''$. This is to say that the formula $\psi$ comes from a formula $\terminateUniv{\,}{\psi}$ in a set constraint of $q''$. For a descent chain $q\descentS\descentT q''$, where the first step does not go through $q'$, it means that $\psi$ comes from $\terminateUniv{\,}{\startUniv{}{\psi}}\in\labelPotential(q'')$. The fact that the other descent chain that we consider from $q$ goes through $q'$ with a \descentS\ step should mean that $\startUniv{}{\psi}$ is in the set constraints of $q'$ but this contradicts the assumption that $\startUniv{}{\neg\psi}$ is there and that there actually exists an $s$ map out of $q$; this contradiction will come syntactically as an inconsistency in the potential canonical labeling of the old \modelH.
\end{example}

To such a descent chain that starts from $q$ we associate a formula as follows: considering $\psi\in\labelPotential(q)$, for $q\descentS q''\descentSTstar$ then $\startUniv{}{\psi}\in\labelPotential(q'')$. We continue until the end of the chain where in the case of a \descentT\ descent step eg.\ $q''\descentT q'''\descentSTstar$ we have $\terminateUniv{\,}{\startUniv{}{\psi}}\in\labelPotential(q''')$. Both these chains reach eventually the same cell in \modelH, cf.\ Corollary~\ref{cor_descents}. Moreover, the chain that reaches \modelH\ faster can continue through the inside of \modelH\ until reaching the end cell of the other chain. This traversing of the old \modelH\ is done under the existential constraints in the potential canonical labeling of \modelH, therefore we can extend the corresponding formula with existential modalities. In the common end cell we have now two formulas, one made only of box modalities and the other which may also have a stack of existential modalities, and both have to be consistent, as being part of the old \modelH. We actually show that these two are inconsistent or cannot be grown to an atom, i.e., lead to an inconsistency in the axiomatic system of \HDML.

Thus we work with two chains starting from $q$ and ending in some common cell in \modelH, and to each chain we associate a formula: one adds modalities to $\psi$ (as being in $\labelPotential(q)$) and the other adds modalities to $\neg\psi$ (as coming from the chain that goes through $q'$ which we assumed to have a formula $\startUniv{}{\neg\psi}$). We denote descent steps that are inside the \modelH, and which are associated with existential modalities, by $\descentSdash$ respectively $\descentTdash$. 
There is an equal number of \descentS\ in each chain (either universal or existential) and hence an equal number of \descentT\ steps also, cf.\ Corollary~\ref{cor_descents}. This translates into the formulas also.
The purpose is to change these chains so that the descent steps match one by one. If one existential matches one universal then we obtain an existential step that leads to the inconsistency more easy.
Also, the purpose is to move as much as possible of the \descentS\ steps to the end of the chain and the \descentT\ steps to the beginning of the chain. In the end we will arrive at a contradiction with the fact that in the initial cell which the algorithm starts with there is the formula $\start{}{\varphi}$ and hence, by the existential constraints in \modelH, all lower cells are consistent with $\start{}{^{k}\varphi}$ depending on the distance from the initial cell.

For example:

\begin{center}
\begin{tabular}{lcl}
$\psi \descentT\descentS\descentSdash$ &  associate & $\start{}{\startUniv{}{\terminateUniv{\,}{\psi}}}$\\
$\neg\psi  \descentS\descentT\descentS$ &  associate  & $\startUniv{}{\terminateUniv{\,}{\startUniv{}{\neg\psi}}}$\\
\end{tabular}
\end{center}
Apply axiom \ref{ax_HDML6} to get $\start{}{\startUniv{}{\terminateUniv{\,}{\psi}}}\wedge \startUniv{}{\startUniv{}{\terminateUniv{\,}{\neg\psi}}}$ which by modal reasoning becomes $\start{}{\startUniv{}{\terminateUniv{\,}{\bottom}}}$, but by modal reasoning and axiom \ref{ax_HDML5} we have as validity $\startUniv{}{^{k}\terminate{}{^{k}\top}}$ where in our case we use it for $k=2$ to get $\start{}{\startUniv{}{\bottom}}$. This results in a contradiction with the fact that we can always start at least to reach a cell with $\varphi$.

For all the patterns that we find in the descent chains there is some axiom associated which helps transform the formulas into the needed ones; we will say that they transform the chains into the proper form. Below we give the patterns with the associated formulas and axioms:

\begin{center}
\begin{tabular}{lclcl}
$\descentS\descentT\descentS$ &  is  & $\startUniv{}{\terminateUniv{\,}{\startUniv{}{}}}\stackrel{\ref{ax_HDML6}}{\rightarrow}\startUniv{}{\startUniv{}{\terminateUniv{\,}{}}}$ & is & $\descentT\descentS\descentS$\\
$\descentT\descentS\descentT$ &  is  & $\terminateUniv{\,}{\startUniv{}{\terminateUniv{\,}{}}}\stackrel{\ref{ax_HDML61}}{\rightarrow}\startUniv{}{\terminateUniv{\,}{\terminateUniv{\,}{}}}$ & is & $\descentT\descentT\descentS$\\
$\descentS\descentTdash$ &  is & $\terminate{}{\startUniv{}{}}\stackrel{\ref{ax_HDML31}}{\rightarrow}\startUniv{}{\terminate{}{}}$ & is & $\descentTdash\descentS$\\
$\descentT\descentSdash$ &  is & $\start{}{\terminateUniv{\,}{}}\stackrel{\ref{ax_HDML3}}{\rightarrow}\terminateUniv{\,}{\start{}{}}$ & is & $\descentSdash\descentT$\\
\end{tabular}
\end{center}
Note that the last pattern does not bring a \descentT\ more close to the beginning of the chain, but does the opposite. This is the case when the other three patterns do not occur but we can match the \descentSdash\ that is brought closer to the beginning to a \descentS, therefore combining the two (by modal reasoning) into a \descentSdash\ applied to $\bottom$ which just makes the descent step disappear into $\bottom$; i.e., $\terminate{}{\bottom}\equivalent\bottom$.

There may be patterns that are not matched by any of the above, like eg.:

\begin{center}
\begin{tabular}{lcl}
$\psi \descentT\descentS\descentS\descentTdash$ &  associate & $\terminate{}{\startUniv{}{\startUniv{}{\terminateUniv{\,}{\psi}}}}$\\
$\neg\psi  \descentS\descentT\descentT\descentS$ &  associate  & $\startUniv{}{\terminateUniv{\,}{\terminateUniv{\,}{\startUniv{}{\neg\psi}}}}$\\
\end{tabular}
\end{center}
The pattern in the lower chain is not matched by any of the four patterns above because there are more than one $\terminateUniv{\,}{}$ stacked on top of each other, i.e., are two consecutive \descentT\ steps surrounded by \descentS\ steps. Nevertheless, such patterns can be \emph{broken} s.t.\ the new chains can be matched by the four main patterns above. Breaking such patterns (also with more than two consecutive \descentT) is done with the use of axioms \ref{ax_HDML5}, \ref{ax_HDML41}, \ref{ax_HDML4}, possibly applied several times. In the particular case above, because we have the first formula then axiom \ref{ax_HDML5} is applied (for $n=1$) to get $\startUniv{}{\terminate{}{\terminate{}{\top}}}$ to which we can apply axiom \ref{ax_HDML41} to get $\startUniv{}{\terminateUniv{\,}{\terminate{}{\top}}}$. This formula now breaks the second chain in the sense that one \descentT\ is transformed into an existential one \descentTdash; i.e., $\startUniv{}{\terminateUniv{\,}{\terminateUniv{\,}{\startUniv{}{\neg\psi}}}}\wedge\startUniv{}{\terminateUniv{\,}{\terminate{}{\top}}}\imply \startUniv{}{\terminateUniv{\,}{\terminate{}{\startUniv{}{\neg\psi}}}}$. To this chain now we can apply the third and then the first pattern from above to obtain $\startUniv{}{\startUniv{}{\terminateUniv{\,}{\terminate{}{\neg\psi}}}}$, i.e., the chain $\neg\psi\descentTdash\descentT\descentS\descentS$. To the first chain we could apply the third pattern two times to get $\startUniv{}{\startUniv{}{\terminate{}{\terminateUniv{\,}{\psi}}}}$, i.e., the chain $\psi \descentT\descentTdash\descentS\descentS$. It is clear that the two formulas contradict with the fact that the current cell of \modelH\ must be consistent, by the existential constraints, with $\start{}{\start{}{\start{}{\varphi}}}$ because $\startUniv{}{\startUniv{}{\terminateUniv{\,}{\terminate{}{\neg\psi}}}}\wedge\startUniv{}{\startUniv{}{\terminate{}{\terminateUniv{\,}{\psi}}}}\imply\startUniv{}{\startUniv{}{\terminate{}{\terminate{\,}{\bottom}}}}\imply\startUniv{}{\startUniv{}{\bottom}}$. In terms of descent chains the two chains match step by step as having the same $s$ or $t$ label and we match either two universal steps, like \descentS, or one universal step with one existential step which yield an existential step, like \descentT\ with \descentTdash. Such match of descent chains yields the inconsistency with the chain $\top\descentSdash\descentSdash\descentSdash$.

For another example of unmatched patterns in chains consider:

\begin{center}
\begin{tabular}{lcl}
$\psi \descentT\descentS\descentS\descentTdash\descentSdash$ &  associate & $\start{}{\terminate{}{\startUniv{}{\startUniv{}{\terminateUniv{\,}{\psi}}}}}$\\
$\neg\psi  \descentS\descentT\descentT\descentS\descentS$ &  associate  & $\startUniv{}{\startUniv{}{\terminateUniv{\,}{\terminateUniv{\,}{\startUniv{}{\neg\psi}}}}}$\\
\end{tabular}
\end{center}
This is the same as the example before only that each chain is extended with one existential \descentSdash\ step respectively an universal \descentS. The same way of breaking the pattern using axioms \ref{ax_HDML5} and \ref{ax_HDML41} is used here also only that because we do not have the formula $\terminate{}{\top}$ we use twice \ref{ax_HDML5} to get $\startUniv{}{\startUniv{}{\terminate{}{\terminate{}{\top}}}}$ which by axiom \ref{ax_HDML41} we obtain the breaking pattern $\startUniv{}{\startUniv{}{\terminateUniv{\,}{\terminate{}{\top}}}}$ with the associated chain $\descentTdash\descentT\descentS\descentS$. This breaks the second chain into $\descentS\descentTdash\descentT\descentS\descentS$ (or the formula becomes $\startUniv{}{\startUniv{}{\terminateUniv{\,}{\terminate{}{\startUniv{}{\neg\psi}}}}}$) to which we can apply the pattern 3 and then 1 to obtain the chain  $\descentTdash\descentT\descentS\descentS\descentS$. To the first one applies pattern 3 two times to obtain $\descentT\descentTdash\descentS\descentS\descentSdash$ so the two chains match step by step. Whenever we are in a situation like this when the two modified chains end up in an existential \descentSdash\ step and a corresponding universal one, we can just remove these two steps because it basically says that there exists this reachable cell where both formulas $\startUniv{}{\startUniv{}{\terminateUniv{\,}{\terminate{}{\neg\psi}}}}$ and $\startUniv{}{\startUniv{}{\terminate{}{\terminateUniv{\,}{\psi}}}}$ hold. These result in an inconsistency with the existential constraints again.
\end{proof}


For a D2 defect, the \textit{lifting construction} lifts the defective cell and all the cells that are connected to it by some $s$ or $t$ map, one level up by adding one new $s$ and $t$ map to each of them. The label of the new $t$ map will be the one repairing the D2 defect. The cubical laws make sure that these new maps reach only new cells; none of the old cells (that are lifted) are involved in these new instances of the cubical laws. We need to be careful how we label all these new cells s.t.\ the canonicity is respected for the new lifted \HDA.

The lifting construction is more involved than the enriching construction. We still label the cells with atoms in the end but during the construction the constraints that the atom has to satisfy are changed. This is why we keep the set of all atoms that satisfy the constraints as possible candidates for the final labeling. This means that we are still working with atoms (i.e., maximal consistent sets of formula) but we do not settle on one particular atom until we have finished the construction.

\begin{lemma}[lifting construction]\label{lemma_lifting}
For a canonical model \modelH, there exists a construction (see Appendix), which we call \emph{lifting of the \modelH\ wrt.\ $q$ and a formula $\terminate{}{\varphi}\in\labelc(q)$}, builds a model $\modelH'$ which is canonical and extends \modelH\ (i.e., $\modelH'\extends\modelH$).
\end{lemma}

\begin{proof}
The lifting construction is the following:

\newpage 
 
{\footnotesize
\begin{lstlisting}[numbers=left,numberstyle=\tiny,numbersep=5pt,mathescape=true,escapechar=\%,morekeywords={return,function,for,if,else}]
function lift(n,q,$\varphi$){
 addTargetMap(n,q,$\{\varphi\}$,$\emptyset$); // add the source map
 addSourceMap(n,q); // add the target map
 for(all cells $q'$ with $q'\in Q_{m}$){
  lift(m,$q'$,$\emptyset$) //lift all other cells
 }}
function addTargetMap(k,q,S1,S2){
 $Q_{k}$:=$Q_{k}\cup\{q_{k}\}$; //fresh cell
 update map $t_{k+1}$ %s.t.% $t_{k+1}(q)=q_{k}$;
 $\label{line_boxConstr1}$add constraints $\labelPotential(q_{k})=$S1$\cup\{\phi\mid\terminateUniv{\,}{\phi}\in\labelPotential(q)\}\cup\{\phi\mid\terminateUniv{\,}{\phi}\in\,$S2$\}$;
 add constraints $\labelPotential(q)=\labelPotential(q)\cup\terminate{}{\widehat{\labelc(q_{k})}}$;
 for(i=1 to k){ 
  $r_{k-1}^{i}$:=addTargetMap(k-1,$s_{i}(q)$,$\emptyset$,$\emptyset$);
  update map $s_{i}$ %s.t.% $s_{i}(t_{k+1}(q))=r_{k-1}^{i}$;
  $\label{line_boxConstrLoop}$add constraints $\labelPotential(q_{k})=\labelPotential(q_{k})\cup\{\phi\mid\startUniv{}{\phi}\in\labelPotential(r_{k-1}^{i})\}$;
  add constraints $\labelPotential(r_{k-1}^{i})=\labelPotential(r_{k-1}^{i})\cup\start{}{\widehat{\labelc(q_{k})}}$;
 }
 for(i=1 to k){ 
  $q_{k-1}^{i}$:=addTargetMap(k-1,$t_{i}(q)$,$\emptyset$,$\labelc(q_{k})$);
  update map $t_{i}$ %s.t.% $t_{i}(t_{k+1}(q))=q_{k-1}^{i}$; 
  add constraints $\labelPotential(t_{k+1}(q))=\labelPotential(t_{k+1}(q))\cup\terminate{}{\widehat{\labelc(q_{k-1}^{i})}}$;
 }
 return $q_{k-1}$;
}
function addSourceMap(k,q){
 $Q_{k}$:=$Q_{k}\cup\{q_{k}\}$; //fresh cell
 update map $s_{k+1}$ %s.t.% $s_{k+1}(q)=q_{k}$;
 add constraints $\labelPotential(q_{k})=\start{}{\widehat{\labelc(q)}}$;
 for(i=1 to k){ //add a fresh $s_{k}$ map to each $t_{i}(q)$
  $r_{k-1}^{i}$:=addSourceMap(k-1,$t_{i}(q)$);
  update map $t_{i}$ %s.t.% $t_{i}(s_{k+1}(q))=r_{k-1}^{i}$;
  add constraints $\labelPotential(q_{k})=\labelPotential(q_{k})\cup\terminate{}{\widehat{\labelc(r_{k-1}^{i})}}$;
 }
 for(i=1 to k){ //add a fresh $s_{k}$ map to each $s_{i}(q)$
  $q_{k-1}^{i}$:=addSourceMap(k-1,$s_{i}(q)$);
  update map $s_{i}$ %s.t.% $s_{i}(s_{k+1}(q))=q_{k-1}^{i}$;
  add constraints $\labelPotential(q_{k-1}^{i})=\labelPotential(q_{k-1}^{i})\cup\start{}{\widehat{\labelc(q_{k})}}$;
 }
 $\label{line_liftSourceLast}$$Q_{k}\setminus\{q\}$; $Q_{k+1}\cup\{q\}$; // move the cell one level up
 return $q_{k-1}$;
}
\end{lstlisting}
}

The proof has several stages:
\begin{enumerate}
\item We need to show that the enriched model is an extension of the old model (i.e., $\modelH'\extends\modelH$), which amounts to:
\begin{enumerate}
\item first showing that the structure of the old \HDA\ is untouched, i.e., all old cells and maps are in place;

\item\label{stage_lift_consistSets} then showing that all set constraints are still consistent sets;

\item\label{stage_lift_existConstr} and third showing that all the new existential constraints do not contradict with the box constraints.
\end{enumerate}
Basically the steps \refeq{stage_lift_consistSets} and \refeq{stage_lift_existConstr} are corresponding to Lemma~\ref{lemma_potentialNotCanon} to show that the new potential labeling is still potential canonical, i.e., that there still exists a way of instantiating the constraints to atoms.

\item The next stage shows that $\modelH'$ is a model indeed, i.e., that all the maps are in place and all necessary cubical laws are respected.

\item The last stage shows that the enriched model does not have the old defect  and that no new defects are introduced in the potential labeling of the initial cell $q$.
\end{enumerate}


First remark that we do not change the initial shape of the original \modelH; we only add fresh cells and fresh maps for these cells; we also add maps to old cells connected to new cells. This concludes the first stage in proving that $\modelH'\extends\modelH$. 
The second stage is proven as Lemma~\ref{lemma_enrich_s2}, whereas the third stage is proven as Lemma~\ref{lemma_enrich_s3}. Therefore, $\modelH'\extends\modelH$.

We show next that we indeed construct a higher dimensional structure. A careful reading of the enriching construction should answer this question in affirmative. We need to make sure that to each new cell we add all the $s$ and $t$ maps according to its dimension and that we link these maps correctly according to the cubical laws.

Note that the algorithm finishes with a completely new layer of cells denoted $Q_{-1}$; in the end of the construction we have to rename all the layers $Q_{i}$ into $Q_{i+1}$ to make justice to the cells that reside there which have now dimension $i+1$ as we added one $s$ and one $t$ map to each. 

Note that the construction terminates iff $q$ is in a hypercube of finite dimension and in this case we ignore all the cells outside this cube. (The construction always terminates when we use it in the repair lemma \ref{lemma_repair}.)

Clearly the two functions do not change the labels nor the shape of the old \modelH\ and hence the lifted $\modelH'$ has all the structure of $\modelH$. 

Now we show that the lifting constructs indeed a \HDA. This means that we must make sure that all the (new) cells have the right number of $s$ and $t$ maps and that all the cubical laws are respected.

The \lstinline[mathescape=true,escapechar=\$]!lift! 
function takes as input the reference cell $q$ and its dimension $n$ together with the formula $\varphi$ that causes the defect (i.e., $\terminate{}{\varphi}\in\labelPotential(q)$). Then the function adds one $t$ map and one $s$ map to $q$ by calling  \lstinline[mathescape=true,escapechar=\$]!addTargetMap! and  \lstinline[mathescape=true,escapechar=\$]!addSourceMap! respectively. These two functions add one new cell and link it with either a $t$ or an $s$ map. 
All other cells that are connected to $q$ must also be lifted, which is done in the loop of the  \lstinline[mathescape=true,escapechar=\$]!lift! function.

Consider now the  \lstinline[mathescape=true,escapechar=\$]!addTargetMap! 
which takes as arguments the cell $q$ (and its dimension $k$) to which the new $t$ map needs to be added. It also takes two sets of formulas which are used to construct the label of the new cell and of the other new cells connected to it recursively. We do not discuss here the labeling because we do this in the Lemmas~\ref{} and \ref{}. The rest of the proof is concerned with the geometric structure of the extended $\modelH'$.

The  \lstinline[mathescape=true,escapechar=\$]!addTargetMap! 
function adds the new $t_{k+1}$ map to $q$, which is the map with the largest index (i.e., the new index showing that the $q$ cell has now dimension one greater, $k+1$). It links this with a new cell $q_{k}$ of dimension one lower than the new dimension of the input cell $q$. The first loop does two operations. First it lifts all the old cells linked to $q$ by an $s$ map (i.e., $s_{i}(q)$) by adding one $t$ map to each; i.e., it invokes \lstinline[mathescape=true,escapechar=\$]!addTargetMap!
recursively. Then, all these cells enter under new cubical laws that involve the $s$ maps of the newly added $q_{k}$ cell. In this way we also add all the necessary $s$ maps of $q_{k}$ and also respect the new cubical laws $s_{i}(t_{k+1}(q))=t_{k}(s_{i}(q))$. 

In the second loop of \lstinline[mathescape=true,escapechar=\$]!addTargetMap! 
we add the new $t_{k}$ map to each old cell linked to $q$ by a $t_{i}$ map; i.e., in the recursive invocation of \lstinline[mathescape=true,escapechar=\$]!addTargetMap!. 
At the same time we add all the $t$ maps for the new $q_{k}$ cell and link these through the cubical laws $t_{i}(t_{k+1}(q))=t_{k}(t_{i}(q))$.

The construction goes recursively at lower levels until reaching cells of dimension $0$. These are the last cells lifted to have dimension $1$. Here the recursion stops.

Consider now the similar function \lstinline[mathescape=true,escapechar=\$]!addSourceMap! 
which adds one $s$ map to the input cell $q$ of dimension $k$ to make it now of dimension $k+1$. Therefore, it adds the map $s_{k+1}(q)=q_{k-1}$. 
This is also the place where the lifted cells are actually moved to the rightful layer $Q_{k+1}$, at the end of the function (i.e., line~\ref{line_liftSourceLast}), after both target and source maps have been added. 

In the first loop \lstinline[mathescape=true,escapechar=\$]!addSourceMap! 
adds a new $s_{k}$ maps to all the old cells linked to $q$ by a $t$ map. This finishes what we started in the second loop of \lstinline[mathescape=true,escapechar=\$]!addTargetMap!, i.e., 
finishes lifting all the $t_{i}(q)$ cells. It also takes care to respect all the new cubical laws $t_{i}(s_{k+1}(q))=s_{k}(t_{i}(q))$ and, hence, to add the $t_{i}$ maps to $q_{k}$.

The second loop complements what we started in the first loop of  \lstinline[mathescape=true,escapechar=\$]!addTargetMap!. 
We finish adding the $s_{k}$ maps to all the $s_{i}(q)$ cells. It also adds all the $s$ maps to $q_{k}$ and respects the new cubical laws $s_{i}(s_{k+1}(q))=s_{k}(s_{i}(q))$.

In conclusion, all the cells of the old \modelH\ have been added one new $t$ and $s$ map, each reaching a new cell. To all these new cells all the $t$ and $s$ maps have been added and linked according to the new cubical laws.

\end{proof}

\begin{lemma}\label{lemma_lift_s2}
The new sets of formulas that are added by the lift algorithm of Lemma~\ref{lemma_lifting} (i.e., at lines~\ref{line_boxConstr1} and \ref{line_boxConstrLoop}) are consistent sets.
\end{lemma}

\begin{proof}
The only place where box constraints are added by the \lstinline[mathescape=true,escapechar=\$]!lift! function is in \lstinline[mathescape=true,escapechar=\$]!addTargetMap!: first at line~\ref{line_boxConstr1} and then repeatedly in the loop at line~\ref{line_boxConstrLoop}.

The set S1 is not empty only when the function is applied to the initial cell $q$ from the statement of the lemma. The lemma assumes that $q\in Q_{n}$ is of dimension $n$, denote it $q_{n}$ for this part of the proof, and it contains $\terminate{}{\varphi}\in\labelPotential(q_{n})$ for which all of its $n$ existing $s$ maps contain $\neg\varphi$. This means that if before $\labelPotential(q_{n})$ was consistent with $\terminate{}{n}$ now we need to write $\terminate{}{n+1}$. Because of axiom~\ref{ax_HDML1} and Lemma~\ref{lemma_atoms_prop}\ref{atoms6} it means that $\labelPotential(q_{n})$ is consistent also with $\terminatei{n+1}{\top}$. (As a side remark, we use Lemma~\ref{lemma_atoms_prop}\ref{atoms6} tacitly in many places during the proofs of the two constructions lemmas.)

The first call to  \lstinline[mathescape=true,escapechar=\$]!addTargetMap(n,$q_{n}$,$\{\varphi\}$,$\emptyset$)! 
makes use only of S1 and constructs the set $\{\varphi\}\cup\{\psi\mid\terminateUniv{\,}{\psi}\in\labelPotential(q_{n})\}$. This set is associated to $q_{n-1}=t_{n+1}(q_{n})$. The proof is easy for this case and uses arguments as in the proof before: if we assume $\psi_{1}\wedge\dots\wedge\psi_{k}\imply\bot$ then we get that $\terminateUniv{\,}{\bot}\in\labelc(q_{n})$ which is a contradiction as $\labelc(q_{n})$ is an atom containing $\terminate{}{\varphi}$; if we assume $\psi_{1}\wedge\dots\wedge\psi_{k}\imply\neg\varphi$ then we get that $\terminateUniv{\,}{\neg\varphi}\in\labelc(q_{n})$ which is again a contradiction.

The second call to  \lstinline[mathescape=true,escapechar=\$]!addTargetMap! is made for each $s$ map of a cell $q$ (in the first loop of the body of the \lstinline[mathescape=true,escapechar=\$]!addTargetMap!) and it uses only the set S2. This means that it labels a cell $q_{n-1}=t_{n+1}(q)$ with a set $\{\psi\mid\terminateUniv{\,}{\psi}\in\labelc(q)\}$. Assume $\psi_{1}\wedge\dots\wedge\psi_{k}\imply\bot$ which means that $\terminateUniv{\,}{\bot}\in\labelc(q)$. This is a contradiction because $\labelc(q)$ is an atom and it contains at least one diamond formula. This is because $q$ has dimension at least $1$ (as it has at least one $t$ map) and we show that any cell of dimension $n$, with $n\geq 1$, has a formula $\terminatei{n}{\top}\in\labelc(q)$. We showed before that the topmost cell $q_{n}$ has the formula $\terminatei{n+1}{\top}$ in its label and hence it is of dimension $n+1$. This means that any cell reached through one of its $t$ maps will have the formula $\terminatei{n}{\top}$ because of axiom~\ref{ax_HDML41} which says that $\terminate{}{\terminatei{n}{\top}}\imply\terminateUniv{\,}{\terminatei{n}{\top}}$ it means that $\terminateUniv{\,}{\terminatei{n}{\top}}\in\labelc(q_{n})$ and by the construction of their labels it means that $\terminatei{n}{\top}\in\labelc(t_{j}(q_{n}))$. This holds for any cell reached through any number of applications of $t$ maps. On the other hand, the cells reached through an $s$ map from $q_{n}$, by canonicity, they contain $\start{}{\terminatei{n+1}{\top}}$, which, by axiom~\ref{ax_HDML51} it means that $\terminatei{n}{\top}\in\labelc(s_{j}(q_{n)})$.


It remains to see that with each iteration of the first loop the updated label remains a consistent set. This update is necessary when we are trying to respect the cubical laws of the form $s_{i}(t_{k+1}(q))=t_{k}(s_{i}(q))$. The proof of this part follows an inductive argument, where the basis was just proven above and the inductive case is for some $i$ iteration, where we consider that the label is a consistent set (and all the other labels that the construction uses have been built already and, hence, are atoms). Assume that for some $\startUniv{}{\psi}\in\labelc(t_{k}(s_{i}(q)))$ there has already been added the $\neg\psi$ to $\labelc(t_{k+1}(q))$. This has happened in two cases: first if $\neg\psi$ comes from $\labelc(q)$, i.e., $\terminateUniv{\,}{\neg\psi}\in\labelc(q)$ which by canonicity it means that $\start{}{\terminateUniv{\,}{\neg\psi}}\in\labelc(s_{i}(q))$. On the other hand we also have that $\terminate{}{\startUniv{}{\psi}}\in\labelc(s_{i}(q))\stackrel{\ref{ax_HDML31}}{\imply}\startUniv{}{\terminate{}{\psi}}\in\labelc(s_{i}(q))$. Together with the above it means that $\start{}{(\terminateUniv{\,}{\neg\psi}\wedge\terminate{}{\psi})}\imply\start{}{\terminate{}{(\neg\psi\wedge\psi)}}\stackrel{\ref{ax_modal1},\ref{ax_modal11}}{\longrightarrow}\bot\in\labelc(s_{i}(q))$ which is a contradiction with the fact that $\labelc(s_{i}(q))$ is an atom. The second case is when $\neg\psi$ has been added in a previous iteration, i.e., $\startUniv{}{\neg\psi}\in\labelc(s_{j}(t_{k+1}(q)))$ with $1\leq j<i$. But this means that each of these two cells must have at least one $s$ map and enter the cubical law $s_{j}(s_{i}(t_{k+1}(q)))=s_{i-1}(s_{j}(t_{k+1}(q)))=q''$. By the canonicity of these lower cells we have that $\start{}{\startUniv{}{\neg\psi}}\in\labelc(q'')$ and $\start{}{\startUniv{}{\psi}}\in\labelc(q'')$. From axiom \ref{ax_HDML21} we have that $\startUniv{}{\start{}{\psi}}\in\labelc(q'')$ and thus $\startUniv{}{\start{}{\psi}}\wedge\start{}{\startUniv{}{\neg\psi}}\imply\start{}{(\start{}{\psi}\wedge\startUniv{}{\neg\psi})}\imply\start{}{\start{}{(\psi\wedge\neg\psi)}}\stackrel{\ref{ax_modal1}}{\imply}\bot\in\labelc(q'')$ which is a contradiction.

The application of \lstinline[mathescape=true,escapechar=\$]!addTargetMap! 
in the second loop uses the S3 set also and we are looking at cubical laws of type $t_{i}(t_{k+1}(q))=t_{n}(t_{i}(q))$ where $S2=\labelc(t_{i}(q))$ and $S3=\labelc(t_{k+1}(q))$. Assume, for the sake of contradiction, that we have $\terminateUniv{\,}{\psi}\in\labelc(t_{k+1}(q))$ and $\terminateUniv{\,}{\neg\psi}\in\labelc(t_{i}(q))$. By canonicity it means that $\terminate{}{\terminateUniv{\,}{\psi}}\in\labelc(q)$ and $\terminate{}{\terminateUniv{\,}{\neg\psi}}\in\labelc(q)$ and from axiom \ref{ax_HDML2} we have $\terminateUniv{\,}{\terminate{}{\neg\psi}}\in\labelc(q)$. This means that $\terminateUniv{\,}{\terminate{}{\neg\psi}}\wedge\terminate{}{\terminateUniv{\,}{\psi}}\imply\terminate{}{(\terminate{}{\neg\psi}\wedge\terminateUniv{\,}{\psi})}\imply\terminate{}{\terminate{}{(\psi\wedge\neg\psi)}}\stackrel{\ref{ax_modal11}}{\imply}\bot\in\labelc(q)$ which is a contradiction.
\end{proof}

\begin{lemma}\label{lemma_lift_s3}
For the enrich algorithm of Lemma~\ref{lemma_lifting} all the new existential constraints that are added to the fresh cells or to cells from the old \HDA\ are consistent with the set constraints of that cell.
\end{lemma}

\begin{proof}
Assume that for the lifted \HDA\ the \textit{second canonicity condition is broken}; i.e., consider $q\in Q_{n}$ and assume $t_{i}(q)=q'$ for which $\varphi\in\labelc(q')$ and $\terminate{}{\varphi}\not\in\labelc(q)$, which is the same as $\neg\terminate{}{\varphi}\in\labelc(q)$. We take cases after $q$. 

First, clearly, if $q,q'\in\modelH$ (meaning that $1\leq i\leq n-1$) then the canonicity is assured by the statement of the lemma (i.e., \modelH\ is canonical). 

Second, $q\in\modelH$ and $q'$ is added by \lstinline[mathescape=true,escapechar=\$]!addTargetMap! as the new cell linked to $q$ by $t_{n}(q)=q'$. Now we take sub-cases depending on where does the $\varphi$ formula come from. 
\begin{itemize}
 \item If $\varphi\in S1$; this is the case when $q$ is the initial cell from the statement of the lemma and hence it cannot be that $\neg\terminate{}{\varphi}\in\labelc(q)$.

\item If $\varphi\in \{\varphi\mid\terminateUniv{\,}{\varphi}\in S2\}$ then $\terminateUniv{\,}{\varphi}\in\labelc(q)$ and the assumption says that $\terminateUniv{\,}{\neg\varphi}\in\labelc(q)$. This is a contradiction as $\terminateUniv{\,}{\varphi}\wedge\terminateUniv{\,}{\neg\varphi}\imply\terminateUniv{\,}{(\varphi\wedge\neg\varphi)}\imply\terminateUniv{\,}{\bot}\in\labelc(q)$ which is not possible because, as we showed before, $\labelc(q)$ contains at least one existential formula, i.e., $\terminatei{k}{\top}$, where $k$ is the dimension of $q$.

\item If $\varphi\in\!\{\varphi\mid\terminateUniv{\,}{\varphi}\in\!S3\}$ then $q'$ is added by the second call to \lstinline[mathescape=true,escapechar=\$]!addTargetMap!, which means that we are respecting the cubical laws $t_{i}(t_{k+1}(q_{k+1}))=t_{k}(t_{i}(q_{k+1}))$, for $1\leq i\leq k$ and for some $q_{n+1}$ for which our $q=t_{i}(q_{k+1})$. Then by the construction of the label it means that $\terminateUniv{\,}{\varphi}\in\labelc(t_{k+1}(q_{k+1}))$ which by the canonicity of these upper cells it means that $\terminate{}{\terminateUniv{\,}{\varphi}}\in\labelc(q_{k+1})$. By axiom \ref{ax_HDML2} it means that $\terminateUniv{\,}{\terminate{}{\varphi}}\in\labelc(q_{k+1})$ and thus, by the canonicity it means that $\terminate{}{\varphi}\in\labelc(q)$ which is a contradiction with our initial assumption as the labels are atoms and hence $\neg\terminate{}{\varphi}$ cannot be in the label $\labelc(q)$.

\item Lastly, assume that $\varphi$ is one of the formulas accumulated in the label of $q'$ as a result of the first loop of \lstinline[mathescape=true,escapechar=\$]!addTargetMap!. This means that we are respecting the cubical laws $s_{i}(t_{k+1}(q))=t_{k}(s_{i}(q))$ and $\startUniv{}{\varphi}\in\labelc(s_{i}(q'))=\labelc(s_{i}(t_{k+1}(q)))=\labelc(t_{k}(s_{i}(q)))$. By canonicity of the other cells it means that $\terminate{}{\startUniv{}{\varphi}}\in\labelc(s_{i}(q))$ which by axiom \ref{ax_HDML31} it means that $\startUniv{}{\terminate{}{\varphi}}\in\labelc(s_{i}(q))$. By canonicity again it means that $\terminate{}{\varphi}\in\labelc(q)$ which is again a contradiction with our initial assumption.
\end{itemize}

Third, both $q$ and $q'$ are newly added by  \lstinline[mathescape=true,escapechar=\$]!addTargetMap!, meaning that we are looking at the second loop. The proof is the same as before as the construction of the label and axiom \ref{ax_HDML2} do all the work.

Forth, both $q$ and $q'$ are newly added by \lstinline[mathescape=true,escapechar=\$]!addSourceMap!, which means that we are in the first loop of \lstinline[mathescape=true,escapechar=\$]!addSourceMap! and there exists a $q_{k+1}$ with $s_{k+1}(q_{k+1})=q$ and $t_{i}(s_{k+1}(q_{k+1}))=q'=s_{k}(t_{i}(q_{k+1}))$ for some $i$. By the construction of the label of $s_{k+1}(q_{k+1})$, i.e., $\labelc(q)$, we have that for our formula $\varphi\in\labelc(q')$ there exists $\terminate{}{\varphi}\in\labelc(q)$ because these are added in the label of $q$ in the $i$ step of the loop. 

Assume that for the lifted \HDA\ the \textit{first canonicity condition is broken}; i.e., consider $q\in Q_{n}$ and assume $s_{i}(q)=q'$ for which $\varphi\in\labelc(q)$ and $\start{}{\varphi}\not\in\labelc(q')$, which is the same as $\neg\start{}{\varphi}\in\labelc(q')$, or, by axiom \ref{ax_modal3}, $\startUniv{}{\neg\varphi}\in\labelc(q')$. We again take cases after $q$. 

Consider that $q\in\modelH$ and $q'$ is added by the function \lstinline[mathescape=true,escapechar=\$]!addSourceMap!. This may be done either in the first or in the second loop, but in any of the cases the construction of the labels ensures that if $\varphi\in\labelc(q)$ then $\start{}{\varphi}\in\labelc(q')$. The same holds for the case when both $q$ and $q'$ are newly added by the second call to \lstinline[mathescape=true,escapechar=\$]!addSourceMap! (in the second loop). 

Consider the case when both $q$ and $q'$ are newly added by the first call to the function \lstinline[mathescape=true,escapechar=\$]!addTargetMap!. Our initial assumption says that $\startUniv{}{\neg\varphi}\in\labelc(q')$ which means, by the iterative construction of the label of $q$ in the loop, that $\neg\varphi\in\labelc(q)$ which is a contradiction with our initial assumption that $\varphi\in\labelc(q)$.

By now we are sure that the labeling of $\modelH'$ is canonical. 

\end{proof}

\begin{lemma}
A finite \HDA\ \modelH\ that is potential canonical can be transformed into a canonical \HDA\ by revealing one labeling that conforms with the potential labeling; this will be canonical. Moreover the way we generate this specific labeling does not introduce defects. 
\end{lemma}

\begin{proof}
Start with the cell that has no existential constraints, but only set constraints. These being consistent sets they can be grown to an atom. Depending on this atom build the rest of the atoms s.t.\ the existential constraints are respected. This can always be done. 

In a finite \HDA\ built using the two enrich and lift constructions starting from the minimal potential canonical \HDA\ as is done in the proof of Theorem~\ref{th_completeness} there always exists a cell with no existential constraints. Order the cells wrt.\ the number of existential constraints that they have. Use this order when building the labeling.
\end{proof}

\begin{lemma}[repair lemma]\label{lemma_repair}
 For any canonical \HDA\ \modelH\ that has a defect we can build a corresponding $\modelH'$ which is canonical and does not have this defect.
\end{lemma}

\begin{proof}
Consider that the canonical \modelH\ from the statement has a defect of type D1. 
Apply the \textit{enriching construction} to \modelH\ wrt.\ the defective cell $q_{n}$ and the formula $\psi$ (where $\start{}{\psi}\in\labelc(q_{n})$). The enriching lemma ensures that the new model $\modelH'$ extends \modelH\ and is canonical. The enriched model $\modelH'$ does not have the defect that \modelH\ had.

Consider that the canonical \modelH\ from the statement has a defect of type D2. 
Apply the \textit{lifting construction} to \modelH\ wrt.\ the defective cell $q_{n}$ (for which $\terminate{}{\psi}\in\labelc(q_{n})$), to obtain, cf. lifting lemma, a canonical $\modelH'$ that extends \modelH. It is clear that the new model does not have the defect that \modelH\ had.
\end{proof}

\begin{theorem}[completeness]\label{th_completeness}
The axiomatic system of Table~\ref{table_HDMLaxioms} is complete; i.e., $\forall\varphi:\ \models\varphi\ \Rightarrow\ \ \prove\varphi$.
\end{theorem}

\begin{proof}
Using the truth lemma \ref{lemma_truth_Pseudo} for pseudo canonical and saturated \HDAs, the proof amounts to showing that for any consistent formula $\varphi$  we can build a pseudo canonical saturated  $\modelH_{\varphi}$ that has a cell labeled with an atom that contains $\varphi$. We construct $\modelH_{\varphi}$ in steps starting with $\modelH_{\varphi}^{0}$ which contains only one cell $q_{0}^{0}$ of dimension $0$. The construction is done in two stages: in the first stage we label the cells with constraints (i.e., we use a potential labeling); and in the second stage we explicit these constraints into corresponding atoms (i.e., we transform the potential labeling into a real labeling).  
The first stage builds the actual \emph{finite} \HDA, $\modelH_{\varphi}$, labeling it with a potential canonical labeling, striving to repair all the defects in the constraints of the cells. The final $\modelH_{\varphi}$ is defect-free. Any finite $\modelH_{\varphi}$ has a cell which will have no existential constraints. We start from this cell to explicit the potential labeling into atoms for each cell. During this second phase only the labels of the $\modelH_{\varphi}$ are affected; i.e., they are transformed into atoms consistent with the potential labeling. This construction does not destroy the property of pseudo canonicity of the model $\modelH_{\varphi}$ that we started with. Moreover, it does not introduce defects. Therefore, in the end we are left with the finite, defect-free and pseudo canonical \HDA\ that we were looking for, where the label of the initial cell contains the initial formula.

Start by labeling $q_{0}^{0}$ with set constraints containing $\varphi$ and all other formulas it implies, i.e., $\labelPotential(q_{0}^{0})=\{\varphi\}\cup\{\psi\mid \varphi\imply\psi\}$. Trivially, $\modelH_{\varphi}^{0}$ is canonical, hence also pseudo canonical and the potential labeling is potential canonical. For each defect in the potential label $\labelPotential(q_{0}^{0})$, i.e., in the set constraints, we apply the repair lemma to obtain a new \HDA\ which does not contain the repaired defect, extends the old defective \HDA, does not introduce new defects into the just repaired potential label $\labelPotential(q_{0}^{0})$ (it may introduce new defects in the new cells), and is pseudo canonical. The algorithm continues repairing $\labelPotential(q_{0}^{0})$ until all defects are removed. It then continues to repair the new cells in the order that they were added, also respecting the order given below.
Note that any atom that is consistent with $\{\varphi\}$ is also consistent with $\{\psi\mid \varphi\imply\psi\}$.

The cells used to construct our model are picked (in the right order) from the following sets  $S_{i}=\{q_{i}^{j}\mid j\in\omega\}$ where $i\in\omega$ corresponds to the dimension $i$. Any of these cells may have defects and thus, we list all the defects, i.e., all the cells, and try to repair them in increasing order (i.e., we treat first defects on level $0$ and continue upwards).
\end{proof}

\begin{theorem}[completeness]
 The axiomatic system of Table~\ref{table_HDMLaxioms} is complete. Formally $\forall\varphi:\ \models\varphi\ \Rightarrow\ \ \prove\varphi$.
\end{theorem}

\begin{proof}
Using the truth lemma \ref{lemma_truth}, the proof amounts to showing that for any consistent formula $\varphi$  we can build a canonical saturated  $\modelH_{\varphi}$ that has a cell labeled with an atom that contains $\varphi$. We construct $\modelH_{\varphi}$ in steps starting with $\modelH_{\varphi}^{0}$ which contains only one cell $q_{0}^{0}$ of dimension $0$ labeled with an atom containing $\varphi$, i.e., $\labelc(q_{0}^{0})=A_{\varphi}$. Trivially, $\modelH_{\varphi}^{0}$ is canonical. The cells used to construct our model are picked (in the right order) from the following sets  $S_{i}=\{q_{i}^{j}\mid j\in\omega\}$ where $i\in\omega$ corresponds to the dimension $i$. Any of these cells may have defects and thus, we list all the defects, i.e., all the cells, and try to repair them in increasing order (i.e., we treat first defects on level $0$ and continue upwards).

At some step $n\geq 0$ in the construction we consider $\modelH_{\varphi}^{n}=(Q^{n},\overline{s^{n}},\overline{t^{n}},l^{n})$ canonical. If $\modelH_{\varphi}^{n}$ is not saturated then pick the smallest defect cell of $\modelH_{\varphi}^{n}$. For a D1 defect, i.e., a cell $q_{k}\in Q_{k}$ and formula $\start{}{\psi}\in\labelc(q_{k})$, apply \lstinline[mathescape=true,escapechar=\$]!enrich(k,$q_{k}$,$\psi$)! and obtain a model $\modelH_{\varphi}^{n+1}$ which is canonical, cf. Lemma~\ref{lemma_enriching}, and does not have the D1 defect, cf. Lemma~\ref{lemma_repair}. For a D2 defect apply the lifting construction to remove the defect. Moreover, any repaired defect will never appear in any extension model, independent of how many times we apply the enriching or lifting constructions.
Both enriching and lifting pick their new cells from $S$ in increasing order.
We obtain $\modelH_{\varphi}$ as a limit construction from all the $\modelH_{\varphi}^{n}$; i.e., $\modelH_{\varphi}=(Q,\overline{s},\overline{t},l)$ as $Q=\bigcup_{n\in\omega}Q^{n}$,\ \  $\overline{s}=\bigcup_{n\in\omega}\overline{s^{n}}$,\ \  $\overline{t}=\bigcup_{n\in\omega}\overline{t^{n}}$,\ \  $l=\bigcup_{n\in\omega}l^{n}$.
\end{proof}

\end{document}